\ifx \isEmbedded \undefined

\usepackage{amssymb}
\usepackage{latexsym}
\usepackage{amsmath} 
\usepackage{subfig}
\usepackage{graphicx}
\usepackage{wrapfig}
\usepackage{verbatim}
\usepackage{chapterfolder}
\usepackage{stmaryrd}
\usepackage{algorithm}
\usepackage[noend]{algpseudocode}
\usepackage{pdfpages}

\usepackage[pdftex,hypertexnames=false,linktocpage=true]{hyperref} 
\hypersetup{colorlinks=true,linkcolor=black,anchorcolor=black, citecolor=black}

\newcommand{\trig}{e_{\mathtt{t}}}
\newcommand{\OR}{{\sf OR} }
\newcommand{\ORi}{\mbox{$\mathsf{OR}$} }
\newcommand{\SR}{{\sf SR}}
\newcommand{\SRi}{\mbox{$\mathsf{SR}$}}

\newcommand{\occurs}[1]{{\sf occurs}(#1)}
\newcommand{\occurred}[1]{{\sf occurred}(#1)}

\newcommand{\ndevent}[1]{{\tt ND}(#1)}
\newcommand{\ndocc}[1]{(\occurred{#1}\wedge\ndevent{#1})}

\newcommand{\vphi}{\ensuremath{\varphi}}
\newcommand{\sat}{\vDash}
\newcommand{\nsat}{\nvDash}
\newcommand{\angles}[1]{\ensuremath{\langle{#1}\rangle}}

\newcommand{\Rrep}{{\cal R}}

\newcommand{\Rmax}{{\Rrep^{max}}}
\newcommand{\fut}{\mathsf{fut}}

\newcommand{\gammax}{\gamma^{\mathsf{max}}}

\newcommand{\imp}{\rightarrow}

\newcommand{\necaff}{\mathsf{\Box aff}}

\newcommand{\necunaff}{\mathsf{\Box unaff}}

\newcommand{\toto}{xxx}
\newenvironment{proofT}{\noindent{\bf Proof }}
{\hspace*{\fill}$\Box_{Theorem~\ref{\toto}}$\par\vspace{3mm}}
\newenvironment{proofL}{\noindent{\bf Proof }}
{\hspace*{\fill}$\blacksquare$\par\vspace{3mm}}

\newtheorem{example}{Example}
\newtheorem{definition}{Definition}
\newtheorem{theorem}{Theorem}
\newtheorem{lemma}{Lemma}

\newcommand{\heb}[1]{\begin{small}\R{#1}\end{small}}

\newcommand{\only}[2]{#2}
\newcommand{\uncover}[2]{#2}
\newcommand{\splus}{\ensuremath{\! + \!}}
\newif\iflazytikz 
\lazytikztrue 
\iflazytikz
	\usepackage{graphicx} 
	\usepackage{tikzexternal}
\else
	\usepackage{tikz}
	\ifx \isEmbedded \undefined
		\input{../tikz_settings}
	\else
		\input{tikz_settings}
	\fi
	\tikzset{external/force remake}	
\fi

\def \cfcurrentfolder{./}

\title{Knowledge and Coordination - version 1}
\begin{document}
\maketitle

\else
	
\fi
\label{ch:coordination}

\section{Studying Coordination via Response Problems}\label{sec:coord_via_response}
Existing literature about causality in distributed systems  deals almost exclusively
with asynchronous systems. This is not really surprising, given that a major
application for Lamport's happened-before is in providing some sort of
synchronized layer on top of asynchronous communication networks, and that in
synchronous systems such a layer is already provided for.

Yet in distributed systems, being able to share a certain global sense of time
is, in most cases, only a means to an end, the real purpose being coordinating
events across remote sites. Is the coordination of events in a synchronous
system as easy as looking at the clock? In some cases yes. Consider a simple
system where \emph{Zoe} and \emph{Xerxes} operate based on a prearranged protocol
that ensures that \emph{Zoe} will pick \emph{Xerxes} up for the movies at
\emph{7:45}. Come \emph{7:45}, \emph{Xerxes} looks at the clock, get's up and
goes outside and into \emph{Zoe}'s car that had just come by.

At other times though, a global clock is not enough to ensure coordination. If
\emph{Zoe}'s arrival hinges upon her getting through all work meetings by
\emph{6:50} (an unpredictable occurrence), then \emph{Xerxes} may find himself
alone at \emph{7:45}. In synchronous systems that allow for nondeterministic occurrences,  a
global clock cannot by itself ensure  proper coordination across sites, if all
events hinge upon some initial nondeterministic occurrence as a trigger. And yet
many coordination tasks depend upon  external input, whether in the form
of timing or of an assignment to an unknown parameter, as a trigger. Such external
input is, for all practical purposes, nondeterministic.

We capture the essence of such coordination tasks in the following manner. 
We identify a particular spontaneous external input  as a {\em triggering
event}, denoted by~$\trig$. A run in which the trigger~$\trig$ occurs is said to
be {\em triggered}. An intended {\em response} to such a trigger is specified by
a pair $\alpha_h=\angles{a_h,i_h}$ with~$a_h$ being an action for
process~$i_h$.\footnote{For simplicity, we assume that~$\trig$ happens at most
once in any given run, as do each one of the actions performed in response to
it.} A response $\alpha_h$ takes place if process~$i_h$ performs the
action~$a_h$. An instance of the Ordered Response problem is parametrized by a
tuple $\angles{\trig,\alpha_1,\ldots,\alpha_k}$, consisting of a trigger and a
sequence of responses. Formally, we define the following class of problems. 

\begin{definition}[Ordered Response] \label{def:OR}
	A protocol~$P$ solves the instance
	$\ORi=\angles{\trig,\alpha_1,\ldots,\alpha_k}$ of the {\em Ordered Response}
	problem if it guarantees that
	\begin{enumerate}
		\item in a triggered run, every response $\alpha_h$, for $h=1,\ldots,k$,
		will occur; moreover, if $h<k$ then $\alpha_h$ will happen before (i.e., no
		later than) $\alpha_{h+1}$ does. Finally, 
		\item  none of the responses $\alpha_h$ occur in runs that
		  are not triggered. 
	\end{enumerate}
\end{definition}

Consider the following simplified scenario, where such a problem is implied.

\begin{example}
\label{ex-intro}
Charlie's bank account is temporarily suspended due to credit problems. Should
Charlie make a sufficient deposit at his local branch, Banker Bob at
headquarters will re-activate the account. Alice holds a cheque from Charlie,
but trying to cash it before the account is re-activated will grant her a fine,
rather than cash. 

Alice, Bob and Charlie can communicate over a communication network as depicted
in Figure~\ref{fig:net1}, where the labels represent maximal transmission
times. In particular, messages from Charlie to Bob and Alice take up to 10 and
12 days to be delivered, respectively. This scenario can be viewed as an
instance of \OR\ in which a deposit by Charlie is the triggering event, and the
responses are the account re-activation by Bob followed by Alice's cashing of
the cheque.

\begin{figure}
	\centering
	\subfloat[The network of Example~\ref{ex-intro}]{
		\label{fig:net1}
		\newcommand{\tikzfname}{tikz_synch_net}
		\tikzsetnextfilename{\cfcurrentfolder Figures/\tikzfname}
		\cfinput{Tikz/\tikzfname}
	}
	\subfloat[Bob is notified by Alice]{
		\label{fig:run-intro}
		\newcommand{\tikzfname}{tikz_synch_chain3}
		\tikzsetnextfilename{\cfcurrentfolder Figures/\tikzfname}
		\scalebox{0.75}{\cfinput{Tikz/\tikzfname}}	
	}
	\caption{Example~\ref{ex-intro}}
	\label{fig:ex-intro}
\end{figure}
\end{example}

Intuitively, we expect \emph{Alice, Bob} and \emph{Charlie} to communicate in
order to ensure a proper ordering of events. Indeed, as shown by Chandy and
Misra, if the network were \emph{asynchronous} there would be no other way to
ensure correct ordering except by a message chain linking \emph{Charlie} to
\emph{Bob}, and then to \emph{Alice}. The synchronous system of
Example~\ref{ex-intro} offers more freedom. For example, Figure~\ref{fig:run-intro} shows
\emph{Charlie} sending word to \emph{Alice}, who sends on the message to
\emph{Bob} and waits for 5 rounds to make sure of its arrival before safely
cashing her check. We use a dashed arrow to denote that a message sent by Alice
at $t+3$ is sure to arrive at Bob's by $t+8$.

The Ordered Response problem captures a natural coordination scenario, and its
precise specification provides us with a clearly defined scope within which
synchronous causality can be investigated. Chapter~\ref{ch:NKG} studies it
further and establishes the exact scope of  flexibility in communication that is
allowed, for establishing ordering such synchronous systems. As we will see in
the following sections of this chapter, knowledge plays a key role in providing
such characterizations.

We now turn to define another type of coordination problem.
In synchronous systems it is often desirable to perform actions
simultaneously at different sites, a classic example being the
\emph{firing squad} problem \cite{FiringSq}.
A natural variant of \OR\ is the {\em Simultaneous Response} problem,
defined as follows.
\begin{definition}[Simultaneous Response]
	Let $\trig$ be an external input.
	Then $\SRi=\angles{\trig,\alpha_1,\ldots,\alpha_k}$ defines an instance of the
	{\em Simultaneous Response} problem.
	A protocol solves the instance $\SRi$ if it guarantees that if the triggering event
	$\trig$ occurs, then at some later point all actions
	$\alpha_1$,\ldots,$\alpha_k$ in the response set of~$\SRi$ will be
	performed simultaneously.
\end{definition}

A causal analysis of the Simultaneous Response problem will be conducted in
Chapter~\ref{ch:CKG}. Note that the Simultaneous Response problem can be
characterized by means of multiple ordering response problems. Let
$\SRi=\angles{\trig,\alpha_1,\ldots,\alpha_k}$. For each pair of required
simultaneous responses $\alpha,\beta \in \{\alpha_1,\ldots,\alpha_k\}$, we define
two ordering response problems: $\ORi^{\alpha\beta}=\angles{\trig,\alpha,\beta}$
and $\ORi^{\beta\alpha}=\angles{\trig,\beta,\alpha}$. A protocol that solves
both $\ORi^{\alpha\beta}$ and $\ORi^{\beta\alpha}$ will solve the simultaneous
subproblem $\SRi^{\alpha\beta}=\angles{\trig,\alpha,\beta}$. A protocol that
solves both  $\ORi^{\alpha\beta}$ and $\ORi^{\beta\alpha}$ for every
$\alpha,\beta \in \{\alpha_1,\ldots,\alpha_k\}$ will solve $\SRi$.

Nevertheless, defining the simultaneous requirement as a separate problem is
worthwhile because solutions to such
problems give rise to tighter epistemic characterizations, in the form of common
knowledge among the set of responding processes.
Chapters~\ref{ch:ordered_group} and~\ref{ch:generalized_ordering} will investigate aspects of generalizing
the  ordering requirements beyond ordered and simultaneous responses.

\section{Knowledge Gain in Synchronous Systems}\label{sec:KG_in_synch}
As noted in Chapter~\ref{ch:model}, Chandy and Misra used \emph{knowledge gain}
to refer to a  scenario wherein process $i$ ``gains knowledge'',
or ``learns'', that some fact $\vphi$ pertaining to process $j$ holds. In an
asynchronous system, as it turns out, the only way for $i$ to gain knowledge
about $j$ is by means of a message chain relating the two. Thus, in the
asynchronous setting, knowledge gain reflects the way information flows in the
system.

Taken at face value, Chandy and Misra's notion no longer captures information flow when
we move to synchronous systems, as process $i$ may learn facts about process $j$
by a mere glance at the clock (for example, by noting that my watch shows
\emph{4pm} I ``learn'' that your watch shows \emph{4pm} too right now, something that I
did not know before, while it was still \emph{3:59}).
In order to maintain the desirable association with information flow we turn, as we did in the 
previous section, to nondeterministic occurrences. 

\begin{definition}[Knowledge gain in synchronous systems]
	We will say that knowledge gain occurs in the interval $[t..t']$ of run $r$
	whenever  a nondeterministic event occurs at some process $j$ no sooner than
	time $t$, and process $i$ knows of this occurrence by time $t'$.
\end{definition}•

At an intuitive level, we expect that knowledge of an ND (nondeterministic)
event is dependent upon communication, and hence knowledge gain of such facts
will reflect information flow. The following chapters will pursue the relations
between knowledge gain and communication. Natural generalizations of the above
notion of knowledge gain are \emph{nested knowledge gain} and \emph{common
knowledge gain}.

Nested knowledge gain occurs within the interval $[t..t']$ of run $r$ if an ND event
$e$ occurs at process $j$ within the interval, and by the end of the interval
$K_{i_{h}}K_{i_{h-1}}\cdots K_{i_{1}}\occurred{e}$ holds for some sequence
of processes $\angles{i_{1},i_{2},..,i_{h}}$. On a similar vein, common
knowledge gain occurs when, by the end of the interval of time,
$C_{G}\occurred{e}$ holds for some group of processes $G$.

The suggestive similarity of nested and common knowledge to
the Ordered and Simultaneous Response problems respectively,  will be examined in 
Section~\ref{sec:response_knowledge}.

\section{Relating Response Problems and Knowledge Gain}\label{sec:response_knowledge}
This section charts out the formal relations between the two previously defined response
problems, and knowledge gain. 

\subsection{Response Problem to Knowledge Gain}
We start by looking at the Ordered Response problem.
Intuitively, the coordination of responses so that they occur  in a particular
sequence suggests that  knowledge gain is involved.
In order to ensure that the events occur in sequence,  each responder $k$ must know that 
the previous $k-1$ responses, as well as the trigger event, had already occurred 
(or are occurring right now). 

We now show that this is indeed the case. A caveat concerning knowledge is
that we must assume of processes that they do not forget that they
had already performed a response. Formally, this property is described as follows. 

\begin{definition}[Response recall]\label{def:response_recall}
	Let \ORi be defined by $\angles{\trig,\alpha_1,..,\alpha_k}$ and assume that
	$\Rrep=\Rrep(P,\gamma)$ is a system of runs for a protocol $P$ where all of
	the responses may occur, and $\gamma$ is any arbitrary context. Protocol
	$P$ recalls responses for $\ORi$ if for all $1\leq h \leq k$,  $r,r'$ in 
	$\Rrep$ and $t'\leq t$,  if $\alpha_h$ occurs at $(i_h,t')$ in $r$ and
	$(r,t)\sim_i(r',t)$, then $\alpha_h$ also occurs at $(i_h,t')$ in $r'$.
\end{definition} 

As we'll show in Section~\ref{sec:K2sync}, the response recall assumption is not
needed when Ordered Response is related directly to communication, rather than
to knowledge. Note that the following theorem, and Theorem~\ref{thm:SRck} as
well, are not dependent on the particular context $\gamma$ being used.

\begin{theorem}
\label{thm:ORnestedK}
	Let \ORi=$\angles{\trig,\alpha_1,\ldots,\alpha_k}$ be an instance of \OR,
	and assume that protocol~$P$ solves~\ORi in~$\gamma$ and that it recalls responses for it.
	Let $r\in\Rrep$ be a run in which~$\trig$ occurs, let $1\le h\le k$,
	and let $t_h$ be the
	time at which $i_h$ performs action~$a_h$ in~$r$. Then
	\[(\Rrep,r,t_h)\sat~K_{i_h}K_{i_{h-1}}\cdots K_{i_1} \ndocc{\trig}.\]
\end{theorem}

\renewcommand{\toto}{thm:ORnestedK}
\begin{proofT}
	We prove the theorem by induction on $k$.
	\begin{itemize}
		\item[] $\mathbf{k=1:}$ By definition of $r$, process $i_1$ performs
		$a_1$ at $t_1$. If $(\Rrep,r,t_1)\not\sat K_{i_1}\occurred{\trig}$ then
		by definition of $\sat$ there exists a run $r'\in \Rrep$ such that
		$(r,t_{1})\sim_{i_{1}}(r',t_{1})$ and where $(\Rrep,r',t_1)\not\sat
		\occurred{\trig}$. Yet as $P$ solves \ORi, in $r'$ action $a_1$ is
		performed only if $\trig$ has occurred, contradiction. Therefore it must
		be that $(\Rrep,r,t_1)\sat K_{i_1}\occurred{\trig}$.

		By definition of $\OR$,  $\trig$ is an external input event and hence
		nondeterministic. This is universally true in the system, and hence
		$(\Rrep,r,t_1)\sat K_{i_1}\occurred{\trig}$ implies $(\Rrep,r,t_1)\sat
		K_{i_1}\ndocc{\trig}$.

		\item[] $\mathbf{k>1:}$ Suppose that it is the case that
		\vspace{-1mm}		
		\[(\Rrep,r,t_k)\not\sat~K_{i_k}K_{i_{k-1}}\cdots K_{i_1}
		\ndocc{\trig}.\] Then by definition of $\sat$ there exists a run $r'\in
		\Rrep$ such that $(r,t_{k})\sim_{i_{k}}(r',t_{k})$ and where
		$(\Rrep,r',t_k)\not\sat~K_{i_{k-1}}K_{i_{k-2}}\cdots K_{i_1}
		\ndocc{\trig}$. Since the protocol recalls responses, we now obtain that
		\vspace{-1mm}		
		\[(\Rrep,r',t_{k-1})\not\sat~K_{i_{k-1}}K_{i_{k-2}}\cdots K_{i_1}
		\ndocc{\trig}.\]

		However, again as $P$ solves \ORi, it also solves the sub-problem \ORi'
		defined by $\angles{\trig,\alpha_1,\ldots,\alpha_{k-1}}$. As
		$\alpha_{k}$ is also performed by $i_k$ in $r'$, it must be that
		$\alpha_{1}..\alpha_{k-1}$ too get performed in $r'$. By the inductive
		hypothesis we get that for all $1\le h\le k-1$
		\vspace{-1mm}		
		\[(\Rrep,r',t_{h})\sat~K_{i_h}K_{i_{h-1}}\cdots K_{i_1} \ndocc{\trig}.\]
		In particular we get that
		$$(\Rrep,r',t_{k-1})\sat~K_{i_{k-1}}K_{i_{k-2}}\cdots K_{i_1}
		\ndocc{\trig}.$$ This contradicts the previous result, and therefore it
		must be the case that 
		\vspace{-1mm}		
		\[(\Rrep,r,t_h)\sat~K_{i_h}K_{i_{h-1}}\cdots
		K_{i_1} \ndocc{\trig}\] for all  $1\le h\le k$, as required.
	\end{itemize}
\end{proofT}

We now turn to consider the Simultaneous Response problem. Here too there is an intuitive
connection with knowledge gain. If all responses are always performed simultaneously,
then every responding site must know that the other sites are responding too. Yet as
the next theorem shows, the simultaneous response requirement implies an even stronger
epistemic condition, in the form of common knowledge. As all responses must occur
simultaneously, there is actually no need here to assume the response recall property from the 
arbitrary protocol.

\begin{theorem}\label{thm:SRck}
	Let $\SR=\angles{\trig,\alpha_1,\ldots,\alpha_k}$,
	and assume that protocol~$P$ solves~$\SR$ in~$\gamma$.
	Moreover, let $G=\{i_1,\ldots,i_k\}$ be the set of processes appearing
	in the response set of~$\SR$.
	Finally, let $r\in\Rrep$ be a run in which~$\trig$ occurs,
	and let $t$ be the time at which the response actions are performed
	in~$r$. Then ~~$(\Rrep,r,t)\sat~C_G \ndocc{\trig}.$
\end{theorem}

\renewcommand{\toto}{thm:SRck}
\begin{proofT}
	Fix $h,g\in\{1..k\}$. We first show that 
	$$\Rrep\sat \occurs{a_{h}}\imp E_G(\occurs{a_{h}}) \wedge \occurred{\trig}).$$ 
	Choose $r',t'$ such that $(\Rrep,r',t')\sat \occurs{a_{h}}$. Note that since $P$ 
	solves $\SR$ and since response actions are performed only upon the occurrence
	of the trigger event $\trig$, we get $(\Rrep,r',t')\sat \occurs{a_{h}}\leftrightarrow \occurs{a_{g}}$ 
	and $(\Rrep,r',t')\sat \occurs{a_{h}}\imp \occurred{\trig}$. From the former equivalence 
	and $(\Rrep,r',t')\sat \occurs{a_{h}}$ we 
	obtain that $(\Rrep,r',t')\sat \occurs{a_{g}}$. Since performing a local action is written, at least 
	for the current round, in the process's local state,  
	we obtain that $(\Rrep,r',t')\sat K_{i_g}\occurs{a_{g}}$. Now using  the former equivalence 
	again we get that $(\Rrep,r',t')\sat K_{i_g}\occurs{a_{h}}$, and using the latter implication 
	we get $(\Rrep,r',t')\sat K_{i_g}\occurred{\trig}$. Putting these results together we conclude 
	that $(\Rrep,r',t')\sat K_{i_g}(\occurs{a_{h}}\wedge \occurred{\trig})$. Since $g$ is arbitrarily 
	chosen in $G$, we get $(\Rrep,r',t')\sat E_G(\occurs{a_{h}}\wedge \occurred{\trig})$, from 
	which it follows that 
	$$(\Rrep,r',t')\sat \occurs{a_{h}}\imp E_G(\occurs{a_{h}}\wedge \occurred{\trig})$$ 
	by our choice of $r',t'$. As false antecedents imply anything, we conclude that
	$\Rrep\sat \occurs{a_{h}}\imp E_G(\occurs{a_{h}}\wedge \occurred{\trig})$. 

    Recall the Knowledge Induction Rule, that derives $\Rrep\sat \vphi\imp C_G\psi$ 
    from $\Rrep\sat \vphi\imp E_G(\vphi \wedge \psi)$. Setting $\vphi=\occurs{a_{h}}$ 
    and $\psi=\occurred{\trig}$ we apply the rule, and based on the above result obtain 
    $\Rrep\sat \occurs{a_{h}}\imp C_G\occurred{\trig}$. We conclude by noting that 
    $(\Rrep,r,t)\sat\occurs{a_{h}}$ by assumption, and hence also $(\Rrep,r,t)\sat~C_G\occurred{\trig}$.
    
	By definition of $\SR$,  $\trig$ is an external input event and hence nondeterministic.
	This is universally true in the system, and hence in particular $(\Rrep,r,t)\sat~C_G\occurred{\trig}$
	implies $(\Rrep,r,t)\sat~C_G\ndocc{\trig}$.    
\end{proofT}

Theorems~\ref{thm:ORnestedK} and~\ref{thm:SRck} show that, in a precise sense,
knowledge gain is a prerequisite for coordinated response. The next section will
show that nested and common knowledge gain indeed characterize ordered
and simultaneous responses, in the sense that they define the minimal 
epistemic prerequisites for such types of coordination. 

\subsection{Knowledge Gain to Response Problem}
It is immediately apparent that no general law exists showing that knowledge gain implies 
a solution to a response problem, for processes are not, in general, required to act
in any way upon the knowledge that they gain. 

In order to bridge the 
gap between knowledge and action, we would need to add requirements on the protocol
being followed by the processes. Since, as stated in Section~\ref{sec:role_of_knowledge},
this thesis is focussed on producing protocol-independent results, we do not delve deeply
into such additions. \footnote{One could try, for example, to characterize those protocols
where knowledge gain \emph{does} imply a solution to the related response problem.} 

Nevertheless, as we will show, there exist protocols where indeed knowledge gain
implies a solution to the response problem. Proving the existence of such a
protocol comes to show that the relevant response problem (say ordered
response), is indeed characterized by the related type of knowledge gain (in
this case, nested knowledge gain). The existence of such a protocol shows that,
in general, no epistemic state stronger than nested knowledge can be gained as a
result of solving the ordering response problem. \footnote{A similar argument is
presented in Sections~\ref{sec:suff_NKG} and~\ref{sec:suff_CKG}, in order to
show that the yet to be defined centipede and centibroom communication patterns
characterize nested and common knowledge gain, respectively.}

To exemplify the existence of protocols where nested knowledge gain implies a solution to $\OR$,
we introduce the following property for protocols.

\begin{definition}[Non-Hesitant Protocol] \label{def:non-hesitant}
	Protocol $P$ is \emph{non-hesitant} with respect to an Ordered Response problem 
	$\OR=\angles{\trig,\alpha_1,\ldots,\alpha_k}$ if for each $h\leq k$, process $i_h$ performs 
	$\alpha_h$ as soon as $K_{i_h}K_{i_{h-1}}\ldots K_{i_1}\ndocc{\trig}$ is established, but no sooner.
\end{definition}

As the following lemma shows,  nested knowledge gain is sufficient for 
solving \OR problems in protocols that are non-hesitant with respect to the problem.
As for the existence of actual protocols that comply with the above definition, we can
make things easy by assuming the context $\gammax$ (as we do in the next two
chapters). Here it is easy to find protocols that satisfy \emph{non-hesitance}, as well 
as \emph{consideracy} (defined below) qualifications by insisting that processes timestamp their messages.

\begin{lemma}\label{lem:suffOR}
    Let $\OR=\angles{\trig,\alpha_1,\ldots,\alpha_k}$ and let $P$ be a non-hesitant 
    protocol with respect to $\OR$. If for every $r\in\Rmax=\Rrep(P,\gammax)$ 
    such that $\trig$ occurs at $(i_0,t)$ in~$r$
    there exists time $t'$ such that 
    $$(\Rmax,r,t')\sat K_{i_k}K_{i_{k-1}}\ldots K_{i_1}\ndocc{e},$$
    then $P$ solves $\OR$.
\end{lemma}

\begin{proofL}
    Assume $r\in\Rmax$ such that $\trig$ occurs at $(i_0,t)$ in~$r$. From 
    $$(\Rmax,r,t')\sat K_{i_k}K_{i_{k-1}}\ldots K_{i_1}\ndocc{e}$$ we obtain the existence of 
    $t_k\leq t'$ such that 
    \begin{itemize}
		\item[(a)] $(\Rmax,r,t_h)\sat K_{i_k}K_{i_{k-1}}\ldots K_{i_1}\ndocc{e}$, and
		\item[(b)] $t_k=0$ or $(\Rmax,r,t_k-1)\nsat K_{i_k}K_{i_{k-1}}\ldots K_{i_1}\ndocc{e}$.
	\end{itemize}
    
    By repeated applications of the Knowledge Axiom we extend this result into a series  
    $t\leq t_1 \leq \cdots \leq t_k \leq t'$ such that for every $h\leq k$
    \begin{itemize}
		\item[(a)] $(\Rmax,r,t_h)\sat K_{i_h}K_{i_{h-1}}\ldots K_{i_1}\ndocc{e}$, and 
		\item[(b)] $t_h=0$ or $(\Rmax,r,t_h-1)\nsat K_{i_h}K_{i_{h-1}}\ldots K_{i_1}\ndocc{e}$.
	\end{itemize}
    
    As $P$ is non hesitant with respect to $\ORi$, we get that for every $h\leq k$ process 
    $i_h$ performs $\alpha_h$ at $t_h$, and we are done.
\end{proofL}

Once again switching to the Simultaneous Response problem, we need a protocol where
processes are more considerate, in order to ensure a solution to the problem.

\begin{definition}[Considerate protocol]\label{def:considerate}
	Protocol $P$ is \emph{considerate} with respect to a Simultaneous Response problem 
	$\SR=\angles{\trig,\alpha_1,\ldots,\alpha_k}$ if for each $h\leq k$, process $i_h$ performs 
	$\alpha_h$ as soon as $C_{i_{1},\ldots,i_{k}}\ndocc{\trig}$ is established, but no sooner.
\end{definition}• 

\begin{lemma}\label{lem:suffSR}
    Let $\SR=\angles{\trig,\alpha_1,\ldots,\alpha_k}$, let
    $G=\angles{i_{1},\ldots,i_{k}}$ and let $P$ be a considerate protocol with
    respect to $\SR$. If for every $r\in\Rmax=\Rrep(P,\gamma)$ such that $\trig$
    occurs at $(i_0,t)$ in~$r$ there exists time $t'$ such that
    $(\Rmax,r,t')\sat C_{G}\ndocc{e}$, then $P$
    solves $\OR$. 
\end{lemma}

The lemma's proof is immediate if we consider that  $C_{G}\vphi \leftrightarrow
K_{g}C_{G}\vphi$ for any $g\in G$.

\ifx \isEmbedded \undefined
\end{document}
\else
\fi


\maketitle

\else
\fi
\label{ch:appendix}

\section{Technical Specification for the Context $\gamma^{bounded}$}\label{app:tech_model}
We now present  the components of the context $\gamma^{bounded}$, at the most explicitly
technical level. A synchronous context $\gamma^{bounded}$ is defined as a tuple
$({\cal G}_0,P_e^{\,\mathsf{s}},\tau)$ where

\paragraph{The environment's state}
Recall that the environment's state keeps track of relevant aspects of
the global state that are not represented in the local states of the
processes. We assume that the environment's state has
three components $\ell_e=(\mathsf{Net},\mathsf{t},\Hist_e, \mathsf{Rate})$, where
\begin{enumerate}
\item $\mathsf{Net}$ is a labelled graph $(\Proc,E,max,min)$ describing the network topology
and bounds on transmission times.
Its nodes are processes, and a directed edge $(i,j)\in E$
captures the fact that there is a channel from~$i$ to~$j$ in the
system. Moreover, the labels $1\le max_{ij}\in\Nat\cup \{\infty\}$ 
and $1\le min_{ij}\in\Nat$ are upper and lower bounds respectively on the
time that a message sent on $(i,j)$ can be in transit. 
The contents of $\mathsf{Net}$ are not affected by~$\tausynch$, and so $\mathsf{Net}$
remains constant throughout the run.

\item The variable~$\mathsf{t}$ keeps track of global time. As we shall see
its value starts at $\mathsf{t}=0$, and advances by~1 following each
round. \textbf{In contexts where the property $global$ holds}, this is a notion of global time 
because we assume that processes have
access to its value as described below. Finally,
\item $\Hist_e$ records the
sequence of joint actions performed so far.
There is no message loss in
$\gamma^{bounded}$, so the $\Hist_e$ component
uniquely determines the contents of all channels.
Indeed, a message $\mu$ is {\em in transit} at a given
global state~$g$ if the $\Hist_e$ component in~$g$ records that $\mu$
has been sent, and does not record its delivery.
\item $\mathsf{Rate}\leq \infty$ is a constraint to which the scheduler 
adheres in the following manner: every process is activated by the scheduler at least once
every  $\mathsf{Rate}$ rounds.
\end{enumerate}

\paragraph{process local states}
We assume that local states have
two components $\ell_i=(\mathsf{t}_i,\mathsf{data}_i)$,
where $\mathsf{t}_i$ is a local counter that is incremented by the transition function
$\tau$ whenever process $i$ is activated by the scheduler. 
The component $\mathsf{data}_i$ serves as the data segment for the process $i$. 
Its contents are a function of the protocol $P$ and the transition function $\tau$. 
\textbf{In contexts where the property $global$ holds}, process local states hold a third
component $\mathsf{Net}_i$, that is a copy of global state's communication graph.

\paragraph{The set ${\cal G}_0$ of initial global states}
We assume that associated with $\gamma^{bounded}$ there is a set~$\Init_i$
of possible initial states for each process~$i\in\Proc$.
We define ${\cal G}_0$ to be the set of global states
$g=(\ell_e,\ell_1,\ldots,\ell_n)$ satisfying:
(1)  $\mathsf{t}_i=0$ for all $i\in\Proc$ and $\mathsf{t}=0$;
(2) for every $i\in\Proc$, ~$\hist_i=\langle\init_i\rangle$,with
$\init_i\in\Init_i$; and
(3) $\Hist_e$ is the empty sequence.

\paragraph{Actions and external inputs}
Associated with the context $\gamma^{bounded}$ are sets $\Intact$
 of internal actions for the processes and sets  $\ExIn$ of external inputs,
respectively. For ease of exposition we assume that $\bot\in\ExIn$,
where $\bot$ stands for the empty external input. Moreover, we
generally assume that $\ExIn\ne\{\bot\}$, so that there is at least one
nontrivial possible external input.
We assume that processes can perform {\sf send} actions and internal
actions. The local action $\sfa_i(k)$ that~$i$
contributes to the joint action in
round~$k+1$ consists of a finite sequence of distinct send and
internal actions.  (Recall that the local action is determined by the
protocol, based on the local state.)
We use external inputs to model spontaneous events. They are generated
by the environment. In addition to external inputs, the environment
is in charge of message delivery. Thus, the environment's action
$\sfa_e(k)$ consists of a finite sequence of external inputs to be delivered to
various individual processes, a subset $\pi$ of $\Proc$ that are activated in the current round,
and a (possibly empty) set of messages
that are to be delivered in the current round.

\paragraph{The environment's protocol $P_e^{\,\mathsf{s}}$}
The environment in $\gamma^{bounded}$ is in charge of delivering external
inputs to processes, activating certain processes for the round and determining message deliveries.
We define $P_e^{\,\mathsf{s}}(g)$ to be the set of actions
$\sfa_e=(\sigma_x,\sigma_d, \pi)$ such that
\begin{enumerate}
\item $\sigma_x\in(\Proc\times\ExIn)$ is a sequence assigning to each
  process $i\in\Proc$ an external input (possibly the empty input
  $\bot$) it receives in the current round, and
\item $\sigma_d$ is a sequence $\langle M_{1}\ldots M_{|\Proc|}\rangle$
where (i) for every $i\in Proc$ the set $M_i$ consists of messages that
are in transit in~$g$,  (ii) $M_i$ contains all messages in transit to~$i$
whose transmission time bounds, as specified in~$\mathsf{Net}$, will be
violated (expire) if the message is not delivered in the current
round, and (iii) none of the messages in $M_{i}$ are such that if delivered in the current
round, will violate the existing minimal transmission time constraints.
\item $\pi\subseteq \Proc$ is the set of processes activated in the current round,
such that if there exists a message $m\in M_{i}$ such that if not delivered
in the current round will violate maximal transmission times, then $i\in \pi$.
\textbf{In contexts where the property $global$ holds}, $\pi=\Proc$ in every round.
\end{enumerate}

Notice that $P_e^{\,\mathsf{s}}$ is genuinely nondeterministic.
Exactly one of the actions in $P_e^{\,\mathsf{s}}(g)$ will be performed in
global state~$g$ in any given instance.
By definition of $\Rrep(P,\gamma^{bounded})$, however,
if $r(k)=g$ then the system contains a run extending the prefix
$r(0),\ldots,r(k)$ for every possible environment action in $P_e^{\,\mathsf{s}}(g)$.
Another point to note is that our definition does not enforce (and
hence does not assume) FIFO transmission; had we done so,
channels would be considered to be queues, and the
nondeterministic choices of messages to deliver would have to obey
FIFO order.
It should also be noted that the scheduler makes sure to comply with
all existing constraints: minimal and maximal transmission times, as well 
as process rate.  
Finally, the fact that external inputs are delivered in a
nondeterministic fashion implies they are not correlated in any way,
and they do not depend on anything that happens
before they are delivered.  This is the sense in which
external inputs can be viewed as independent, ``spontaneous'' events.

\paragraph{The transition function~$\tau$}
The transition function $\tau$ implements
the joint actions in a rather straightforward manner.
In every round:
(i) the global clock variable $\mathsf{t}$ and the local variables $\mathsf{t}_i$ of 
all $i\in \pi$ are advanced
by one;
(ii) a copy of the joint action is added to the environment's
history log $\Hist_e$; and
(iv) For every process $i$, a record of all current round message deliveries and external inputs 
to the process is written in $\mathsf{data}_i$. Note that this record is overwritten  on every round, 
so that a protocol must take special measures in order to maintain a persistent copy of these contents.

\ifx \isEmbedded \undefined


\maketitle

\else
\fi
\label{ch:model}

\section{Causal Analysis in Distributed Systems}\label{sec:lamport_analysis}
In distributed systems, a group of autonomous processes with limited means of communication
are typically set to cooperate and coordinate their local actions in order to achieve a system-wide global 
requirement. In general, the less processes know of actions and of occurrences at remote sites,
the more difficult the task of coordination becomes. 

Of particular difficulty is achieving coordination in asynchronous systems, where no guarantees are given
regarding the rate at which each process proceeds, and message delivery can be indefinitely postponed.
In a seminal paper \cite{Lamclocks}, Lamport proposed the \emph{happened-before} relation $\Lam$ between 
events in asynchronous systems, and based on this relation a mechanism for logical clocks that 
allow processes to exercise some control over the ordering of events. 

Lamport takes his lead from the theory of special relativity, where the way by
which light dissipates in space over time determines upper bounds on the spread
of information and of causality in general, as nothing can get from source to
target faster than light itself. Applying this analogy to distributed systems,
Lamport notes that in asynchronous systems, causality and information cannot
travel faster than the messages that are sent and received between the
processes.  This suggests that the following relation on events applies to $e$
and $e'$ whenever the occurrence of $e'$ is causally dependent upon occurrence of $e$.\footnote{The
formulation differs slightly from that  of \cite{Lamclocks} as we do not impose
irreflexivity, for the sake of a simpler formulation.}

\begin{definition}[Happened-before]
	Fix an execution $r$ of the system. The \emph{happened-before} relation $\Lam$ over events of $r$ is 
	the smallest relation satisfying the following  conditions:
	\begin{enumerate}
		\item If $e$ and $e'$ are events in the same process, and $e$ comes before or with  $e'$, 
		then $e\Lam e'$.
		\item  If $e$ is the sending of a message by one process and $e'$ is the receipt of the same 
		message by another process, then $e\Lam e'$.
		\item If $e\Lam e''$ and $e''\Lam e'$ then $e\Lam e'$.
	\end{enumerate}•
\end{definition}•

As messages are never delivered before they are sent, happened-before implies
that whenever $e\Lam e'$, the occurrence of event $e$ temporally precedes (or is
simultaneous with) the occurrence of $e'$, even if the events occur at distinct
sites.  Thus, a partial ordering that is implicit in every  execution of a
distributed system is made explicit.

The event ordering that is determined by the happened-before relation is
sometimes referred to as a \emph{causal ordering}. Causality is an elusive
concept whose nature has been widely contested over the centuries. Lamport's
relation circumvents these sticky philosophical issues, in the following sense.
Whenever events $e,e'$ do occur  in an execution and $e \not\Lam e'$, then event
$e$ cannot be a cause of event $e'$ under any interpretation of causality, as
its effect has not reached the site of $e'$ by the time it occurs. Note that,
strictly speaking, the converse scenario where $e\Lam e'$ \emph{does} hold can
only mean that $e$ is a \emph{potential cause} of $e'$, as the occurrence of
$e'$ may have been nondeterministic, or based on the occurrence of events other
than $e$.

Lamport offers an immediate application for causal ordering. Logical clocks are
defined as local counters ($Clock_{i}$ for each process $i$), that assign a
number to each local event. By timestamping each message sent with the current
value of the sender's counter, a simple mechanism is devised to make sure for
all events $e,e'$ occurring at sites $i,j$ respectively, that
$Clock_{i}(e)<Clock_{j}(e')$ whenever $e \Lam e'$ and $i\neq j$.

The immense import of Lamport's paper on the development of theoretical and
practical distributed systems cannot be overestimated. For us it is important
to summarize by saying that Lamport defined a relation, based on
communication patterns (the inter-process message chains that establish the
$\Lam$ relation), that  traces the dissemination of causal effect in a
system. Moreover, he  showed how this causal ordering can be used to establish a
temporal ordering on events.

Chandy and Misra's follow up paper \cite{ChM} explicitly relates Lamport's
relation to knowledge. This paper offers a reinterpretation of Lamport's ideas
in terms of knowledge, rather than of coordination. A more in-depth coverage of
knowledge in distributed systems is offered in Section~\ref{sec:framework},
while the current discussion will be kept to an intuitive level. In distributed
systems each process is immediately acquainted only with its own local state.
Thus, facts that pertain to local states of remote sites may be hidden from it.
Now consider a run (or execution) of the system where, at the current time,
process $i$'s local state is $\ell$. If, when one looks at all possible runs, an
arbitrary fact $\vphi$ holds true of the system whenever $i$'s local state is
$\ell$, then  $i$ is said to ``know'' that $\vphi$: there is simply no
way that, given its local state $\ell$, fact $\vphi$ could fail to hold. How
would process $i$ come to know that, say, the value of process $j$'s local variable $X_{j}$
is $10$?

A simple answer can be given if $X_{j}=10$ is an invariant specified by the
protocol. To filter away such ``uninteresting'' cases, what if at time $t$
process $j$ itself does not know that $X_{j}=10$, and at time $t'>t$ process $i$
knows that process $j$ knows that $X_{j}=10$?\footnote{Since $X_{j}$ is a part
of $j$'s local state, by definition process $j$ will know its value at all
times, so if $j$ doesn't know that $X_{j}=10$, it must be that $X_{j}\neq 10$.}
Chandy and Misra call such a development \emph{knowledge gain}, where  process
$i$ comes to gain new knowledge about the state of process $j$. They surmise,
and then prove, that in such a case it must be that process $j$ at time $t$ is
happened-before related to process $i$ at time $t'$. We will denote such a
relation with $(j,t)\Lam(i,t')$. More generally, Chandy and Misra show that if
at time $t'$ process $i_{k}$ knows that process $i_{k-1}$ knows that... process
$i_{1}$ knows that process $j$ knows that $X_{j}=10$, then it must be that there
are times $t=t_{0}<t_{1}< \cdots < t_{k}=t'$ such that
$(j,t_{0})\Lam(i_{1},t_{1})\Lam \cdots \Lam (i_{k},t_{k})$.

While Lamport relates communication to coordination, Chandy and Misra relate 
it to knowledge gain. In both cases the happened-before relation can be seen to
give as good a characterization as can be achieved of the spread of causal effect in the system. 
However,  once we have formalized the notions of knowledge and of coordination
with which the thesis deals, we will show in Section~\ref{sec:response_knowledge}  that
knowledge gain is a necessary condition for coordination, and thus provides
a ``closer to home'' approximation of causality than coordination.  As such, we will 
study it extensively in the thesis, with the aim of giving a precise understanding
of causality in synchronous systems.

\section{Causality, Knowledge, Coordination}\label{sec:role_of_knowledge}
Roughly sketched, the scenery drawn out by Lamport and by Chandy and Misra
for asynchronous systems shows that communication is prerequisite for 
knowledge gain and that, similarly,  knowledge gain is necessary for 
the coordinated ordering of events. These relations justify
associating Lamport's happened-before relation with causality in such settings. 

In this thesis we will investigate causality as it manifests itself in
\emph{synchronous} settings.  Example~\ref{ex-intro}, presented in the next
chapter, will show us that the happened-before relation no longer
characterizes causal relations in their entirety under synchrony. Our main goals
will be to identify the communication patterns that \emph{do} characterize
causality here.

Our method is to define various scenarios where knowledge gain, as a rigorously
defined approximate for causality, takes place. Each of the following chapters
is dedicated to such a scenario. In Chapters~\ref{ch:NKG} and~\ref{ch:CKG} we
provide the Ordered and Simultaneous Response problems as motivating leads.
Given the necessity of nested and common knowledge gain for the $\OR$ and $\SR$
problems respectively, characterizing solutions to these coordination problems
in terms of causality pretty much reduces to an analysis of knowledge gain in
such terms.

The study of causal relations leading to knowledge gain is thus relevant in 
the context  two differing  
research programmes:

\begin{itemize}
	\item•Focussing on the relations between knowledge gain and causality, we
	hope to make the thesis results instrumental in the widely defined field of
	epistemic analysis in multi agent systems. The thesis results may be
	applicable in the linguistic study~\cite{Lascarides1993}, as well as in game
	theoretic analysis of interactive epistemics~\cite{Au,Chwe2000}, and possibly also in the
	philosophical analysis of causality~\cite{Pearl,Williamson2007}.
	\item•By encompassing also the relations between knowledge and coordination,
	we relate  coordination directly to communication.  Unlike knowledge,
	coordination and communication are both tangible, and results characterizing
	one in terms of the other would be easier to apply.

	Thus, even in the context of more applicative study of distributed systems,
	knowledge based analysis can be made to play a subtle, if highly beneficial,
	role. Knowledge is a powerful tool for  extracting underlying
	generalizations in such systems and is our best approximation for causal
	phenomena. Once these generalizations have been properly characterized,
	direct connections between them can be drawn out, pretty much obsoleting the
	interpretive epistemic layer.
\end{itemize}•

One final guiding principal for the inquiries made in this thesis needs to be
mentioned. It is widely understood that different process protocols lead to
widely varying characteristics for the system as a whole. Nevertheless, our key
results are not protocol dependent, and in this sense they characterize
\emph{all} synchronous systems. We adhere to the idea of characterizing systems
rather than protocols throughout the thesis, and even where protocol specific
results are given, they bear significance for all systems (by showing that our
definitions are tight). The one exception to this guideline is made in
Chapter~\ref{ch:ignorance}, where gaining knowledge of ignorance is discussed.

\section{The Interpreted Systems Framework}\label{sec:framework}
Results pertaining to knowledge gain in distributed systems provide the main
formal backbone of this thesis. For this reason we utilize the 
\emph{interpreted systems} framework of  Fagin,
Halpern, Moses, and Vardi~\cite{FHMV}.
We shall simplify its exposition somewhat here, and review
just enough of the details to support the
formal analysis.
Essentially all of the definitions in this section are taken
from~\cite{FHMV}.

Informally, we view a multi-process system as consisting of a set
$\Proc=\{1,\ldots,n\}$  of
processes connected by a communication network.
We assume that, at any given point in time,
each process in the system is in some {\em local state}.
A {\em global state\/} is just a tuple
$g=\langle\ell_e, \ell_1, \ldots, \ell_n\rangle$
consisting of local states of the processes,
together with the state~$\ell_e$ of the {\em environment}.
The environment's state accounts for everything that is relevant to
the system that is not contained in the state of the processes.

A {\em run\/} is a function from time to global states.
Intuitively, a run is a complete description of what happens over time
in one possible execution of the system.  A {\em point\/} is a pair
$(r,t)$ consisting of a run $r$ and a time $t$.
If $r(t) = \langle\ell_e, \ell_1, \ldots, \ell_n\rangle$,
then we use $r_i(t)$ to
denote process $i$'s local state $\ell_i$ at the point $(r,t)$,
for  $i = 1,\ldots, n$,
and $r_e(t)$ to denote~$\ell_e$.  For
simplicity, time here is taken to range over the natural numbers
rather than the reals
(so that time is viewed as discrete, rather than dense or continuous).
{\em Round\/} $t$ in run $r$ occurs between time $t-1$ and $t$.

We identify a {\em protocol\/} for a process~$i$ with a function from
local states of~$i$ to nonempty sets of actions.
(We mostly consider {\em deterministic\/}  protocols,
in which each local state is mapped to a singleton set of actions.
Such a protocol essentially maps local states to actions.)
A joint protocol is just a sequence of protocols $P=(P_1,\ldots,P_n)$,
one for each process.

We generally study knowledge in runs of a given protocol~$P$ in a
particular setting of interest.  To do this, we separately describe the setting, or {\em
  context\/}, in which $P$ is being executed.
Formally, a context $\gamma$ is a tuple $(\Gz,P_e,\tau)$,
where $\Gz$ is a set of initial global states,
$P_e$ is a protocol for the environment,
and $\tau$ is a {\em transition function}.%
\footnote{Depending on the application, a context can include additional
components, to account for fairness assumptions, probabilistic
assumptions, etc.
Moreover, additional aspects of a context that are usually suppressed from
the notation are nonempty sets $\Intact$
and $\ExIn$ of internal actions for the processes and external inputs,
respectively.}
The environment is viewed as running a protocol (denoted by~$P_e$)
just like the processes; its protocol is used to capture
nondeterministic aspects
of the execution, such as the actual transmission times, external
inputs into the system, etc.
The transition function $\tau$
describes how the actions performed by the processes and the
environment change the global state.
Thus, if $g$ is a global state and
$\vec{\sfa}=\langle \sfa_e,\sfa_1,\ldots,\sfa_n\rangle$ is a {\em
  joint action\/}
(consisting of an action for the environment and one for each of the
processes), then $\tau(\vec{\sfa},g)=g'$ specifies that~$g'$ is the state
that results when $\vec{\sfa}$ is performed in state~$g$. When 
modeling asynchronous systems, we assume that some processes 
will be executing a $NULL$ action, of which they are not even aware
(their local states are left unaltered).

A run $r$ is consistent with a protocol $P$ if it could have been
generated when running protocol $P$.  Formally,
run $r$ is {\em consistent with joint
protocol $P$ in context $\gamma$\/} if
\begin{enumerate}
\item $r(0)\in\Gz$, so that it starts from a $\gamma$-legal initial global
  state, and
\item for all~$t\ge 0$, the transition
from global state $r(t)$ to $r(t+1)$ is the result of performing
one of the joint actions specified by~$P$ and the environment protocol $P_e$
(the latter is specified in~$\gamma$) in the global state $r(m)$.
That is, if $P = (P_1, \ldots, P_n)$, $P_e$ is the environment's
protocol in context $\gamma$, and
$r(m) = \langle\ell_e, \ell_1, \ldots, \ell_n\rangle$, then
there must be a joint action $\vec{\sfa}=\langle\sfa_e, \sfa_1,\ldots,
\sfa_n\rangle$ such that $\sfa_e \in P_e(\ell_e)$, $\sfa_i \in
P_i(\ell_i)$ for $i = 1, \ldots, n$, and $r(m+1) =
\tau(\vec{\sfa},r(m))$ (so that $r(m+1)$ is the result of applying the
joint action~$\vec{\sfa}$ to $r(m)$).
\end{enumerate}
We use $\Rrep(P,\gamma)$ to denote the set of all
runs of~$P$ in~$\gamma$, and call it the {\em system representing~$P$ in
context~$\gamma$}.

A description of the specific context $\gammab$ that we deal with throughout the thesis is found
in Section~\ref{sec:gamma_bounded}.

\section{Defining Knowledge in a Distributed System}
We aim at a logical analysis of gained knowledge regarding the occurrence of events. 
The interpreted systems framework~\cite{FHMV} provides us with much of the necessary machinery here.
We focus on a
simple logical language in which the set~$\Phi$ of primitive
propositions consists of propositions of the form $\occurred{e}$,
$\ndevent{e}$ and $\ttime=t$ for all events~$e$ and times $t$. To
obtain the logical language~$\Lang$, we close $\Phi$ under
propositional connectives and knowledge formulas. Thus,
$\Phi\subset\Lang$, and if $\varphi\in\Lang$, $i\in\Proc$ and
$G\subseteq\Proc$, then $\{K_i\varphi,E_G\varphi,
C_G\varphi\}\subset\Lang$. The formula $K_i\varphi$ is read {\em process
  $i$ knows~$\varphi$}, $E_G\varphi$ is read {\em everyone in~$G$
  knows~$\varphi$}, and $C_G\varphi$ is read {\em $\varphi$ is common
  knowledge to~$G$}. In addition,
we add a timestamping operator as
well. Thus, if $\varphi\in\Lang$ and $t\in\Nat$, then
$\atn{t}{\varphi}\in\Lang$.\footnote{In this thesis we do not
investigate complexity and decidability issues pertaining to the use
of explicitly timestamped formulas. The system's existing constraints
on transmission times require some sort of temporal metric on
formulas, and we opt for this choice based on the clarity and
conciseness that it offers.}      

The truth of a formula is defined with respect to a triple $(R,r,t)$. We write
~~$(R,r,t)\sat\varphi$ ~to state that $\varphi$ holds at time~$t$ in run~$r$,
with respect to system~$R$. It is always assumed that $r\in R$ in a triple
$(R,r,t)$. The precise meaning of nondeterminism in this system is given in
Section~\ref{sec:gamma_bounded} below. Denoting by $r_i(t)$ process~$i$'s local
state at time~$t$ in~$r$, we inductively define
\begin{itemize} 
\item[$~$] $(R,r,t)\sat \ndevent{e}$ \quad iff \quad event~$e$ occurs in $r$
  and is nondeterministic there;
\item[$~$] $(R,r,t)\sat\occurred{e}$ \quad iff \quad  event~$e$ occurs at
  time~$t'$ in~$r$ such that $t'\leq t$;
\item[$~$] $(R,r,t)\sat\ttime=t'$ \quad iff \quad  $t'=t$;
\item[$~$] $(R,r,t)\sat\atn{t'}{\varphi}$
  \quad iff \quad $(R,r,t')\sat\varphi$;
\item[$~$] $(R,r,t)\sat K_i\varphi$ \quad iff \quad 
\raisebox{-5pt}{\begin{tabular}{l}$(R,r',t')\sat \varphi$
  for every run~$r'$ satisfying\\ $r_i(t)=r'_i(t')$;
\end{tabular}}
\item[$~$] $(R,r,t)\sat E_G\varphi$ \quad iff \quad $(R,r,t)\sat K_i\varphi$
  for every $i\in G$; ~~~and
\item[$~$] $(R,r,t)\sat C_G\varphi$ \quad iff \quad $(R,r,t)\sat
  (E_G)^k\varphi$ for every $k\ge 1$.
\end{itemize}
Propositional connectives are handled in the standard way, and their
clauses are omitted above. In some cases it will be convenient to also
syntactically derive the proposition 
$(R,r,t)\sat \occurs{e}$ iff  $(R,r,t)\sat \occurred{e} \wedge (\ttime=0 \vee \atn{t-1}{\neg\occurred{e}})$, 
so $(R,r,t)\sat \occurs{e}$ holds true if $e$ occurs exactly at time $t$ in $r$.

By definition, $K_i\varphi$ is satisfied at $r\in R$ and  time $t$ if
$\varphi$ holds at all
points at which~$i$ has the same local state as in $(r,t)$. Thus, given~$R$, the
local state determines what processes know. 
Intuitively, a fact~$\varphi$ is common knowledge to~$G$ if
everyone in~$G$ knows~$\varphi$, everyone knows that everyone
knows~$\varphi$, and so on {\em ad infinitum}.
In particular, if $(R,r,t)\sat C_G\varphi$ then
$(R,r,t)\sat K_{i_h}K_{i_{h-1}}\cdots K_{i_1}\varphi$, for every
string $K_{i_h}K_{i_{h-1}}\cdots K_{i_1}$ and~$h>0$.

We write $R\sat\varphi$ and say that ``$\varphi$ is {\em valid} in~$R$''
if $(R,r,t)\sat\varphi$ holds for all $r\in R$ and $t\ge 0$.
A formula is valid, written $\sat\varphi$, if it is valid in all
systems~$R$. 

It is convenient to treat boundary cases for some of the operators in the following
way: for $|G|=0$ we have $R\sat E_G\varphi$ for all $\vphi$, and hence also
$R\sat C_G\varphi$. For all $G$, we say that $(R,r,t)\sat 
(E_G)^0\varphi$ iff $(R,r,t)\sat \varphi$.

In the context of distributed systems, the knowledge operator embodies an
important function that is often left unstated. Intuitively, a process ``knows"
$\vphi$ if it is in possession of ample evidence that $\vphi$ is true.
Essentially, all such evidence must be based on the local state of the process.
The local states of other processes are not immediately available for it to
inspect. We consider the local state $\ell$ of process $i$ as ``ample evidence
that $\vphi$'', if at every possible  point in the execution of  the distributed
system where the local state of $i$ is $\ell$, $\vphi$ holds.  Thus knowledge
can be seen as a \emph{localizing} qualifier for the inner  formula $\vphi$.  To
see this, consider the two statements below.

\begin{itemize}
    \item $(R,r,t)\sat \vphi$, and
    \item $(R,r,t)\sat K_i\vphi$
\end{itemize}

The former statement is straightforward: $\vphi$ obtains at the distributed
system in run $r$ at time $t$. The latter statement makes a stronger claim: not
only does $\vphi$ obtain  in $r$ at $t$, but it also holds in every possible
point in which  $i$'s local state is the same as it is now (at the point
$(r,t)$).

Formulas pertaining to nested knowledge, such as $(R,r,t)\sat K_jK_i\vphi$, can
now also be given a formal interpretation. What it means is that process $j$'s
local state at $(r,t)$ provides ample evidence to support the claim that process
$i$'s local state at $(r,t)$ provides ample evidence that $\vphi$ is true at $(r,t)$.

\section{The Synchronous Model}\label{sec:gamma_bounded}
\subsection{The  Synchronous Context $\gammab$}\label{ssec:gammab}
We are interested in characterizing the effect of synchronous constraints on
distributed systems. We therefore define a class of synchronous contexts $\gammab$ that ensure
the following properties for all systems defined on top of them:

\begin{itemize}
\item The set of processes is denoted by~$\Proc$. These are connected by a
network of weighted channels. For each pair of processes $i,j$ connected by a
communication channel, the weights $min_{ij}$ and $max_{ij}$  denote the minimal
and maximal transmission times for messages over the channel, respectively. In
all cases $min_{ij}\geq 1$. Whenever there is no upper bound on transmission we
have that $max_{ij}=\infty$.
\item We assume that processes can receive external inputs from the outside
world. These are determined in a genuinely nondeterministic fashion,
and are not correlated with anything that comes before in the
run, or with external inputs currently received by other processes.
\item The scheduler, which we typically call the {\em environment}, is in
charge of choosing the external inputs, and
of determining message transmission times. The latter are also
determined in a nondeterministic fashion, subject to the 
delivery time constraints as detailed by the weights on the channels.
\item Time is identified with the natural numbers $\Nat$, and each process is
assumed to take a step at each time~$t\in \Nat$.
For simplicity, the processes follow deterministic protocols. Hence,
a given protocol~$P$ for the processes and a given behavior of the
environment completely determine the run.
\item Events are sends, receives, external inputs and internal
  actions. All events in a run are distinct, and we denote a generic
  event by the letter~$e$. For simplicity, events do not take time to
  be performed. At a given time point a process can perform an
  arbitrary finite set of actions.
\end{itemize}

We shall, for the most part, be concerned with contexts that are more
restrictive than $\gammab$. Thus, any $\gammab$ context where $1=min_{ij}\leq
max_{ij}<\infty$ for all channels $(i,j)$ will be called a $\gammax$ context , it is a 
context whose systems are characterized by the existence of finite upper bounds on
delivery time. Similarly, a $\gammab$ context where $1\leq
min_{ij}<max_{ij}=\infty$ for every channel $(i,j)$ is a $\gammin$ context. This is a 
context where there are only lower bounds on transmission times.

Whichever superscript $\alpha$ is used to denote a particular context (as in
$\gamma^{\alpha}$), will also be made use of to denote a system
$\Rrep^{\alpha}=\Rrep(P,\gamma^{\alpha})$ where $P$ is any arbitrary protocol.

\subsection{Detailed Specification for $\gammab$}
A synchronous context $\gammab$ is defined as a tuple
$({\cal G}_0,P_e^{\,\mathsf{b}},\tau)$ where

\paragraph{The environment's state}
Recall that the environment's state keeps track of relevant aspects of
the global state that are not represented in the local states of the
processes. We assume that the environment's state has
three components $\ell_e=(\mathsf{Net},\mathsf{t},\Hist_e)$, where
\begin{enumerate}
\item $\mathsf{Net}$ is a labelled graph $(\Proc,E,max,min)$ describing the network topology
and bounds on transmission times.
Its nodes are processes, and a directed edge $(i,j)\in E$
captures the fact that there is a channel from~$i$ to~$j$ in the
system. Moreover, the labels $1\le max_{ij}\in\Nat\cup \{\infty\}$ 
and $1\le min_{ij}\in\Nat$ are upper and lower bounds respectively on the
time that a message sent on $(i,j)$ can be in transit. 
The contents of $\mathsf{Net}$ are not affected by~$\tau$, and so $\mathsf{Net}$
remains constant throughout the run.

\item The variable~$\mathsf{t}$ keeps track of global time. As we shall see
its value starts at $\mathsf{t}=0$, and advances by~1 following each
round. Finally,
\item $\Hist_e$ records the
sequence of joint actions performed so far.
The $\Hist_e$ component
uniquely determines the contents of all channels.%
\footnote{This holds true even if we allow message loss by setting $bij=\infty$.} 
Indeed, a message $\mu$ is {\em in transit} at a given
global state~$g$ if the $\Hist_e$ component in~$g$ records that $\mu$
has been sent, and does not record its delivery.
\end{enumerate}

\paragraph{Process local states}
We assume that local states have three components
$\ell_i=(\mathsf{Net}_i,\mathsf{t}_{i},\mathsf{data}_i)$, where $\Net_i$ and
$\Time_i$ are copies of the $\Net$ and $\Time$ values from the environment's
state.. The component $\mathsf{data}_i$ serves as the data segment for the
process $i$. Its contents are a function of the protocol $P$ and the transition
function $\tau$.

\paragraph{The set ${\cal G}_0$ of initial global states}
We assume that associated with $\gammab$ there is a set~$\Init_i$
of possible initial states for each process~$i\in\Proc$.
We define ${\cal G}_0$ to be the set of global states
$gl=(\ell_e,\ell_1,\ldots,\ell_n)$ satisfying:
(1)  $\mathsf{t}_i=0$ for all $i\in\Proc$ and $\mathsf{t}=0$;
(2) the network components $\mathsf{Net}$ and $\mathsf{Net}_i$ are all identical;
(3) for every $i\in\Proc$, ~$\hist_i=\langle\init_i\rangle$,with
$\init_i\in\Init_i$; and
(4) $\Hist_e$ is the empty sequence.

\paragraph{Actions and external inputs}
Associated with the context $\gammab$ are sets $\Intact$
 of internal actions for the processes and sets  $\ExIn$ of external inputs,
respectively. For ease of exposition we assume that $\bot\in\ExIn$,
where $\bot$ stands for the empty external input. Moreover, we
generally assume that $\ExIn\ne\{\bot\}$, so that there is at least one
nontrivial possible external input.
We assume that processes can perform {\sf send} actions and internal
actions. The local action $\sfa_i(k)$ that~$i$
contributes to the joint action in
round~$k+1$ consists of a finite sequence of distinct send and
internal actions.  (Recall that the local action is determined by the
protocol, based on the local state.)
We use external inputs to model spontaneous events. They are generated
by the environment. In addition to external inputs, the environment
is in charge of message delivery. Thus, the environment's action
$\sfa_e(k)$ consists of a finite sequence of external inputs to be delivered to
various individual processes, a subset $\pi$ of $\Proc$ that are activated in the current round,
and a (possibly empty) set of messages
that are to be delivered in the current round.

\paragraph{The environment's protocol $P_e^{\,\mathsf{b}}$}
The environment in $\gammab$ is in charge of delivering external
inputs to processes and determining message deliveries.
For every global state $gl$ we define $P_e^{\,\mathsf{b}}(gl)$ to be the set of actions
$\sfa_e=(\sigma_x,\sigma_d)$ such that
\begin{enumerate}
\item $\sigma_x:\Proc\imp\ExIn$ is a sequence assigning to each
  process $i\in\Proc$ an external input (possibly the empty input
  $\bot$) it receives in the current round, and
\item $\sigma_d$ is a sequence $\langle M_{1}\ldots M_{|\Proc|}\rangle$
where (i) for every $i\in Proc$ the set $M_i$ consists of messages that
are in transit in~$gl$,  (ii) $M_i$ contains all messages in transit to~$i$
whose transmission time bounds, as specified in~$\mathsf{Net}$, will be
violated (expire) if the message is not delivered in the current
round, and (iii) none of the messages in $M_{i}$ are such that if delivered in the current
round, will violate the existing minimal transmission time constraints.
\end{enumerate}

Notice that $P_e^{\,\mathsf{b}}$ is genuinely nondeterministic.
Exactly one of the actions in $P_e^{\,\mathsf{b}}(gl)$ will be performed in
global state~$gl$ in any given instance.
By definition of $\Rrep(P,\gammab)$, however,
if $r(k)=gl$ then the system contains a run extending the prefix
$r(0),\ldots,r(k)$ for every possible environment action in $P_e^{\,\mathsf{b}}(gl)$.
Another point to note is that our definition does not enforce (and
hence does not assume) FIFO transmission; had we done so,
channels would be considered to be queues, and the
nondeterministic choices of messages to deliver would have to obey
FIFO order.
It should also be noted that the scheduler makes sure to comply with
all existing constraints: minimal and maximal transmission times, as well 
as process rate.  
Finally, the fact that external inputs are delivered in a
nondeterministic fashion implies they are not correlated in any way,
and they do not depend on anything that happens
before they are delivered.  This is the sense in which
external inputs can be viewed as independent, ``spontaneous'' events.

\paragraph{The transition function~$\tau$}
The transition function $\tau$ implements
the joint actions in a rather straightforward manner.
In every round:
(i) the global clock variable $\mathsf{t}$ and the local variables $\mathsf{t}_i$ of 
all $i\in \pi$ are advanced
by one;
(ii) a copy of the joint action is added to the environment's
history log $\Hist_e$; and
(iii) For every process $i$, a record of all current round message deliveries and external inputs 
to the process is written in $\mathsf{data}_i$. Note that this record is overwritten  in every round, 
so that a protocol must take special measures in order to maintain a persistent copy of these contents.

\section{Road Map}\label{sec:roadmap}
This chapter has provided an outline of the 
necessary background upon which it is built: the causal analysis of 
distributed systems introduced by Lamport, and the knowledge-based framework
of Fagin et. al. The rest of the thesis describes novel results obtained as part of our research. 
We conclude it with a roadmap that offers a general outline of what the thesis is all about. 

Chapter~\ref{ch:coordination} defines the formal and conceptual ``playground'' within which 
our research is conducted. We start by introducing the \emph{Ordered} and the \emph{Simultaneous Response} problems: two generic coordination problems that set constraints on the temporal 
ordering of events. The problem definitions involve a set of required responses to a 
spontaneous non-deterministic event. As we argue there,  spontaneity is a required ingredient
if we want to study those cases of coordination that necessitate information flow in the system.
This rather abstract notion of information flow is also given a formal interpretation, in 
terms of \emph{knowledge gain}.  Apart from definitions, the chapter also provides 
initial claims and their proofs. We study and prove the relations between the types of response
problems and correlated epistemic states of gained knowledge: 
nested knowledge is necessary and sufficient for ensuring correct solutions to the 
Ordered Response problem. Similarly, common knowledge is necessary and sufficient
for  Simultaneous Response. A discussion on the role
of knowledge as an intermediate layer between causality and coordination concludes this chapter.

Chapter~\ref{ch:NKG} is the first in a series of four chapters that each
study and characterize a particular coordination problem. This chapter studies
the Ordered Response problem, but it also introduces several key notions that
are utilized in the following chapters. We use the set of process-time pairs as
the domain in which causal relations are  defined. This domain is more suitable
than the set of events, or of processes, given the synchronous characteristics
of the system. The two most basic causal relations that we use are \emph{timing
guarantees} and \emph{syncausality}, the latter being a generalization of
Lamport's happened-before. 

In asynchronous systems,  the correct ordering of more than two events requires
the  repeated application of the happened-before relation to each pair of
subsequent events. A careful analysis of solutions to the most simple cases of
Ordered Response reveals that such ordering in synchronous  systems requires
complex relations between all of the related process-time nodes. We define a
causal structure, the \emph{centipede}, that combines both syncausality and
timing guarantees, and prove the Centipede Theorem, showing that the existence
of a centipede is necessary for ensuring the correct ordering of a sequence of
events. For ease of exposition, the formal results of this chapter (as well as
those of Chapters~\ref{ch:CKG} through~\ref{ch:generalized_ordering}) are given
using the $\gammax$ context, in which only upper bounds are defined. We show
however that the theorem also applies in two complementing boundary cases: one, where
there are no upper bounds on delivery, and the other where delivery times are
fixed. In the former case, the centipede structure is trivialized into a
Lamport-style message chain. We conclude this chapter by showing that our
definition is tight, in the sense that under some protocols the existence of the
centipede is also sufficient for proper event ordering. We suggest the
\emph{Full Information Protocol} ($\fip$) for synchronous systems for this purpose.

Moving on to the next type of response problem, Chapter~\ref{ch:CKG} investigates
the causal structures necessary for ensuring the simultaneous happening
of events. As such, it constitutes a complete break with existing analysis 
of asynchronous causality, where no such constraint can be ensured. We introduce
the \emph{centibroom} structure, a variant of the centipede, and prove the Centibroom 
Theorem, an analog to the Centipede Theorem that shows the existence of a centibroom
to be necessary for ensuring simultaneous actions. Sufficiency of the centibroom for 
such coordinated responses is also proved, under $\fip$. 
As an application of the Centibroom Theorem we suggest two novel variants of 
a global snapshot algorithm for synchronous systems, one of which
is shown to provide optimal time complexity.   

The particular form of the centibroom, and the  results of Chapter~\ref{ch:CKG}
showing that it is a prerequisite for common knowledge, provide a clear and
graphic demonstration that the nature of common knowledge is finitistic, despite
its familiar definition being based on an infinite conjunction of facts. Further
investigation into the properties of the centibroom and of common knowledge is
used to show that, roughly speaking,  it takes time to obtain deeply nested
knowledge without ``collapsing'' into common knowledge. Nevertheless, this
result is shown to be dependent upon the protocol being followed. A 
counterexample is suggested where every level of nested knowledge may be achieved
without common knowledge ensuing.

Chapters~\ref{ch:ordered_group} and~\ref{ch:generalized_ordering} deal with
generalizations of both the Ordered Response and the Simultaneous Response
problems. First, the \emph{Ordered Group Response} problem is defined, which can
be seen as an immediate ``merge'' of Ordered and Simultaneous Response
requirements. In analogy, the \emph{generalized centipede} structure is defined,
and is shown to be necessary in solutions to the  Ordered Group Response
problem. Then we take the generalization even further and define the
\emph{Generalized Response} Problem, where the required temporal ordering of
events can be specified using any partial order. Characterization is provided in
terms of sets of generalized centipedes. Our understanding of common knowledge
is advanced further by showing how such an epistemic state is dependent upon the
joint histories of the processes in the group.

Chapter~\ref{ch:ignorance} takes a different stance from the one followed to this point.
In epistemic terms, each of the response problems we considered thus far is reduced into a
rather complex requirement concerning knowledge about knowledge, which is then
reduced further into a causal communication condition that is highly dependent upon
the existence of upper bounds on message delivery times. In
Chapter~\ref{ch:ignorance} we consider the complementary approach: we ask what
is the causal condition that will ensure knowledge about ignorance rather than
knowledge about knowledge. This leads us  into a more detailed discussion of causal
cones of influence, from which the conditions for such ignorance are then distilled. 
As it turns out, it is the existence of \emph{lower bounds} 
on transmission times that makes such knowledge, which is of value in 
competitive settings, possible.

Chapter~\ref{ch:conclusions} brings the thesis to conclusion and discusses possible
further research and various open questions.

\section{Related Work}\label{sec:related_work}
A great abundance of work pertaining to Lamport's happened-before relation has
been collected over the years, and we shall make no attempt to provide a survey
of this work. A thorough report can be found in \cite{mattern94}. 
Two of the most widely known works that build atop it are Mattern's
generalization of Lamport's scalar clocks into vector clocks in
\cite{mattern1988,fidge1988} and Chandy and Lamport's utilization in the
snapshot algorithm \cite{ChL} which is further discussed in
Chapter~\ref{ch:CKG}. Another related work is the Chandy and Misra paper \cite{ChM}
discussed in Section~\ref{sec:lamport_analysis}.

Formal study of knowledge, and knowledge about knowledge, touches on many
fields, ranging from philosophy~\cite{Lew} and psychology~\cite{CM81}, to
linguistics~\cite{Grice,Parikh92}, economics~\cite{Au}, AI~\cite{MH},
cryptography~\cite{DiffieHellman,RabOT,GMR} and distributed
systems~\cite{HM1,ChM,PK92}. The interpreted systems framework, epitomized in
\cite{FHMV}, stands at the base of related research in to distributed
computing~\cite{MT,DM,HalMoWa2001}.

Explicit and implicit use of time bounds, introduced in Chapter~\ref{ch:NKG},
for coordination and improved efficiency is ubiquitous in distributed computing.
An elegant example of its use is made by Hadzilacos and Halpern in \cite{HH}.
That knowledge can be gained by way of $\NULL$  messages when timing guarantees
are available has been part of the folklore for decades. To our knowledge,
Lamport \cite{LamSM} was first to explicitly explore the use of $\NULL$ 
messages beyond their customary timeout semantics. In effect Lamport's state
machine protocols in that paper are based on an  implicit  notion of
causality which we will later (see Chapter~\ref{ch:NKG}) define 
as syncausality, yet no attempt at rigorous formalization is made there, and the
general role of time bounds is not developed. A tutorial by Moses~\cite{Mo92}
suggests as a viable topic for future work performing an explicit analysis of
the effect of $\NULL$  messages on knowledge gain. He also presents an example
in which communication can be saved by using timeouts. However, \cite{Mo92} does
not suggest modifying Lamport causality to suit synchronous systems, and none of
the new notions or technical results in this thesis were suggested
in~\cite{Mo92}. Krasucki and Ramanujam in~\cite{Krasucki94} study of the
interaction between knowledge and the ordering of events in a distributed
system. They consider concurrency in a rather abstract setting, where they show
that causality is related to the existence of particular partially ordered sets.
They do not explicitly study the synchronous model, however, and do not
explicitly consider synchronous time bounds on channels. Moses and
Bloom~\cite{MB} perform a knowledge-based analysis of clock synchronization in
the presence of bounds on transmission times. They generalize Lamport's relation
by defining a notion of timed causality $e\linkalpha e'$ that corresponds to $e$
taking place {\em at least}~$\alpha$ time units before~$e'$. It appears that
`$\linkalpha$' is a quantitative generalization of Lamport causality for the
purpose of determining relative {\em timing} of events.
A similar notion appears in the work of Patt-Shamir and
Rajsbaum~\cite{PSR}.

Chapter~\ref{ch:CKG} deals extensively with relations between nested knowledge,
common knowledge and time. The growing body of research dealing with the
dynamics of interactive epistemology  has brought to light some of the intricate
relations here \cite{HM1,FHMV,Ditmarsch2007}. Halpern and Moses \cite{HM1}
proved that common knowledge cannot be \emph{gained} in the face of unreliable
or asynchronous communication.  Parikh and Ramanujam~\cite{PR} investigate
nested knowledge in connection to formal languages. Common knowledge is
typically perceived in terms of an infinite conjunction of $E^k$, for $k>0$.
There are also definitions of common knowledge in terms of a fixed point (see,
e.g., \cite{Lew,FHMV,Bar1}). Fischer and Immerman \cite{FI1} first showed that
the level of nested knowledge that can be achieved without ``collapsing'' into
common knowledge are bounded in finite state systems. The combinations of nested
and common knowledge that are discussed in Chapter~\ref{ch:ordered_group} are,
interestingly, somewhat similar to those  found in Chwe's~\cite{Chwe2000} game
theoretic analysis of coordination.

\ifx \isEmbedded \undefined
\bibliography{../../z}


\maketitle

\else
\fi
\label{ch:ordered_group}

\section{Introduction}\label{sec:intro_GO}
The Ordered Response problem deals with a totally ordered sequence of response,
and the Simultaneous Response problem with groups of responses that must be
enacted in unison. As seen in Section~\ref{sec:coord_via_response}, we can look
at the required time ordering in an instance
$\SR=\angles{\trig,\alpha_1,\ldots,\alpha_k}$ as the set of requirements
$\{Time(\trig)\leq Time(\alpha_{g})\leq Time(\alpha_{h})| g,h\in 1,..,k\}$.

Two possible extensions of problem specifications come to mind.
The first extension, that we call the \emph{Ordered Group Response} problem,
is an immediate generalization of the $\OR$ and $\SR$ problems.

\begin{definition}[Ordered Group Response] 
	Let $\trig$ be an external input and
	let $A^{h}=\angles{\alpha_{1}^{h},..,\alpha_{\ell_{h}}^{h}}$ be a set of
	responses of length $\ell_{h}$, for every $h=1,..,k$. A protocol~$P$ solves the
	instance $\OGR=\angles{\trig,A^{1},\ldots,A^{k}}$ of the {\em Ordered Group Response}
	problem if it guarantees that 
	\begin{enumerate} 
		\item in a triggered run, for every $h=1,\ldots,k$ all of the actions in the response set
		$A^{h}$ are performed simultaneously; moreover, if
		$h<k$ then the actions in $A^h$ will happen before (i.e., no later than)
		those of $A^{h+1}$. Finally,
		\item  none of the responses, in any of the sets  $A^{1},\ldots,A^{k}$,
		occurs in runs that are not triggered.
	\end{enumerate} 
\end{definition}

We will use  $I^{h}$ to denote the set of processes $\{i\in\Proc|\angles{i,a}\in A^{h}\}$
for every $h<k$. 

It is easy to see that every instance of $\OR$ can be rewritten as an instance
of $\OGR$ where all sets of responses are singletons. Similarly, every instance of $\SR$ can
be rewritten as an instance of $\OGR=\angles{\trig,A^{1}}$ where all simultaneous
responses are members of $A^{1}$.

The second extension, which is even  wider scoped than $\OGR$, is
a problem specification where the required event ordering is given by any arbitrary partial order.
The \emph{Generalized Ordering} problem will be defined and studied in the next chapter.
In this chapter we will focus on the $\OGR$ problem. We defer giving a leading example
until the next chapter, which will make use of this chapter's results. Apart from providing 
the foundational results necessary for the next chapter, this chapter can also be seen
as providing a unifying account that merges the thus-far separately treated theories
that surround $\OR$ and $\SR$. 

\section{Relating Ordered Group Response and Nested Common Knowledge}\label{sec:ogr_nck}
In order to relate the new ordering problem to a causal structure, we will first identify
an epistemic condition that is implied by protocols solving the problem.
As solutions to $\OR$ require nested knowledge and those of $\SR$ imply 
common knowledge, we expect that solutions to $\OGR$ will necessitate a little
of both kinds of epistemic states.

We will say that nested common knowledge of $\vphi$ obtains at run $r\in \Rrep$ and time $t$
with respect to groups of processes $G_{1},..,G_{k}$ if
$$(\Rrep,r,t)\sat C_{G_{k}}C_{G_{k-1}}\cdots C_{G_{1}}\vphi$$
holds. As we will show, nested common knowledge is a necessary requirement
in protocols that solve the $\OGR$ problem. In fact, we show that nested common knowledge
is necessitated even if the protocol only \emph{weakly solves} $\OGR$, according to
the following definition.

\begin{definition}[Weakly solving Ordered Group Response] \mbox{}\\
	Let $\OGR=\angles{\trig,A^{1},\ldots,A^{k}}$ be an instance of the ordered group response
	problem. A protocol~$P$ \emph{weakly solves} $\OGR$ if it guarantees that
	for every $h=1,\ldots,k$  and $\alpha \in A^{h}$
	\begin{enumerate}
		\item for every $r\in \Rmax$, time $t$ and $\alpha'\in A^{h}$, 
		$\alpha$ occurs at $(r,t)$ iff $\alpha'$ occurs at $(r,t)$
		\item for every $r\in \Rmax$, time $t$ and $\alpha'\in A^{h'}$ where $h'\leq h$,
		if $\alpha$ occurs at $(r,t)$ then $\alpha'$ occurs at $(r,t')$ where $t'\leq t$.		
	\end{enumerate}•
\end{definition}

Note that every protocol that solves $\OGR$ also weakly solves it, but that the opposite
implication does not hold. A protocol that weakly solves $\OGR$ does not necessitate
that any of the responses occur in a triggered run.

Recall that in order to prove the relation between solutions to $\OR$ and nested 
knowledge in Theorem~\ref{thm:ORnestedK}, we had to assume that processes
can recall responses that they had performed. We now define a stronger
recall requirement for nested common knowledge.

\begin{definition}[Group response recall]\label{def:group_response_recall}
	Let $\OGR=\angles{\trig,A^{1},\ldots,A^{k}}$ and assume that
	$\Rrep=\Rrep(P,\gamma)$ is a system of runs for a protocol $P$ where all of
	the responses may occur, and $\gamma$ is any arbitrary context . Protocol
	$P$ \emph{recalls group responses} for $\OGR$ if for all $\alpha\in A^{h}$
	where $1\leq h \leq k$, $r\in\Rrep$, $t'\leq t$ and $i\in \Proc$,  
	if $(\Rrep,r,t)\sat K_{i}\occurred{\alpha}$ then $(\Rrep,r,t')\sat K_{i}\occurred{\alpha}$.
\end{definition} 

A protocol recalls group responses if processes, once they know that a response has
taken place, never forget this fact. We are now ready to prove that $\OGR$
requires nested common knowledge. 

\begin{theorem}\label{thm:OGRnck}
	Let $\OGR=\angles{\trig,A^{1},\ldots,A^{k}}$, and assume that protocol~$P$
	weakly solves~$\OGR$ in~$\gamma$ and that it recalls group responses for $\OGR$.
	Let $r\in\Rrep$ be a run in
	which~$\trig$ occurs at time $t_{0}$, and where the processes in $I^{h}$ perform the
	actions~$a^{h}_{1}$ to $a^{h}_{\ell_{h}}$ simultaneously at $t_h\geq t_{h-1}$, for
	every $1\le h\le k$.
	
	\noindent Then ~~$(\Rrep,r,t_{h})\sat~C_{I^{h}}C_{I^{h-1}}\cdots C_{I^{1}} \ndocc{\trig}$ for every $1\le h\le k$.
\end{theorem}•

\begin{proofT}
	We proceed by  induction on $k$.
	\begin{itemize}
		\item[] $\mathbf{k=0:}$ $(\Rrep,r,t_{0})\sat~ \ndocc{\trig}$ by definition of $r$.
		\item[] $\mathbf{k>0:}$ We will use the Induction Rule for common
		knowledge to prove the inductive step. Recall that
		$A^{k}=\angles{\alpha_{1}^{k},..,\alpha_{\ell_{k}}^{k}}$. Fix
		$h,g\in\{1,..,\ell_{k}\}$. We first show that $\Rrep\sat
		\occurs{a^{k}_{h}}\imp E_G(\occurs{a^{k}_{h}} \wedge C_{I^{k-1}}\cdots
		C_{I^{1}} \ndocc{\trig})$.
		Note that as all responses in $A^{k}$ are performed simultaneously, we
		get $\Rrep\sat \occurs{a^{k}_{h}}\leftrightarrow \occurs{a^{k}_{g}}$.
		Since whether the (deterministic) protocol $P$ performs the action
		${a^{k}_{g}}$ is a function of $i_g$'s local state, we have that
		$\Rrep\sat \occurs{a^{k}_{g}}\imp K_{i_g}\occurs{a^{k}_{g}}$. Now using
		the former equivalence we get that
		
		$$(*)\qquad \Rrep\sat \occurs{a^{k}_{h}}\imp K_{i_g}\occurs{a^{k}_{h}}.$$
		
		Now choose arbitrary $r',t'$. Suppose that  $(\Rrep,r',t')\sat
		\occurs{a^{k}_{h}}$. Note that since $P$ weakly solves $\OGR$, we have both
		\begin{itemize}
			\item[(i)] when $a^{k}_{h}$ is performed, the responses in $A^{k-1}$ have
			already been performed (or are being performed). Say that these have
			been performed at a time $t'_{k-1}\leq t'$. And,
			\item[(ii)] protocol $P$ also weakly solves the sub-problem\\
			$\OGR'=\angles{\trig,A^{1},\ldots,A^{k-1}}$. By the inductive
			hypothesis, we have $$(\Rrep,r',t'_{h-1})\sat~C_{I^{h-1}}\cdots
			C_{I^{1}} \ndocc{\trig}$$ for all $h<k$.
		\end{itemize}•
		
		As processes recall group responses and $t'_{k-1}\leq t'$, we obtain from
		$(ii)$ that $(\Rrep,r',t')\sat~C_{I^{k-1}}\cdots C_{I^{1}}
		\ndocc{\trig}$. Given this and the fact that if
		$(\Rrep,r',t')\nsat\occurs{a^{k}_{h}}$ then  $(\Rrep,r',t')\sat
		\occurs{a^{k}_{h}}\imp \alpha$ for any $\alpha$, we conclude that
		$$\Rrep\sat~\occurs{a^{k}_{h}}\imp C_{I^{k-1}}\cdots C_{I^{1}}
		\ndocc{\trig}.$$ Combined with $(*)$, we obtain  
		$$\Rrep\sat
		\occurs{a^{k}_{h}}\imp K_{i_g}C_{I^{k-1}}\cdots C_{I^{1}}
		\ndocc{\trig}.$$
		Since $g$ was arbitrarily chosen, we get $\Rrep\sat
		\occurs{a^{k}_{h}}\imp E_{I^{k}}\occurs{a^{k}_{h}}$, and also $\Rrep\sat
		\occurs{a^{k}_{h}}\imp E_{I^{k}}C_{I^{k-1}}\cdots C_{I^{1}}
		\ndocc{\trig}$, ending up with 
		$$\Rrep\sat \occurs{a^{k}_{h}}\imp
		E_{I^{k}}(\occurs{a^{k}_{h}} \wedge C_{I^{k-1}}\cdots C_{I^{1}}
		\ndocc{\trig},$$
		 as required.

		Let $\Phi=\occurs{a^{k}_{h}}$ and $\Psi=C_{I^{k-1}}\cdots C_{I^{1}}
		\ndocc{\trig}$. Applying the Knowledge Induction Rule we get $\Rrep\sat
		\Phi\imp C_G\Psi$ from $\Rrep\sat \Phi\imp E_G(\Phi \wedge \Psi)$,
		giving us that $$\Rrep\sat \occurs{a^{k}_{h}}\imp
		C_{I^{k}}C_{I^{k-1}}\cdots C_{I^{1}}	\ndocc{\trig}.$$ Recalling that
		$(\Rrep,r,t)\sat\occurs{a^{k}_{h}}$ by assumption, we obtain that
		$\Rrep\sat  C_{I^{k}}C_{I^{k-1}}\cdots C_{I^{1}}	\ndocc{\trig}$.
	\end{itemize}
\renewcommand{\toto}{thm:OGRnck}
\end{proofT}

An immediate corollary is that nested common knowledge is also necessitated 
in protocols that solve (not weakly solve) $\OGR$.
 
\begin{corollary}\label{cor:OGRnck}
	Let $\OGR=\angles{\trig,A^{1},\ldots,A^{k}}$, and assume that protocol~$P$
	solves $\OGR$ in~$\gamma$ and that it recalls group responses for $\OGR$.
	Let $r\in\Rrep$ be a run in
	which~$\trig$ occurs at time $t_{0}$, and where the processes in $I^{h}$ perform the
	actions~$a^{h}_{1}$ to $a^{h}_{\ell_{h}}$ simultaneously at $t_h\geq t_{h-1}$, for
	every $1\le h\le k$.
	
	\noindent Then  for every $1\le h\le k$
	$$(\Rrep,r,t_{h})\sat~C_{I^{h}}C_{I^{h-1}}\cdots C_{I^{1}} \ndocc{\trig}.$$
\end{corollary}

As mentioned above, both nested knowledge and common knowledge are specific
cases of nested common knowledge. As such, Theorems~\ref{thm:ORnestedK}
and~\ref{thm:SRck} can be derived as further corollaries from the above
one.

To complete the picture, we briefly point out that there are protocols for which
nested common knowledge gain implies a solution to $\OGR$.

\begin{definition}[Group considerate protocol]\label{def:considerate}
	Let $\OGR=\angles{\trig,A^{1},\ldots,A^{k}}$, where 
	$A^{h}=\angles{\alpha_{1}^{h},..,\alpha_{\ell_{h}}^{h}}$ for every $h=1,..,k$.
	Protocol $P$ is \emph{group considerate} with respect to $\OGR$ 
	 if for each $h\leq k$ and $m\leq \ell_{h}$, response $\alpha_{m}^{h}$ 
	 is carried out by its respective process $i_{m}^{h}$  
	 as soon as $C_{I^{h}}\ndocc{\trig}$ is established, but no sooner.
\end{definition}• 

\begin{lemma}\label{lem:suffOGR}
    Let $\OGR=\angles{\trig,A^{1},\ldots,A^{k}}$,  and let $P$ be a group considerate protocol with
    respect to $\OGR$. If for every $r\in\Rmax=\Rrep(P,\gamma)$ such that $\trig$
    occurs at $(i_0,t)$ in~$r$ there exists time $t'$ such that
    $$(\Rmax,r,t')\sat C_{I^{k}}C_{I^{k-1}}\cdots C_{I^{1}}\ndocc{e},$$ then $P$
    solves $\OGR$. 
\end{lemma}

The proof repeats the one of Lemma~\ref{lem:suffOR}, with nested common knowledge
replacing nested knowledge. That a process will know immediately that common
knowledge has been achieved is given by the validity  $C_{G}\vphi
\leftrightarrow K_{g}C_{G}\vphi$ for any $g\in G$.

\section{Generalized Centipedes}\label{sec:gcenti}
We expect that just as $\OGR$ generalizes both $\OR$ and $\SR$, a
characterizing causal structure will generalize both the centipede
and the centibroom. The \emph{generalized centipede}, defined below, offers just
this kind of generalization.

\begin{definition}[Generalized Centipede]\label{def:gcentipede}
	Let $r\in\Rmax$, let $I^{h}\subseteq\Proc$  for $1\le h\le k$  and let and $t\leq t_{1}\leq \cdots \leq t_{k}$.
	A {\em generalized centipede} for $\langle \theta_{0},I^{1},\ldots,I^{k}\rangle$ 
	in $\rmmp$ is a 
	sequence of nodes ~~$\theta_0\spc\theta_1\spc\cdots\spc\theta_k$
	such that $\theta_0=(i_0,t)$, 
	and ~$\theta_h\tspc (i^{h}_{m},t')$ ~holds for all $h=1,\ldots, k$ and $i^{h}_{m}\in I^{h}$.
\end{definition}•

\begin{figure}[!h]
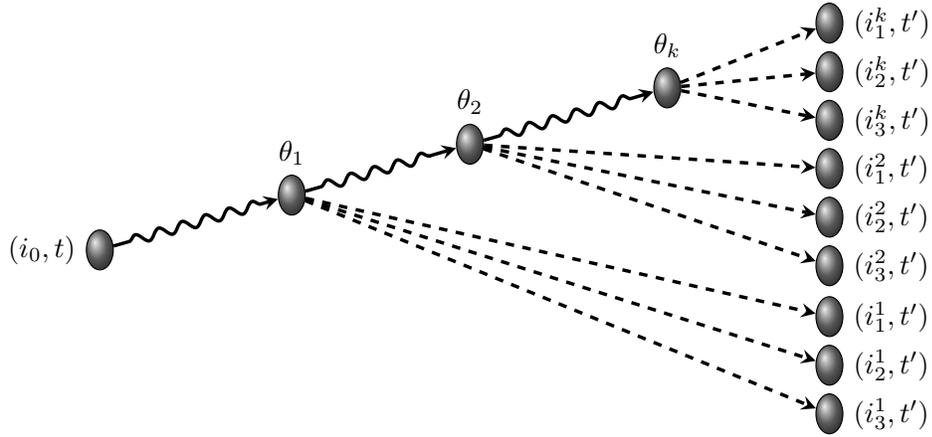

	\centering
		\newcommand{\tikzfname}{tikz_gencenti}
		\tikzsetnextfilename{\cfcurrentfolder Figures/\tikzfname}
		\resizebox{1\textwidth}{!}{\cfinput{Tikz/\tikzfname}}	
	\caption{A generalized centipede}
	\label{fig:nd_variations}
\end{figure}

In chapters~\ref{ch:NKG} and~\ref{ch:CKG} we found it
convenient to consider special kinds of centipedes and centibrooms, namely
centinodes and bridging centibrooms, respectively. Once again, the following
definition extends both of these special kinds.

\begin{definition}[Bridging Generalized Centipede]
	A generalized centipede $\angles{\theta_{0},\ldots,\theta_{k}}$ for $\langle
	(i_{0},t), I^{1},\ldots,I^{k}\rangle$ in $\rmmp$ is bridging if $\theta_{h}$
	is a bridging centibroom for $\angles{(i_{0},t), I^{h}}$ in $\rmmp$ for all
	$h=1\ldots k$.
\end{definition}

As the following lemma shows, generalized centipedes and their bridged sub-kind
may be freely interchanged.
 
\begin{lemma}\label{lem:bridge_gcent}
	A generalized centipede for $\langle \theta_{0}, I^{1},\ldots,I^{k}\rangle$ exists in $\rmmp$ 
	\qquad iff \qquad
	a bridging generalized centipede  for $\langle \theta_{0}, I^{1},\ldots,I^{k}\rangle$ exists in $\rmmp$.
\end{lemma}•

\begin{proofL}
	That the existence of a bridging generalized centipede implies that of a generalized
	centipede is immediate. We now prove the other direction.
	Assume that
	$\cC=\angles{\theta_0,..,\theta_k}$ is a generalized centipede for $\langle
	I^{0},\ldots,I^{k}\rangle$ in $\rmmp$. We define by induction on $h\le k$ generalized centipedes
	$\cC_h=\angles{\theta'_0,..,\theta'_h,\theta_{h+1},..,\theta_k}$ in $\rmmp$, in
	which the nodes $\theta'_0$ to $\theta'_h$ are bridging centibrooms for 
	$\angles{I^{0},\ldots,I^{h}}$, respectively. 

	\begin{description}
	    \item[$h=0:$] By definition, $\theta_{0}=(i_{0},t)$. As $I^{0}=\{i_{0}\}$,
	    $\theta_{0}$ is a trivial bridging centibroom for $I^{0}$ in $\rmmp$, and hence 
	    a bridging generalized centipede for $\angles{I^{0}}$.
	    \item[$h>0:$] Assume that a bridging generalized centipede\\
	    $\cC_{h-1}= \angles{\theta'_0,..,\theta'_{h-1},\theta_{h},..,\theta_k}$ as
	    described above has been constructed. By Lemma~\ref{lem:bridges-exist} there
	    exists a node $\theta'_h$ bridging $\theta'_{h-1}$ and $\theta_h$. Since
	    $\theta_h\tspc (i^{h},t')$ for all $i^{h}\in I^{h}$ we get that  $\theta'_h$ is a bridging centibroom
	     for $\angles{I^{0},\ldots,I^{h}}$ in $\rmmp$. Define
	    $\cC_h=\angles{\theta'_0,..,\theta'_{h},\theta_{h+1},..,\theta_k}$. If $h=k$
	    then we are done. Otherwise, since
	    $\theta_{h'}\tspc\theta_h\spc\theta_{h+1}$ and $\cC_{h-1}=
	    \angles{\theta'_0,..,\theta'_{h-1},\theta_{h},..,\theta_k}$ is a generalized centipede
	    for $\langle i_0,\ldots,i_k\rangle$ in $\rmmp$, we obtain that $\cC_h$ is
	    also such a generalized centipede, as required.
	\end{description}
\end{proofL}

We now formulate the Generalized Centipede Theorem, which will be proved
in the next section. The theorem is proved for protocols weakly solving
the $\OGR$ problem. An immediate corollary gives us that the same conditions
hold for protocols that (non-weakly) solve the $\OGR$. This later corollary can be used
to derive Theorems~\ref{thm:centi2OR} and~\ref{thm:centibroom} as immediate further corollaries. 

\begin{theorem}\label{thm:OGRgcpede}
	Let $\OGR=\angles{\trig,A^{1},\ldots,A^{k}}$, and assume that protocol~$P$
	weakly solves~$\OGR$ in~$\gamma$. 
	Let $r\in\Rrep(P,\gammax)$ be a run in
	which~$\trig$ occurs at time $t=t_{0}$, and where the processes in $I^{h}$ perform the
	actions~$a^{h}_{1}$ to $a^{h}_{\ell_{h}}$ simultaneously at $t_h\geq t_{h-1}$, for
	every $1\le h\le k$, with $t'=t_{k}$.

	\noindent Then there is a generalized centipede for
	$\angles{I^{1},\ldots,I^{k}}$ in $\rmmp$.
\end{theorem}

\subsection{Nested Common Knowledge Gain Requires Generalized Centipedes}\label{sec:NCKgcenti}
We start by revisiting the relation between $\pas$, the past causal cone, and
knowledge. We repeat here the definition of $\pas$ and $\fut$ cones, as we  now
wish to make use of the complete data that is encoded in the definitions.

\renewcommand{\toto}{def:cones}
\begin{definitionOf}
	We define the {\em future causal cone} of a
	node~$\alpha$ (in run $r$) to be $$\fut(r,\alpha)=
	\left\{\angles{\theta,ND_{\theta}}:
	\begin{array}{l}\alpha\spc\theta\ \mbox{ in } r \text{ and }ND_{\theta}\\
	\text{is the set of ND events and initial states in }\theta\text{ in }r
	\end{array}\right\}.$$  
	
	Similarly, the {\em past causal cone} of~$\alpha$ is
	$$\pas(r,\alpha)=
	\left\{\angles{\theta,ND_{\theta}}:
	\begin{array}{l}\theta\spc\alpha\ \mbox{ in } r \text{ and }ND_{\theta}\\
	\text{is the set of ND events and initial states in }\theta\text{ in }r
	\end{array}\right\}.$$  
\end{definitionOf}

Although $\pas$ and $\fut$ are sets that contain pairs of node and event-set, 
we will frequently treat them simply as sets of nodes,
when the second component of the pair is irrelevant in the context.

Lemma~\ref{lem:past-local}  showed that the local state of a
process, and hence also its knowledge state, is determined by its past causal cone. 
A straight forward extension shows that common knowledge of a group of processes 
is determined by the union of their past cones.

\begin{definition}[Group past cones]
	For every $G\subseteq\Proc$ and time $t$, we will write 
	\begin{enumerate}
		\item $\uPas(r,G,t)$ to denote the set $\bigcup_{g\in G}\pas(r,(g,t))$, and
		\item $\iPas(r,G,t)$ to denote the set $\bigcap_{g\in G}\pas(r,(g,t))$.
	\end{enumerate}•
\end{definition}•

\begin{lemma}\label{lem:past_CK}
	Fix $r,r'\in \Rmax$, $G\subseteq \Proc$ and time t, such that
	$\uPas(r,G,t)=\uPas(r',G,t)$.\\
	For every $\vphi\in \Lang$, if $(\Rmax,r,t)\sat C_{G}\vphi$ then
	$(\Rmax,r',t)\sat C_{G}\vphi$.
\end{lemma}•

\begin{proofL}
	Suppose that $(\Rmax,r',t)\nsat C_{G}\vphi$. Then there exists some sequence 
	$\angles{g_{1},g_{2},..,g_{k}}$ such that $(\Rmax,r',t)\nsat K_{g_{1}}K_{g_{2}}\cdots K_{g_{k}}\vphi$.
	Write $\vphi'=K_{g_{2}}\cdots K_{g_{k}}\vphi$. We get $(\Rmax,r',t)\nsat K_{g_{1}}\vphi'$.
	From $\uPas(r,G,t)=\uPas(r',G,t)$ we obtain, in particular, that $\pas(r,(g_{1},t))=\pas(r',(g_{1},t))$.
	By Lemma~\ref{lem:past-local} we get that $(\Rmax,r,t)\nsat K_{g_{1}}\vphi'$, which gives us
	$(\Rmax,r,t)\nsat K_{g_{1}}K_{g_{2}}\cdots K_{g_{k}}\vphi$, contradicting the assumption that
	$(\Rmax,r,t)\sat C_{G}\vphi$.
\end{proofL}

The next two lemmas give us an even greater focus on 
the effect of the causal past upon the current epistemic state.
The first lemma points out that the causal past itself is 
fully determined by those nondeterministic events that occur in it.
We add the following definition.

\begin{definition}[Nondeterministic past]
	For every $r\in\Rmax$,node $\theta$ and $G\subseteq \Proc$,
	\begin{enumerate}
		\item Define $\ndpas(r,\theta)=
		\{\angles{\psi,ND_{\psi}}\in \pas(r,\theta):ND_{\psi}\neq \emptyset\}$.
		This is the set of nodes  $\psi$
		such that $\psi\in \pas(r,\theta)$ and either an ND event occurs at
		$\psi$ in $r$ or $\psi$ is an initial state.
		\item Define $\undPas(r,G,t)=\bigcup_{g\in G}\ndpas(r,(g,t))$.
		\item Define $\indPas(r,G,t)=\bigcap_{g\in G}\ndpas(r,(g,t))$.
	\end{enumerate}•	
\end{definition}•

\begin{lemma}\label{lem:nd_past}
	Fix $r,r'\in\Rmax$ and node $(i,t)$. \\
	If $\ndpas(r,(i,t))=\ndpas(r',(i,t))$ 
	then $\pas(r,(i,t))=\pas(r',(i,t))$.
\end{lemma}

\begin{proofL}
	We prove the claim by induction on $t$.
	\begin{description}
		\item[$t=0$]\quad $\pas(r,(i,0))=\{\angles{\ell s_{i}, ND_{\ell s_{i}}}\}$, the singleton initial
		local state of $i$ in $r$. By assumption it is the same state as in
		$r'$, and hence $\pas(r,(i,0))=\pas(r',(i,0))$.
		
		\item[$t>0$]\quad Suppose that $r_{i}(t)\neq r'_{i}(t)$. Then, wlog, by
		Lemma~\ref{lem:past-local} there exists some $\angles{\theta,ND_{\theta}}\in \pas(r,(i,t))$
		such that $\angles{\theta,ND_{\theta}}\notin \pas(r',(i,t))$. If $\theta\spc(i,t)$ in $r'$ then 
		it must be that $ND_{\theta}$ is different in  $r$ and $r'$, in which case we get that
		$\ndpas(r,(i,t))\neq\ndpas(r',(i,t))$, contra the lemma's assumptions. Hence
		 there must exist some $\theta$
		such that $\theta\spc(i,t)$ in $r$ but $\theta\not\spc(i,t)$ in $r'$.
		
		By
		Lemma~\ref{lem:bridges-exist} there exists a bridge $\psi$ such that
		$\theta\spc\psi\tspc(i,t)$. As $\theta\notin \pas(r',(i,t))$, it must be
		that $\theta\not\tspc (i,t)$, and hence that $\psi\neq \theta$.

		We now consider two cases:
		\begin{description}
			\item[$\psi=(i,t)$:] In this case, as
			$\ndpas(r,(i,t))=\ndpas(r',(i,t))$, $\theta\spc \psi$ and an ND
			event occurs at $\theta$ in $r$, it must be that
			$\theta\in\pas(r',(i,t))$, contradicting the assumption that
			$\theta\notin \pas(r',(i,t))$.

			\item[$\psi\neq (i,t)$:] In this case from
			Lemma~\ref{lem:step-synch}, it must be that $\psi=(j,t')$ for some
			$t'<t$. From $\ndpas(r,(i,t))=\ndpas(r',(i,t))$ and
			$\psi\in\pas(r,(i,t))$ we obtain that
			$\ndpas(r,(j,t'))=\ndpas(r',(j,t'))$. By the inductive hypothesis
			$\pas(r,(j,t')=\pas(r',(j,t'))$ and hence, as
			$\theta\in\pas(r,(j,t'))$ it must also be that
			$\theta\in\pas(r,(i,t))$, again contra the assumption that
			$\theta\notin \pas(r',(i,t))$.
		\end{description}•
	\end{description}•
\end{proofL}

Proving the next lemma is done by  composing the two previous lemmas. 

\begin{lemma}\label{lem:nd_past_CK}
	Fix $r,r'\in\Rmax$, time $t$ and $G\subseteq \Proc$. 
	If $\undPas(r,G,t)=\undPas(r',G,t)$ , then
	if  $(\Rmax,r,t)\sat C_{G}\vphi$ then
	$(\Rmax,r',t)\sat C_{G}\vphi$.
\end{lemma}•

%

\begin{definition}[Centibroom past]
	Let $r\in\Rmax$, $G\subseteq \Proc$ and fix time $t$.
	The \emph{centibroom past} of $G$ in $r$ at time $t$ is the set 
	$$\BPas{r,t,G}=\left\{\angles{\theta,ND_{\theta}}\in \undPas(r,G,t)|
	\begin{array}{l}\text{there exists a centibroom for}\\ \angles{\theta,G} \text{ in }(r,0..t)\end{array}
	\right\}.$$
\end{definition}•

As before, in the case of $\pas$ and $\fut$, we will often  treat $\BPas{r,t,G}$ as a set of nodes,
rather than a set of pairs.
Note that $\BPas{r,t,G}\subseteq \indPas(r,G,t) \subseteq \undPas(r,G,t)$. 

When
$G=\{i,j\}$, we partition the nodes in $\undPas(r,G,t)$ based on the existence
of bridging nodes. Recall that a  node $b$ bridges $\theta$ and $\psi$ if
$\theta\spc b \tspc \psi$ and there is no node $b'$ such that $b'\neq b$ and
$\theta\spc b' \tspc b$. Note that there may exist more than one bridge node
connecting $\theta$ and $\psi$, if there is more than one syncausal path between
the nodes. In the following partition, for each node $\theta\in \undPas(r,G,t)$
we look at the nodes bridging $\theta$ and $(i,t)$, as well as those nodes
bridging $\theta$ to $(j,t)$. Moreover, we focus on those bridging nodes that
are \emph{earliest}: $(i,t)$ is an earliest node bridging $\theta$ and $\psi$ if
it bridges the two nodes and if there is no alternate bridging node $(j,t')$ for
$\theta$ and $\vphi$ such that $t'<t$.

\begin{definition}[Partitioning  $\undPas(r,G,t)$]\label{def:past_partition}
	Given a run $r\in\Rmax$, a time $t$ and  $G=\{i,j\}$, for each $\angles{\theta,ND_{\theta}}\in
	\undPas(r,G,t)$ we have $\theta\spc (i,t)$ or $\theta\spc (j,t)$. Let
	$t^{\theta}_{i}$ denote the time of the earliest bridge nodes between $\theta$ and $(i,t)$, 
	with $t^{\theta}_{i}=\infty$ if there are no bridges between $\theta$ and $(i,t)$ (i.e. $\theta\not\spc (i,t)$).
	Similarly define $t^{\theta}_{j}$. The set $\undPas(r,G,t)$ can be partitioned into the
	following subsets:
	
	\begin{enumerate}
		\item $\BPas{r,t,G}$
		\item $\bri=\{\angles{\theta,ND_{\theta}}\in 
		\undPas(r,G,t)- \BPas{r,t,G}| t^{\theta}_{i}< t^{\theta}_{j}\}$
		\item $\brj=\{\angles{\theta,ND_{\theta}}\in 
		\undPas(r,G,t)- \BPas{r,t,G}| t^{\theta}_{j}< t^{\theta}_{i}\}$
		\item $\brn=$\\$\{\angles{\theta,ND_{\theta}}\in 
		\undPas(r,G,t)- \BPas{r,t,G}| t^{\theta}_{j}= t^{\theta}_{i}<\infty\}$
	\end{enumerate}•

\end{definition}

Notice that Definition~\ref{def:past_partition} first constructs the cell for all
pairs $\angles{\theta,ND_{\theta}}\in \undPas(r,\{i,j\},t)$ for which there exists a centibroom for
$\angles{\theta,\{i,j\}}$ in $r$, and then takes all the remaining nodes
in $\undPas(r,\{i,j\},t)$ and partitions them further into $\bri,\brj$ and $\brn$. In particular, this
means that in case of some $\theta\in \brn$, where $t_{i}=t_{j}$, the nodes bridging $\theta$ to $(i,t)$ will be distinct from those bridging $\theta$ to $(j,t)$, or otherwise we would have that $\theta\in \BPas{r,t,G}$. 

Once again using $t^{\theta}_{i}$ to denote the time of the earliest bridge
nodes between $\theta$ and $(i,t)$, with $t^{\theta}_{i}=\infty$ if there are no
bridges, we divide the cells of Definition~\ref{def:past_partition} further, by
``slicing'' each cell according to the time associated with the earliest
bridging nodes.

\begin{definition}[slicing the partition cells of $\undPas(r,\{i,j\},t)$]\mbox{}\\
	Given a run $r\in\Rmax$, a time $t$ and  $G=\{i,j\}$, let 
	$\BPas{r,t,G}, \bri, \brj$ and $\brn$ form the
	partition in Definition~\ref{def:past_partition}.
	
	We divide some of the partition cells further into time-slices in the following
	way
	
	\begin{enumerate}
		\item $\bri=\bigcup _{t'\leq t}\bri(t')$, 
		where $\bri(t')=\{\angles{\theta,ND_{\theta}}\in \bri| t^{\theta}_{i}=t'\}$, 
		\item $\brj=\bigcup _{t'\leq t}\brj(t')$, 
		where $\brj(t')=\{\angles{\theta,ND_{\theta}}\in \brj| t^{\theta}_{j}=t'\}$, and
		\item $\brn=\bigcup _{t'\leq t}\brn(t')$, where\\
		$\brn(t')=\{\angles{\theta,ND_{\theta}}\in \brn| t^{\theta}_{j}=t^{\theta}_{i}=t'\}$. 
	\end{enumerate}•
\end{definition}•

For every $G\subseteq \Proc$, we will say that runs $r$ and $r'$ are $G$-reachable
at time $t$ if there exists a sequence $\angles{r_{0}, \ldots r_{k}}$ such that
$r=r_{0}, r'=r_{k}$, and for each $h<k$ there exists $i_{h}\in G$ such that
$r_{h}\sim_{(i_{h},t)}r_{h+1}$. The next lemma shows that the nodes 
in cells $\bri, \brj$ and $\brn$ 
are not essential for determining $G$-reachability for groups of size $2$.

\begin{lemma}\label{lem:nd_cleanup}
	Fix $r,r'\in \Rmax$, time $t$, $G=\{i,j\}\subseteq \Proc$. 
	There exists $C\geq 0$ and a sequence of runs $\angles{r_{1},r_{2},..,r_{2C}}$
	such that $r$ and $r_{2C}$ are $G$-reachable, 
	$\BPas{r_{2C},t,G}=\BPas{r,t,G}$, 
	and 
	$\bri[r_{2C}]=\brj[r_{2C}]=\brn[r_{2C}]=\emptyset.$
\end{lemma}

\begin{proofL}
	We prove by induction that for each  $d=0,..,t$ that there exists a run
	$r_{2d}$ such that
	\begin{enumerate}
		\item $r$ and $r_{2d}$ are $G$-reachable,
		\item $\BPas{r_{2d},t,G}=\BPas{r,t,G}$,
		\item $\bri[r_{2d}]=\bigcup_{1\leq h \leq  t-d}\bri(h)$ ,
		\item $\brj[r_{2d}]=\bigcup_{1\leq h \leq  t-d}\brj(h)$, and
		\item $\brn[r_{2d}] = \bigcup_{1\leq h \leq  t-d}\brn(h)$.
	\end{enumerate}•
	
	\begin{description}
		\item[$d=0$]\quad In this case $r_{2d}=r_{0}=r$, and all requirements
		trivially hold (for example, $\bri[r_{0}]=\bri=\bigcup_{1\leq h
		\leq  t}\bri(h)$).

		\item[$d>0$]\quad Inductively assume the existence of a run $r_{2d-2}$ 
		satisfying all requirements. Note that  all $\theta$ in
		$\bri[r_{2d-2}]$, in $\brj[r_{2d-2}]$ and in
		$\brn[r_{2d-2}]$ occur no later than at $t-d+1$. Let $r_{2d-1}$
		be a run identical to $r_{2d-2}$, except that for every
		$\theta\in \bri[r_{2d-2}](t-d+1) \cup \brn[r_{2d-2}](t-d+1)$,
		$\theta\spc(i,t)$ but $\theta\not\spc(j,t)$.
		
		We now show that such a run exists. Iterating over every node 
		$\theta\in \bri[r_{2d-2}](t-d+1) \cup \brn[r_{2d-2}](t-d+1)$,
		we examine two possible cases:
		\begin{description}
			\item[{$\theta\in \brn[r_{2d-2}](t-d+1)$}] \quad For every node
			$b^{\theta}_{j}$ bridging $\theta$ and $(j,t)$, arbitrarily choose
			enough early events on the path $\theta \spc b^{\theta}_{j}$ that
			occur in $r_{2d-2}$, and cancel their occurrence in $r_{2d-1}$, so
			as to make sure that $\theta \not\spc b^{\theta}_{j}$ in $r_{2d-1}$.
			This is possible since $\theta\not\tspc b^{\theta}_{j}$, or we
			would have that $\theta\in \brj$ or $\theta\in \BPas{r,t,G}$.

			Since by assumption  $\theta\notin \BPas{r_{2d-2},t,G}$ , changing
			occurrences in $\fut(r_{2d-1},\theta)$ does not alter the set\\
			$\BPas{r_{2d-1},t,G}$. Moreover, as there exists a bridge
			$b^{\theta}_{i}\neq b^{\theta}_{j}$ for all $b^{\theta}_{j}$, making
			changes in the nodes of\\ $\fut(r_{2d-1},\theta) \cap \pas(r_{2d-1},(j,t))$ does not affect
			$\ndpas(r_{2d-2},(i,t))$.
			
			\item[{$\theta \in \bri[r_{2d-2}](t-d+1)$}]\quad 
			If $\theta\not\spc(j,t)$ then $\theta$ does not require that we alter $r_{2d-1}$
			with respect to the current $r_{2d-2}$. Otherwise, $\theta\spc(j,t)$, but for every
			node $(b_{j},t_{j})$ bridging $\theta$ and $(j,t)$ in $r_{2d-2}$, there exists some 
			$(b_{i},t-d+1)$ bridging $\theta$ and $(i,t)$ in $r_{2d-2}$ such that $t-d+1<t_{j}$.
			It could be that $(b_{i},t-d+1)\spc(b_{j},t_{j})$ for some such bridges, or it could be
			that there is no causal relation between the bridges. 
			
			In either case, $(b_{i},t-d+1)\not\tspc(b_{j},t_{j})$ or we would
			have that $\theta\in \BPas{r,t,G}$. In the first case, we choose
			$r_{2d-1}$ such that enough early receive events are canceled along
			every path from $(b_{i},t-d+1)$ to  $(b_{j},t_{j})$, so that
			$\theta\not\spc (b_{j},t_{j})$ for every such bridge. In the second
			case we choose $r_{2d-1}$ such that early receive events can be
			cancelled anywhere along the path from $\theta$ to $(b_{j},t_{j})$,
			once again resulting in $\theta\not\spc (b_{j},t_{j})$. Having gone
			over all nodes bridging $\theta$ and $(j,t)$ and removed bridges
			from $r_{2d-1}$, we end up with $\theta\not\spc(j,t)$ in $r_{2d-1}$.
			
			We get that $\BPas{r_{2d-1},t,G}=\BPas{r_{2d-2},t,G}$, for the same
			reasons as in the previous case. Moreover, the early receives cancelled in 
			${r_{2d-1}}$ with respect to ${r_{2d-2}}$ are either not in $\pas(r_{2d-1},(i,t))$, or
			are in nodes $\psi$ such that $\psi\tspc(i,t)$. In either case then, canceling early
			receives does not alter the set $\ndpas(r_{2d-2},(i,t))$ and we have\\
			$\ndpas(r_{2d-1},(i,t))=\ndpas(r_{2d-2},(i,t))$.			
	\end{description}

	In both cases examined we get that 
	$$\ndpas(r_{2d-1},(i,t))=\ndpas(r_{2d-2},(i,t))$$
	and hence that ${r_{2d-1}}_{i}(t)={r_{2d-2}}_{i}(t)$. We also get that
	$$\BPas{r_{2d-1},t,G}=\BPas{r_{2d-2},t,G}.$$ Therefore, based on the inductive
	hypothesis we obtain that 
	\begin{itemize}
		\item $\{r_{2d-1}\}_{i}(t)=\{r\}_{i}(t)$ - the local states of $i$ at time $t$ in runs $r_{2d-1}$ 
		and $r$ are identical, and
		\item $\BPas{r_{2d-1},t,G}=\BPas{r,t,G}$. 
	\end{itemize}
	Moreover,
	$\bri[r_{2d-1}](t-d+1)=\emptyset$, and again based on induction that
	$\bri[r_{2d-1}]=\bigcup_{1\leq h \leq  t-d}\bri(h)$. Finally we have
	for every $\psi\in \brn[r_{2d-1}](t-d+1)$ that $\psi\not\spc(j,t)$ and that, as
	we only remove early receives, $\brn[r_{2d-1}](t-d+1)\subseteq \brn[r_{2d-2}](t-d+1)$.

	We now apply the same arguments in order to choose the run $r_{2d}$,
	replacing $j$ with $i$ whenever possible. For every $\theta\in \brn[r_{2d-1}](t-d+1)$,
	we cancel enough early receives so as to make sure that $\theta\not\spc(i,t)$.
	As we also had that $\theta\not\spc(j,t)$ for every $\theta\in \brn[r_{2d-1}](t-d+1)$, 
	we end up with $\brn[r_{2d}](t-d)=\emptyset$
	Summing up, we get that 
	\begin{itemize}
		\item[] $r_{2d-1}\sim_{(i,t)}r_{2d-2}$ and 	$r_{2d}\sim_{(j,t)}r_{2d-1}$,
		\item[] $\bri[r_{2d}]=\bigcup_{1\leq h \leq  t-d}\bri(h)$,
		\item[] $\brj[r_{2d}]=\bigcup_{1\leq h \leq  t-d}\brj(h)$, and finally that
		\item[] $\brn[r_{2d}] = \bigcup_{1\leq h \leq  t-d}\brj(h) $.
	\end{itemize}•
	
	and the induction is complete.
	\end{description}
	
	In particular, for $C=t$ we get the lemma's required result.
\end{proofL}

We are now ready to prove the following theorem, 
showing that the state of common knowledge in a group of processes $G$
is characterized in a precise sense by those nodes of their 
pasts in which an ND event occurs and which are centibroom-related
to all of the processes in $G$.

\begin{theorem}\label{thm:CP_CK}
	Fix $r,r'\in\Rmax$, time $t$ and $G\subseteq \Proc$. \\
	If  $\BPas{r,t,G}=\BPas{r',t,G}$ then \\
	$(\Rmax,r,t)\sat C_{G}\vphi$,
	~~iff~~ $(\Rmax,r',t)\sat C_{G}\vphi$.
\end{theorem}•

\begin{proofT}
	Cases where $|G|=0$ are trivial, and those where  $|G|=1$ are solved using 
	Lemma~\ref{lem:nd_past}. So assume that $|G|\geq 2$. Suppose that 
	$(\Rmax,r',t)\nsat C_{G}\vphi$. Then there exists some formula 
	$\vphi'=K_{i_{1}}K_{i_{2}}\cdots K_{i_{k}}\vphi$ and some $G'=\{i,j\}\subseteq G$ such that
	$(\Rmax,r,t)\sat C_{G'}\vphi'$ but $(\Rmax,r',t)\nsat C_{G'}\vphi'$.
	
	From Lemma~\ref{lem:nd_cleanup}
	we get that there exists a sequence of runs $\angles{r_{1},..,r_{k}}$ such that
	\begin{itemize}
		\item[(i)] $r\sim_{(i,t)}r_{1}\sim_{(j,t)}r_{2}\sim_{(i,t)}\cdots
			\sim_{(i,t)}r_{k-1}\sim_{(j,t)}r_{k}$, and 
		\item[(ii)]$\BPas{r_{k},t,G'}=\BPas{r,t,G'}$, and
		\item[(iii)] $\bri[r_{k}]=\brj[r_{k}]=\brn[r_{k}]=\emptyset$.
	\end{itemize}
		
	From $(i)$ above and from $(\Rmax,r,t)\sat C_{G'}\vphi$ 
	we get that $(\Rmax,r_{k},t)\sat C_{G'}\vphi$.		
		
	Again applying Lemma~\ref{lem:nd_cleanup} we obtain a sequence of runs 
		$\angles{r'_{1},..,r'_{m}}$ such that
		\begin{itemize}
			\item[(i')] $r'\sim_{(i,t)}r'_{1}\sim_{(j,t)}r'_{2}\sim_{(i,t)}\cdots
				\sim_{(i,t)}r'_{m-1}\sim_{(j,t)}r'_{m}$, and 
			\item[(ii')]$\BPas{r'_{m},t,G'}=\BPas{r',t,G'}$, and
			\item[(iii')] $\bri[r'_{m}]=\brj[r'_{m}]=\brn[r'_{m}]=\emptyset$.
		\end{itemize}
		
	From $(i')$ above and from $(\Rmax,r',t)\nsat C_{G'}\vphi$ 
	we get that $(\Rmax,r'_{m},t)\nsat C_{G'}\vphi'$.
		
	Note that given $(ii),(iii),(ii'),(iii')$ and from $$\BPas{r,t,G'}=\BPas{r',t,G'}$$ 
	we get that $\undPas(r_{k},G',t)=\undPas(r'_{m},G',t)$. By using the Theorem's assumptions
	regarding ND events and initial states and Lemma~\ref{lem:nd_past_CK}, we obtain
	that $(\Rmax,r_{k},t)\sat C_{G'}\vphi'$ implies $(\Rmax,r'_{m},t)\sat C_{G'}\vphi'$,
	contradicting the above result $(\Rmax,r'_{m},t)\nsat C_{G'}\vphi'$.
\renewcommand{\toto}{thm:CP_CK}
\end{proofT}

Theorem~\ref{thm:CP_CK} can be weakened into the following 
useful corollary. Considering that $\BPas{r,t,G}\subseteq \indPas(r,G,t)$, we get 

\begin{corollary}
	Fix $r,r'\in\Rmax$, time $t$ and $G\subseteq \Proc$. 
	If $\indPas(r,G,t)=\indPas(r',G,t)$ 
	then
	if  $(\Rmax,r,t)\sat C_{G}\vphi$ then
	$(\Rmax,r',t)\sat C_{G}\vphi$.
\end{corollary}

At long last we are ready to prove that nested common knowledge gain 
necessitates the existence of a generalized centibroom that relates
the process-groups with the triggering node. The proof proceeds
much in the same fashion as that of Theorem~\ref{thm:cpede}, but
process groups have replaced individual processes. Thus,
Theorem~\ref{thm:nckg} generalizes both Theorems~\ref{thm:cpede} and~\ref{thm:CKgain}.

\begin{theorem}[Nested Common Knowledge Gain]\label{thm:nckg}
    Let~$P$ be a deterministic protocol, $I^{h}\subseteq\Proc$ for $h=1\ldots
    k$, and let $r\in\Rmax=\Rrep(P,\gammax)$. Assume that   $e$ is an ND event
    at $(i_0,t)$ in~$r$. If $$(\Rmax,r,t')\sat C_{I^{k}}C_{I^{k-1}}\cdots
    C_{I^{1}}\ndocc{e},$$ then there is a generalized centipede for
    $\angles{I^{1},\ldots,I^{k}}$ in $\rmmp$.
\end{theorem}

\begin{proofT}
	We shall prove the contrapositive form: if no bridging generalized centipede 
	for $\angles{I^{1},\ldots,I^{k}}$ exists in $\rmmp$, then
	$$(\Rmax,r,t')\nsat C_{I^{h}}C_{I^{h-1}}\cdots C_{I^{1}}\ndocc{e}.$$ We reason by
	induction on $k\ge 1$:
	
	\begin{description}
		\item[$k=1$]\quad By assumption, there is no generalized centipede for
		$\angles{I^{1}}$ in $\rmmp$. Hence, by definition of generalized centipede
		there is no centibroom for $\angles{(i_{0},t), I^{1}}$ in $\rmmp$. By
		Theorem~\ref{thm:CKgain} it follows that $(\Rmax,r,t')\nsat
		C_{I^{1}}\ndocc{e}$, as claimed.
			
		\item[$k\ge 2$]\quad Assume inductively that the claim holds for $k-1$.
		Moreover, assume that no bridging generalized centipede for
		$\angles{I^{1},..,I^{k}}$ exists in $\rmmp$. For every $r'\in\Rmax$ let
		$$C^{r'}=\left\{\psi_{k-1}\;\middle\vert\; 
		\begin{array}{l}
		\angles{\psi_{0},\ldots,\psi_{k-1}} \text{ is a bridging generalized centipede}\\
		\text{for } \langle I^{1},\ldots,I^{k-1}\rangle\text{ in }(r',t..t')
		\end{array}
		\right\}.$$
		
		Observe that  for every $\psi'_{k-1}\in C^{r}$, there is no centibroom
		node $\vphi$ for $\angles{\psi'_{k-1},I^{k}}$ in $\rmmp$. Otherwise, by
		Lemma~\ref{lem:bribroom_exists}, there would exist a bridging centibroom
		$\vphi'$ for $\angles{\psi'_{k-1},I^{k}}$, and
		$\angles{\psi'_{0},\ldots,\psi'_{k-1},\vphi'}$ would be a bridging
		generalized centipede for  $\angles{ I^{1},\ldots,I^{k}}$, contradicting
		our assumption. Thus, $C^{r} \cap \BPas{r',t,G}=\emptyset$.
			
		Choose $r'\in\Rmax$ such that
		\begin{itemize}
			\item [(i)] the environment's actions at all nodes in
		    $\pas(r,\theta)$ for every $\theta\in \BPas{r,t',G}$ are identical to those in~$r$; and
		    \item[(ii)] all messages delivered to nodes not in 
		    $\pas(r,\theta)$ for any of the  $\theta\in \BPas{r,t',G}$, are
		    delivered at the maximal possible transmission time according
		    to the bounds~$\bij$.
		\end{itemize}•
		\vskip -3mm
		To see that such a run $r'$ indeed exists in $\Rmax$, we note that clauses
		(i) and (ii) relate to different sets of nodes, that it is impossible by definition of $\BPas{r,t',G}$ 
		that there exists some $\theta\notin \BPas{r,t',G}$ such that $\theta\in \pas(r,\psi)$
		for some $\psi\in \BPas{r,t',G}$, and that by definition all early message receives can
		be delayed, independent of the run's past or concurrent events. Since
		$\Rmax$ contains all runs of~$P$ in~$\gammax$, it must include~$r'$.
	
		Notice that by construction of~$r'$ we have that
		$\alpha\spc\alpha'$ holds in $r'$ only if $\alpha\spc\alpha'$ in $r$, and
		that  every early receive in~$r'$ is an early receive in~$r$. Considering
		that bounds are universal in all runs, we obtain that every bridge node in
		$r'$ is also a bridge node in $r$, and hence that $C^{r'}\subseteq C^r$. By
		definition of~$r'$, and since $C^{r} \cap \BPas{r,t',G}=\emptyset$, 
		none of the nodes in the set $C^r$, and hence also in 
		$C^{r'}$, experiences an early receive in~$r'$. Yet from
		Lemma~\ref{lem:bridges-early}  and from $(i_0,t)\notin C^{r'}$ it follows
		that every node $\theta'\in C^{r'}$ must be a nontrivial bridge node
		in~$r'$,  thus experiencing an early receive. We therefore conclude that
		$C^{r'}=\emptyset$.
		
		By on the inductive hypothesis we obtain from this that 	
		$(\Rmax,r',t')\not\sat C_{I^{k-1}}\cdots
		C_{I^{1}}\ndocc{e}$, and  using the Knowledge Axiom we get that
		$(\Rmax,r',t')\not\sat C_{I^{k}}C_{I^{k-1}}\cdots	C_{I^{1}}\ndocc{e}$. 
		By applying Theorem~\ref{thm:CP_CK} we get that 
		$$(\Rmax,r,t')\not\sat C_{I^{k}}C_{I^{k-1}}\cdots	C_{I^{1}}\ndocc{e},$$
		and we are done.
	\end{description}
\renewcommand{\toto}{thm:nckg}
\end{proofT}

Using Theorems~\ref{thm:OGRnck} and~\ref{thm:nckg}, we can now prove
Theorem~\ref{thm:OGRgcpede}. The proof repeats that of Theorem~\ref{thm:centi2OR}
almost to the letter so it will not be repeated here.

\section{Conclusions}\label{sec:NCK_conclusions}
This chapter introduced the $\OGR$ problem, along with nested common knowledge
and the generalized centipede. These provide a unifying theory for the concepts and results
presented in Chapters~\ref{ch:coordination}, \ref{ch:NKG} and~\ref{ch:CKG}. 

But these concepts also provide important infrastructure for solutions 
to the \emph{Generalized Ordering} problem, which will be discussed in the next chapter. In
particular, the notion of \emph{weakly solving} $\OGR$ will play a central part.

An interesting result presented in this chapter is Theorem~\ref{thm:CP_CK}, that 
characterizes common knowledge among a group $G$, based on a subset of nodes
in their shared pasts. Thus, neither complete information about the local states of these processes
nor  about their causal past are needed, in order to settle
the scope of common knowledge among the group's members.

\ifx \isEmbedded \undefined


\maketitle

\else
\fi
\label{ch:NKG}
\section{Introduction}\label{sec:NKG_intro}
This chapter investigates the minimal communication needed in order to achieve
nested knowledge gain in synchronous systems. As such, its results provide an
immediate generalization of previous findings pertaining to
asynchronous ones. 

We have argued elsewhere (see Chapter~\ref{ch:coordination}) that knowledge gain
can be seen as a close approximation of the spread of causal effect in
distributed systems. Yet even though it is more rigorously defined than
causality, knowledge is still a rather abstract notion. Thus, we motivate our
investigation by  studying the more concrete Ordered Response problem. We will
show that in order for a protocol to solve the problem, a certain generalization
of message chains must relate the trigger and responding sites in every
triggered run.

Sections~\ref{sec:NKG_intro} through~\ref{sec:K2sync} will introduce and discuss
the new concepts involved in the analysis, and informally sketch out the results
in terms of a necessity relation tying in communication to Ordered
Response solutions. Sections~\ref{sec:K2sync} to~\ref{sec:suff_NKG} will then
retrace our steps and provide the necessary formal underpinnings that uphold these
results. The methodology, as discussed in Section~\ref{sec:role_of_knowledge},
will be to prove that certain communication patterns are necessary in order for
knowledge gain to arise, and then to use Theorem~\ref{thm:ORnestedK} to similarly
relate these patterns  to an Ordered Response.   

For the sake of clear presentation, we assume throughout this chapter and the
next one that all examples and proofs take place over a synchronous system in
which upper bounds are given for every existing communication channel. We have
denoted contexts that generate such systems by $\gammax$ (see
Section~\ref{sec:gamma_bounded} for more). We start
by scrutinizing the frozen account example first shown in
Section~\ref{sec:coord_via_response}.

\begin{example}
\label{ex1}
Charlie's bank account is temporarily suspended due to credit
problems. Should Charlie make a sufficient deposit at his local
branch, Banker Bob at headquarters will re-activate the account.
Alice holds a cheque from Charlie, but trying to cash it before the
account is re-activated will grant her a fine, rather than cash.
Alice, Bob and Charlie can communicate over a communication network as
depicted in Figure~\ref{fig:net1}. In particular, messages from
Charlie to Bob and Alice take up to 10 and 12 days to be delivered,
respectively.
This scenario can be viewed as an instance of \ORi\ in which a deposit
by Charlie is the event, and the responses are the account
re-activation by Bob followed by Alice's cashing of the cheque.

\begin{figure}
	\centering
	\subfloat[The network of Example~\ref{ex1}]{
		\label{fig:net1}
		\newcommand{\tikzfname}{tikz_synch_net}
		\tikzsetnextfilename{\cfcurrentfolder Figures/\tikzfname}
		\cfinput{Tikz/\tikzfname}
	}
	%
	\subfloat[Cheque cleared at $t+10$]{
		\label{fig:run1}
		\newcommand{\tikzfname}{tikz_synch_chain1a}
		\tikzsetnextfilename{\cfcurrentfolder Figures/\tikzfname}
		\scalebox{0.75}{\cfinput{Tikz/\tikzfname}}
	}
	\caption{Example~\ref{ex1}}
	\label{fig:ex1}
\end{figure}

In a particular instance, depicted in Figure~\ref{fig:run1}, Charlie makes a
deposit at time~$t$, and immediately broadcasts a message stating this to both
Alice and Bob. The message reaches Bob in~2 days and Alice in~4.  Bob
immediately re-activates Charlie's account \footnote{For ease of exposition, we
assume throughout the thesis that actions are performed instantaneously;
alternative assumptions would not significantly affect the analysis.}  at time
$t+2$.  When can Alice deposit the cheque? The cheque would be cashed
successfully at any time after $t+2$. However, Alice only knows about Charlie's
deposit at $t+4$. But even at that point, she must keep waiting. In the absence
of additional information indicating when Bob actually received Charlie's
message, she is only guaranteed that this will happen by time $t+10$. Knowing
Bob's protocol, she can safely submit the cheque at or after time $t+10$, but
not sooner.  \hfill$\Box$
\end{example}

In this example, Alice acts after Bob does. While in an asynchronous setting she
would need to obtain explicit notification that Bob acted, in the synchronous
setting considered in Example~\ref{ex1} she can base her action on the
information that Charlie sent Bob the message at time~$t$, combined with the
bound determining when this message will arrive, and her knowledge of Bob's
protocol, which ensures that Bob will act immediately upon receiving Charlie's
message. Her action, which  clearly depends on Bob's action having taken place,
can be performed without an explicit message chain from Bob. Example~\ref{ex2}
illustrates a variation on Example~\ref{ex1}, in which Alice can do slightly
better.

\begin{example}
\label{ex2}
In a setting similar to Example~\ref{ex1}, Susan is Bob's supervisor at the
bank. The network is now as depicted in Figure~\ref{fig:net2}. Suppose that
Charlie broadcasts his deposit to all three, and that communication is delivered
as in Figure~\ref{fig:run2}. In this case Alice can, as before, submit her
cheque at $t+10$. But she can do even better. Since she receives a message from
Susan at $t+8$ that was sent at $t+3$, the bound on the Susan-Bob channel
ensures her that Bob is to be notified of Charlie's deposit by time $t+7$. Thus
Charlie's account will also be solvent as of time $t+7$, and Alice can safely
cash her cheque in this case upon receiving Susan's message.
\hfill$\Box$
\end{example}

\begin{figure}
\centering
\subfloat[The network of Examples~\ref{ex2} and~\ref{ex3}]{
	\label{fig:net2}
		\newcommand{\tikzfname}{tikz_synch_net2}
		\tikzsetnextfilename{\cfcurrentfolder Figures/\tikzfname}
		\cfinput{Tikz/\tikzfname}
}
\subfloat[Susan speeds up Alice's response]{
	\label{fig:run2}
		\newcommand{\tikzfname}{tikz_synch_chain2y}
		\tikzsetnextfilename{\cfcurrentfolder Figures/\tikzfname}
		\scalebox{0.75}{\cfinput{Tikz/\tikzfname}}	
}
\caption{Examples~\ref{ex2} and~\ref{ex3} }
\label{fig:ex2}
\end{figure}

In both examples, the timing of Alice's action depends on the time
bounds, but Example~\ref{ex2}  shows a more complex interaction
between message arrivals and time bounds. In Example~\ref{ex1}
Alice combines information gained by means of a message chain from
Charlie to her with the known time bound on the Charlie-Bob
channel. Charlie's message to her serves both to notify her about the
occurrence of a deposit event, and as a temporal anchor for a
timing argument that allows her to properly coordinate her response
with Bob's action.

Example~\ref{ex2} starts out the same for Alice. She is still notified of the
deposit event by a message from Charlie at $t+4$. As before, this message can
also be used to coordinate her response to follow Bob's by clearing the cheque
at $t+10$. But then the message from Susan arrives, and a second message chain
between Charlie and Alice  is completed. Alice already knows about Charlie's
deposit based on the earlier message. The new message serves her to coordinate a
temporally ``tighter'' response to Bob's action at $t+8$ rather than $t+10$. For
the purpose of coordination, Susan's message plays a similar role in
Example~\ref{ex2} to that played by Charlie's in Example~\ref{ex1}.

Had the triggering event $\trig$ been unconditionally guaranteed to take place
at some time $t_0$, then the protocol could directly specify the times $t_k\ge
t_{k-1}\ge \cdots\ge t_1\ge t_0$ at which the response actions could be
performed with proper coordination. However, since $\trig$ is a spontaneous
event, knowledge about its occurrence must ``flow'' from~$\itrig$ to the
responding sites. With respect to coordination, however, the above examples
demonstrate that coordination between the sites does not necessary require
explicit communication between successive responses. A site~$i_{h+1}$ may be
able to know that $\alpha_h$ has taken place by combining {\em a priori} 
knowledge regarding timing guarantees, information it has regarding other
processes' protocols, and information it obtains via explicit communication. 

Examples~\ref{ex1} and~\ref{ex2} illustrate how a process can come to coordinate
its response with another process despite the lack of explicit communication
between the sites, based on knowledge of existing upper bounds on communication.
But bounds can be used in an additional fashion. Namely, if by time $t+\bij$
process~$j$ receives no message sent by~$i$ at time~$t$, then~$j$ can discover
that no such message was sent \cite{LamSM}. Depending on $i$'s protocol, this
can provide~$j$ with information about $i$'s state at time~$t$.  Consider the
following refinement of Example~\ref{ex2}.

\begin{example}\label{ex3}
In the network of Example~\ref{ex2} depicted in Figure~\ref{fig:net2},
suppose that Susan sends Alice a message in every round as long as
Susan has {\em not} heard from Charlie about an appropriate deposit.
In this particular instance, Susan receives a message from Charlie at
time $t+2$, at which point she stops sending her update messages.
At time $t+7$ Alice will be able to ``time-out'' on Susan's
time~$t+2$ message. She then knows that Susan  heard from Charlie at
$t+2$. Moreover, knowing that Susan
relays information to Bob as before, Alice knows that Bob heard about
the deposit no later than time $t+6$. Hence, Alice can
safely cash her cheque at time $t+7$ rather than $t+10$.
\hfill$\Box$
\end{example}

In Example~\ref{ex3} Alice learns of Charlie's deposit without receiving {\em
any message whatsoever}. She clearly receives no message chain originating from
Charlie. Nevertheless, it seems instructive to think of Susan as sending Alice a
``NULL message'' in the sense of \cite{LamSM} at time $t+2$, carrying relevant
information, by {\em not} sending an actual message. Under this interpretation,
Example~\ref{ex3} contains a message chain from Charlie to Alice that consists
of Charlie's concrete message to Susan, followed by Susan's $\NULL$  message to
Alice.

Lamport utilized $\NULL$  messages in \cite{LamSM} for his algorithms
implementing the state machine model in the synchronous setting. In the current
paper however, $\NULL$  messages will be used to define a relation called  {\em
syncausality}, extending happened before. A syncausal chain will then be a chain
consisting of a sequence of concrete and $\NULL$  messages. Syncausal chains are
required for information flow regarding nondeterministic events such as
spontaneous external inputs.

\section{Bound Guarantees}\label{sec:Tbounds}
In the synchronous model we consider, every directed communication
link between adjacent processes $i$ and $j$ provides a bound $\bij$ on
the maximal transmission time of messages. These local bounds  naturally
induce more general  bounds, or guarantees as we call them,  for any pair of
(not necessarily directly) connected sites.

\begin{figure}[h!]
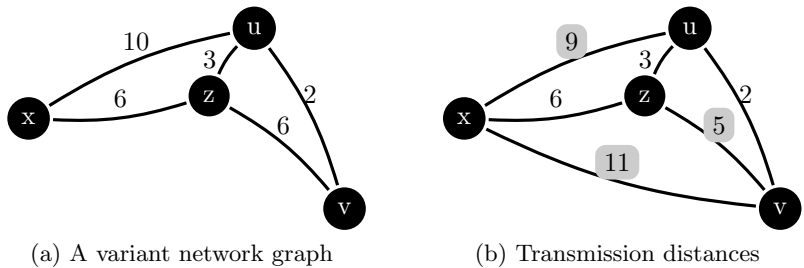

\centering	
\subfloat[A variant network graph]{
		\label{fig:net3}
		\newcommand{\tikzfname}{tikz_guarantee_demo1}
		\tikzsetnextfilename{\cfcurrentfolder Figures/\tikzfname}
		\cfinput{Tikz/\tikzfname}
}
\quad
\subfloat[Transmission distances]{
		\label{fig:tran_dis}
		\newcommand{\tikzfname}{tikz_guarantee_demo2}
		\tikzsetnextfilename{\cfcurrentfolder Figures/\tikzfname}
		\cfinput{Tikz/\tikzfname}
}
\caption{Timing guarantees}
\label{fig:bounds}
\end{figure}

Consider Figure~\ref{fig:net3}, showing a variant of the network
graph shown in Figure~\ref{fig:net2}. Assuming that all of the sites
involved are fully cooperative in relaying messages, a message sent by
Charlie with destination Bob can be guaranteed to arrive after 9 rounds,
if it travels from Charlie to Susan and then to Bob, rather than from
Charlie to Bob directly. Similarly, a message from Susan to Alice can
be guaranteed to arrive after no more than 5 rounds, if it makes the
roundabout trip through Bob. Charlie and Alice may also communicate,
with a bound of 11 rounds, if they use Susan and Bob as a relays.

Thus, the naturally induced {\em   transmission distance} between
processes~$h$ and~$k$, which we denote by $\sD(h,k)$, is the minimal
distance between~$h$ and~$k$ in the 
weighted directed network graph, in which  the bounds $\bij$ on transmission
times are the edge weights. In particular, $\sD(i,i)=0$ for all
$i\in\Proc$. Figure~\ref{fig:tran_dis} shows the induced transmission
distances for the network of Figure~\ref{fig:net3}.  Values that
differ from the ones in (a) are shown in a gray box. 

We find it convenient to represent that time instant~$t$ on process~$i$'s
timeline by the pair $(i,t)$, called a {\em process-time node}, or {\em node}
for short. Based on the transmission time bounds for the channels, we define the
following {\em bound guarantee} relation among process-time nodes:

\begin{definition}[Bound Guarantee]
\label{def:guar}
	We say that  $(i,t)$ and $(j,t')$ are  {\em related by a bound guarantee},
	and write ~$(i,t)\tspc(j,t')$, ~iff ~$t+\sD(i,j)\,\le\, t'$.
\end{definition}

Observe that bound guarantees are  independent of the speed at which messages
{\em actually} arrive at a particular run; they depend only on the weighted
network topology. If $(i,t)\tspc(j,t')$ then it is possible to guarantee that a
message sent by $i$ at time $t$ will arrive at $j$ by time $t'$, assuming that
relay is instantaneous. Since the bound-weighted network is assumed to be known
to the processes, the passage of time can allow a process to obtain knowledge
about remote events that would not be available, say, in an asynchronous
setting.

The next sections will explore
the ways by which this knowledge can be exploited. 

\section{Syncausality}\label{sec:syncausality}
In Example~\ref{ex3} Alice learns of Charlie's deposit without a message chain
from Charlie reaching her. A message chain of a slightly more general type does
exist there, however, in which Susan's not sending a message to Alice at time
$t+3$ is a $\NULL$ message. More formally, consider a network in which $i$ and
$j$ are directly connected by a communication link with bound $\,\bij$. Then~$i$
can be thought of as ``sending'' a $\NULL$ message over this channel at $(i,t)$
if it sends no physical message over the channel at time~$t$. This message is
considered as being ``delivered'' to $j$ at $(j,t+\bij)$ (see \cite{LamSM}). In
the presence of clocks and bound guarantees, $\NULL$ messages can serve to
transfer information between processes. By identifying that no message was sent
at $(i,t)$, process~$j$ may be able to draw nontrivial conclusions about $i$'s
state and~$i$'s knowledge there.

We now formally define {\em Syncausality}, a generalization of
Lamport's\\ happened-before relation that accounts for $\NULL$
messages and is thus based on ``generalized'' message chains that can
contain $\NULL$ messages as well as normal messages. 
The relation is defined over process-time nodes rather than events,
since defining not sending and not receiving messages as explicit
events would be cumbersome. 

\begin{definition}[Syncausality]
\label{def:syncau}
	Fix a run $r$. The {\em syncausality} relation $\spc$ over nodes
	of~$r$ is the smallest relation satisfying the following four conditions:
	\begin{enumerate}
	    \item If ~$t\leq t'$, then $(i,t)\spc (i,t')$;
	    \item If some message is sent at $(i,t)$ and received at $(j,t')$
	      then $(i,t)\spc (j,t')$;
	    \item If no message is sent at $(i,t)$ to $i$'s neighbor~$j$
	     ~then $(i,t)\spc (j,t+\bij)$; and
	    \item If ~$(i,t)\spc (h,\hat{t})$ and ~$(h,\hat{t})\spc (j,t')$,
	      then ~$(i,t)\spc (j,t')$.
	\end{enumerate}
\end{definition}

Clauses (1), (2) and (4) correspond to the local precedence, message precedence
and transitivity clauses that define the happened-before relation. Syncausality
thus refines (and hence directly generalizes) happened-before. The third clause
corresponds to  {\em timeout precedence}, capturing the case of a $\NULL$ 
message being sent by $(i,t)$ and eventually received at $(j,t+\bij)$. We can
thus view syncausality as being based on {\em syncausal chains}, consisting of a
chain of actual and $\NULL$ messages.

Syncausality is also a generalization of bound guarantees. Note that nodes
$(i,t)$ and $(j,t+\bij)$ will always be syncausally related: either $(i,t)$ does
send $j$ a message, in which case the message is received by $(j,t+\bij)$ and
$(i,t)\spc(j,t+\bij)$ will hold by message and local precedence, or no such
message is sent, which case $(i,t)$ and  $(i,t)\spc(j,t+\bij)$ will hold by
timeout precedence. Indeed, a straightforward induction on the number of edges
in the shortest path of length $D(i,j)$ between nodes~$i$ and~$j$ in the
network immediately yields:
\footnote{In fact, given the natural definition of $\Lam$ over
  process-time nodes, $\spc$ is the coarsest common refinement
  of $\Lam$ and $\tspc$.}    
\begin{lemma}\label{lem:guarantee2sync}
	If $(i,t)\tspc(j,t')$ then $(i,t)\spc (j,t')$ in every
        run~$r$.
\end{lemma}

Notice that the bound guarantee relation depends only on the network
and the transmission bounds $\bij$. We view it as being given {\em a
  priori} as part of the context. 
By contrast, clause (2) of the syncausality relation, capturing
message precedence, depends on the actually realized message
transmission times in a given run. Therefore, syncausality is run-dependent. 

We will show in Section~\ref{sec:K2sync} that knowledge about
the occurrence of nondeterministic events can be obtained only by way of
syncausal chains.  
Consequently, for the Ordered Response problem we can show the following: 

\begin{theorem}\label{thm:1R2syncausality}
    Let $P$ be a protocol solving
    $\OR=\angles{\trig,\alpha_1,\ldots,\alpha_k}$. 
    If $\trig$ occurs at $(i_0,t)$ in $r\in\Rmax$, and $\alpha_h$
    does at $(i_h,t_h)$, then  $(i_0,t)\spc(i_h,t_h)$ in $r$.
\end{theorem}

The formal proof of Theorem~\ref{thm:1R2syncausality} can be found in
Section~\ref{sec:K2sync}.  It is obtained by first showing that indeed
$i_h$ needs to know that $\trig$ occurred at $(i_h,t_h)$, and then
showing that without the syncausal connection process~$i_h$ can not
know this. 

\section{Double Response}\label{sec:DResponse}
The examples in the introduction all involve a simple problem of the form $\OR =
\angles{\trig,\alpha_1,\alpha_2}$. We call this form a {\em double response}. We
can view a double response as incorporating two single responses to the
triggering event, with an added  coordination requirement to ensure that  that
$\alpha_2$ does not occur before $\alpha_1$.

If the first response is performed by~$i_1$ at time~$t_1$, and the second by
$i_2$ at $t_2$, then Theorem~\ref{thm:1R2syncausality} implies that in a
triggered run of any protocol solving $\OR$, necessarily $(i_0,t)\spc(i_1,t_1)$
and $(i_0,t)\spc(i_2,t_2)$ must hold. A number of
ways by which the required coordination between $i_1$ and $i_2$ may be
achieved have already been informally considered in
Sections~\ref{sec:NKG_intro} and~\ref{sec:coord_via_response}.

\begin{wrapfigure}{r}{0.33\textwidth}
 \vspace{-6mm}
  \begin{center}
  		\newcommand{\tikzfname}{tikz_double_response0}
		\tikzsetnextfilename{\cfcurrentfolder Figures/\tikzfname}
	    \resizebox{0.30\textwidth}{!}{\cfinput{Tikz/\tikzfname}}
  \caption{}
  \label{fig:4a}
  \end{center}
  \vspace{-6mm}
\end{wrapfigure}

First, if $(i_0,t)\spc(i_1,t_1)\spc(i_2,t')$, as seen in
Figure~\ref{fig:4a}, then it is easy to see that $t_1\leq t'$. Wavy
arrows have replaced the straight ones used in
Section~\ref{sec:NKG_intro}, to denote the possibility of  non-trivial
syncausal chains connecting the nodes. This case echoes the chain
structures prevalent in the asynchronous model, with syncausal chains
replacing the pure message chains. Yet as our previous examples have
shown, there are other possible means for coordination. 

\begin{wrapfigure}{r}{0.33\textwidth}
 \vspace{-6mm}
  \begin{center}
    		\newcommand{\tikzfname}{tikz_double_response1}
		\tikzsetnextfilename{\cfcurrentfolder Figures/\tikzfname}
	    \resizebox{0.30\textwidth}{!}{\cfinput{Tikz/\tikzfname}}
  \caption{}
  \label{fig:4b}
  \end{center}
 \vspace{-6mm}
\end{wrapfigure}

In Example \ref{ex1} we had $i_2$ (played out by Alice) waiting until
a time $t'=t+10$ such that $(i_0,t)\tspc(i_1,t')$. So  by time
$t'$ when $a_2$ is performed, $i_1$ must surely have gotten the
message about the occurrence of $\trig$ and have performed $a_1$.
This possibility is schematically shown in Figure~\ref{fig:4b}. The
figure shows the syncausal relations that are necessary in order for
$i_2$ to coordinate its action with $i_1$. As $i_2$ need not be aware
of the realized connection between $i_0$ and $i_1$ and the time $t_1$
when $a_1$ was actually performed, this connection is not shown.

\begin{wrapfigure}{r}{0.33\textwidth}
 \vspace{-6mm}
  \begin{center}
    		\newcommand{\tikzfname}{tikz_double_response2}
		\tikzsetnextfilename{\cfcurrentfolder Figures/\tikzfname}
    		\resizebox{0.30\textwidth}{!}{\cfinput{Tikz/\tikzfname}}
  \caption{}
  \label{fig:4c}
  \end{center}
  \vspace{-6mm}
\end{wrapfigure}

In Examples~\ref{ex2} and \ref{ex3} there exists some node
$(i_3,t_3)$, representing Susan at times $t+3$ and $t+2$ respectively,
that is tied into the coordination process as depicted in
Figure~\ref{fig:4c}.

In this case $a_2$ is set by $i_2$ to a time $t'$ such that
$(i_3,t_3)\tspc(i_1,t')$.  Observe that the shadowed lines in
Figure~\ref{fig:run2} outline an underlying formation identical to
the one shown in Figure~\ref{fig:4c}.

Another formation that can be used to ensure
coordination had been brought up in Example~\ref{ex-intro}. Consider
Figure~\ref{fig:4d}. Here $i_2$ serves as a relay for $i_1$, and
performs $a_2$ at a time $t'$ such that $(i_2,t_2)\tspc(i_1,t')$. In
terms of the cheque clearance scenario from Section~\ref{sec:NKG_intro},
this would be equivalent to Alice getting a message from Charlie at
time $t_2$ and then forwarding it on to Bob.

\begin{wrapfigure}{r}{0.33\textwidth}
  \vspace{-6mm}
  \begin{center}
        	\newcommand{\tikzfname}{tikz_double_response3}
		\tikzsetnextfilename{\cfcurrentfolder Figures/\tikzfname}
    		\resizebox{0.30\textwidth}{!}{\cfinput{Tikz/\tikzfname}}
  \caption{}
  \label{fig:4d}
  \end{center}
  \vspace{-6mm}
\end{wrapfigure}

Knowing that her message will not take longer than 5 days to arrive,
she waits until day $t_2+5$ and then clears her cheque.

What about other formations? Consider the scenario depicted in
Figure~\ref{fig:4e}. It can be associated with the situation in
Example~\ref{ex1}, at time $t'=t+6$. At this stage, both Alice and Bob
have received a message from Charlie, but Alice cannot be sure that
that the message to Bob has indeed arrived, because the transmission
distance between Charlie and Bob is $10$.
\begin{wrapfigure}{r}{0.33\textwidth}
 \vspace{-6mm}
  \begin{center}
        	\newcommand{\tikzfname}{tikz_double_response4}
		\tikzsetnextfilename{\cfcurrentfolder Figures/\tikzfname}
        \resizebox{0.30\textwidth}{!}{\cfinput{Tikz/\tikzfname}}
  \caption{}
  \label{fig:4e}
  \end{center}
  \vspace{-6mm}
\end{wrapfigure}

\noindent Returning to the figure, at $t'$ process $i_2$ cannot
perform $\alpha_2$ without risking the possibility that $\alpha_1$ has
not yet been performed. A protocol in which $i_2$ \emph{does} do
$\alpha_2$ at $t'$ is one that does not solve the instance of the \OR
problem in cases where the message to $i_1$ takes longer than $6$ time
steps to arrive. As we are assuming a protocol that solves the
ordering problem in all runs, such a scenario is impossible.

\begin{figure}[h!]
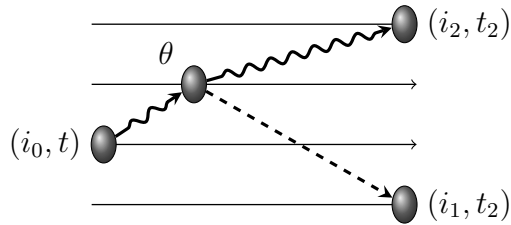

	 \begin{center}
		\newcommand{\tikzfname}{tikz_one_legged-n}
		\tikzsetnextfilename{\cfcurrentfolder Figures/\tikzfname}
		\cfinput{Tikz/\tikzfname}
	\caption{Coordination in Double Response}
	\label{fig:4f}
    \end{center}
\end{figure}

We can show that the formations in Figures~\ref{fig:4a} to
\ref{fig:4d} exhaust the possible causal formations in a protocol that
solves the double response \OR problem. All of the possible
configurations considered above are described by the generalized
Figure~\ref{fig:4f}, where $\theta$ is a parameterized node on a
syncausal path between $(i_0,t)$ and $(i_,t_2)$. If we set
$\theta=(i_1,t_1)$ we get the setting shown in
Figure~\ref{fig:4a}. Similarly, setting $\theta$ to $(i_0,t)$, $
(i_3,t_3)$ and $(i_2,t_2)$ gives us Figures~\ref{fig:4b}, \ref{fig:4c}
and \ref{fig:4d} respectively.\footnote{Recall that both $\spc$ and
  $\tspc$ are reflexive relations, so for example setting
  $\theta=(i_0,t)$ is not at odds with $\theta\spc(i_0,t)$.}

\section{Centipedes}\label{sec:centipedes}
The double response problem incorporates two aspects required for ordering
responses: notification of the responding sites regarding the occurrence of the
trigger event, and coordination of the responses between these sites. This
section tackles the \OR problem in its most general form, where any number of
responses may be required to the trigger's occurrence.

Comparing the  individual instances in Figures~\ref{fig:4a} to
\ref{fig:4d} to their generalization in Figure~\ref{fig:4f}, we see a
pattern emerging, where the actual node standing in for the
parameterized $\theta$ gets informed of the occurrence of the trigger
event, and serves to split the path and route the new information to
both $(i_1,t')$ and $(i_2,t')$. The condition on this splitting node
is that it must be able to guarantee the arrival of information at
$t_1$ by time $t'$. This promise can then be used by $i_2$ to
coordinate its own action with $i_1$'s.

Intuitively, we would expect then to see similar split-and-promise
mechanisms crop up further down the line when multiple responses are
required. This idea gives rise to the 
\emph{centipede} structure, defined below. 

\begin{figure}[h!]
	\centering
		\newcommand{\tikzfname}{tikz_general_OR3}
		\tikzsetnextfilename{\cfcurrentfolder Figures/\tikzfname}
		\cfinput{Tikz/\tikzfname}
	\caption{A centipede}
	\label{fig:centipede}
\end{figure}

\begin{definition}[Centipede]
\label{def:cpede}
Let $r\in\Rmax$, let $i_h\in\Proc$ for $0\le h\le k$ and let $t\le t'$.
A {\em centipede} for $\langle i_0,\ldots,i_k\rangle$ in the interval
$\rmmp$ is a
sequence of nodes ~~$\theta_0\spc\theta_1\spc\cdots\spc\theta_k$
such that $\theta_0=(i_0,t)$, $\theta_k=(i_k,t')$,
and ~$\theta_h\tspc (i_h,t')$ ~holds for $h=1,\ldots, k-1$.
\end{definition}

A centipede for $\langle i_0,\ldots,i_k\rangle$ in the interval $\rmmp$ is
depicted in Figure~\ref{fig:centipede}. Extending Figure~\ref{fig:4f},
Figure~\ref{fig:centipede} shows a syncausal chain extending between $(i_0,t)$
and $(i_k,t')$, and along this chain a sequence of ``route splitting" nodes
$\theta_1,\theta_2,$ etc. such that each $\theta_h$ can guarantee the arrival of
a message to $i_h$ by time $t'$. Such a message can serve to inform $i_h$ of the
occurrence of the trigger event, and as the set of previously made guarantees
gets shuffled on to the next splitting node, each responding site $i_h$ can be
confident that all previous sites $\angles{i_{1},\ldots,i_{h-1}}$ had already
responded.

We remark that, since both $\spc$ and $\tspc$ are reflexive, it is possible for
adjacent $\theta_j$'s to coincide. Moreover, it is possible (in fact, probably
quite common) that $\theta_h=(i_{h},t_{h})$ for some $t_{h}\leq t'$. 
Observe that every simple (Lamport-style) message
chain gives rise to a centipede of a simple form in which all body
nodes~$\theta_h$ are co-located in this sense with their respective leg nodes. 
It follows that a centipede is a natural, albeit nontrivial,
generalization of a Lamport-causal chain.

While the centipede structure may seem rather intuitive, it is not at
all clear that such a structure  should
exist whenever the \OR problem is solved. Nevertheless,
Theorem~\ref{thm:centi2OR} below shows that this is the case, thus
providing a concise statement of the  communication structures that are
required for the ordering of events in every synchronous system.

\begin{theorem}[Centipede Theorem]
\label{thm:centi2OR}
Let $P$ be a protocol solving the $\OR=\OR\angles{\trig,\alpha_1,\ldots,\alpha_k}$ 
in $\gammax$, and assume
that $\trig$ occurs at $(i_0,t)$ in $r\in\Rmax$.  If $i_k$ performs
$a_k$ at time $t'$ in~$r$ then there is a centipede for $\langle
i_0,\ldots,i_k\rangle$ in~$\rmmp$.
\end{theorem}

Note however the subtle relation between the Ordered Response
requirement and the  structures that it necessitates. Suppose
that $\trig$ occurs at time $t$ in a run $r\in \Rrep(P,\gammax)$
where $P$ solves $\OR=\angles{\trig,\alpha_1,\ldots,\alpha_k}$. Then there
exist times $t_1\leq t_2\leq\cdots\leq t_k$ where the actions
$\alpha_1,\alpha_2,..\alpha_k$ are performed respectively. The theorem
implies that for each $h\leq k$ there exists a centipede for $\angles{
  i_0,\ldots,i_h}$ in $(r,t..t_h)$. Such a centipede could be used to
inform all  $\angles{ i_0,\ldots,i_h}$ processes of the occurrence of
$\trig$, and would also serve to coordinate response $\alpha_h$ so as
not to occur before any of the responses
$\alpha_1...\alpha_{h-1}$. However, a centipede for $\angles{
  i_0,\ldots,i_h}$ in $(r,t..t_h)$ need not be a sub-structure in a
centipede for $\angles{ i_0,\ldots,i_{h+1}}$ that occurs in
$(r,t..t_{h+1})$. 

\begin{figure}
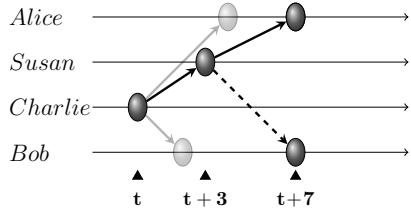

	\centering
		\newcommand{\tikzfname}{tikz_synch_chain2}
		\tikzsetnextfilename{\cfcurrentfolder Figures/\tikzfname}
		\scalebox{0.75}{\cfinput{Tikz/\tikzfname}}	
	\caption{A centipede in Example~\ref{ex2}}
	\label{fig:ex2_again}
\end{figure}

Recall Example~\ref{ex2} in the introduction, here redrawn in
Figure~\ref{fig:ex2_again}. The centipede for $\angles{Charlie,Bob}$
is the sequence $\angles{(Charlie,t),(Bob,t+2)}$. The
centipede for $\angles{Charlie,Bob,Alice}$ that is used by Alice to
coordinate her actions with those of Bob is the sequence\\
$\angles{(Charlie,t), (Susan,t+3),(Alice,t+7)}$.

\section{Knowledge Requires Syncausality}\label{sec:K2sync}
This section begins the second part of the chapter, wherein we review
the findings described thus far in light of the knowledge-based
analysis paradigm. This will provide us with
formal proofs to all quoted theorems, as well as a deeper
understanding of the forces at play. 

A word on terminology. Despite not having proved
Theorems~\ref{thm:1R2syncausality} and~\ref{thm:centi2OR}
just yet, we will take to referring to syncausality and the centipede 
as \emph{causal structures}, anticipating the results of the coming sections.

We start by considering the basic syncausality relation~$\spc$. 
In Section~\ref{sec:syncausality} we stated that every response event 
must be syncausally related to the trigger~$e_t$. In this section we
prove this claim formally. 

A simple and useful property of syncausality is a slight
extension of the idea that if one node syncausally
affects another, then the former must have happened before the latter
one: 
\begin{lemma}
\label{lem:step-synch}
If $(i,t)\spc (j,t')$ then $t\le t'$, with $t=t'$ holding only if
$i=j$.
\end{lemma}

\begin{proofL}
  All $(i,t)\spc (j,t')$ causality instances generated by clause 1
  have the property that $i=j$ and $t\le t'$.  By definition of the
  synchronous context $\gammax$, messages take at least one time
  step to be delivered. Moreover, the upper bounds on message
  transmission times are assumed to satisfy $\bij\ge 1$.  Therefore,
  all $(i,t)\spc (j,t')$ causality instances generated by clauses 2
  and~3 have the property that $i\ne j$ and $t<t'$.  As a result, a
  straightforward induction on the number of times the transitivity 
  clause~4 is applied
  in a derivation of $(i,t)\spc (j,t')$ yields the desired claim.
\end{proofL}

Lamport relates the happened-before relation to light cones in
Minkowski space-time \cite{Lamclocks}.  In the same vein, it is
natural to consider past and future causal ``cones'' induced by
syncausality.  

\begin{definition}[$\pas$ and $\fut$ cones]\label{def:cones}
	We define the {\em future causal cone} of a
	node~$\alpha$ (in run $r$) to be $$\fut(r,\alpha)=
	\left\{\angles{\theta,ND_{\theta}}:
	\begin{array}{l}\alpha\spc\theta\ \mbox{ in } r \text{ and }ND_{\theta}\\
	\text{is the set of ND events and initial states in }\theta\text{ in }r
	\end{array}\right\}.$$  
	
	Similarly, the {\em past causal cone} of~$\alpha$ is
	$$\pas(r,\alpha)=
	\left\{\angles{\theta,ND_{\theta}}:
	\begin{array}{l}\theta\spc\alpha\ \mbox{ in } r \text{ and }ND_{\theta}\\
	\text{is the set of ND events and initial states in }\theta\text{ in }r
	\end{array}\right\}.$$  
\end{definition}

We will often treat the sets $\pas$ and $\fut$ more simply as sets of nodes, rather than sets 
of pairs, when the related ND events are immaterial in the context. 

Observe
that the cones induced by syncausality in synchronous systems are
significantly larger than the ones that follow just from Lamport's
happened-before relation.  Moreover, just as the future and past cones
meet at the current point in space-time for light cones, we can use
Lemma~\ref{lem:step-synch} to show:
\begin{lemma}
\label{lem:cones}
 For all runs $r\in\Rmax$ and nodes $\alpha$ and~$\beta$:
 \begin{enumerate}
 \item $\fut(r,\alpha) \cap \pas(r,\alpha)=\{\alpha\}$, ~~and
 \item $\alpha\spc\beta$ ~iff~~$\fut(\alpha)\cap\pas(\beta)\ne \emptyset$.
\end{enumerate}
\end{lemma}

 \begin{proofL}
   Let $\alpha=(i,t)$ be a process-time node.  For part~(1), observe
   that $\alpha\spc\alpha$ since ~`$\spc$' is reflexive by clause (1)
   of Definition~\ref{def:syncau}.  Hence $\alpha\in \fut(r,\alpha)
   \cap \pas(r,\alpha)$.  By Lemma~\ref{lem:step-synch}, if
   $\alpha\ne\beta=(j,t')\in\pas(\alpha)$, then $t'<t$.  Similarly, if
   $\alpha\ne\theta=(h,t'')\in\fut(\alpha)$ then $t<t''$. It follows
   that ~~$\fut(r,\alpha) \cap\pas(r,\alpha)=\alpha$.

 For part (2), assume that $\alpha\spc\beta$. Then
 $\alpha\in\pas(\beta)$ by definition of~$\pas$. 
 Since $\alpha\in\fut(\alpha)$ by
 part (1), it follows that
 $\alpha\in\fut(\alpha)\cap\pas(\beta)\ne\emptyset$. For the other
 direction, suppose that $\theta\in\fut(\alpha)\cap\pas(\beta)$. 
Then by definition we have that $\alpha\spc\theta$ and
$\theta\spc\beta$. By transitivity of $\spc$ (clause 4) we have that
$\alpha\spc\beta$, and we are done. 
  \end{proofL}

 The next step in relating knowledge to syncausality in synchronous
 systems comes from the observation that the events that occur in the
 past (syncausal) cone of a node completely determine the local state
 at the node. A proof by induction on all nodes $(j,t')$ with $0\le
 t'\le t$ shows:
\begin{lemma}
\label{lem:past-local}
Let $r,r'\in\Rmax$.\\If $\pas(r,(i,t))=\pas(r',(i,t))$ 
then $r_i(t)=r'_i(t)$.
\end{lemma}

\begin{proofL}
A straightforward proof by induction on $t'$ in the range
    $0\le t'\le t$ shows that, for all $j\in\Proc$, if
    $(j,t')\in\pas(i,t)$ then $r_j(t')=r'_j(t')$.
    By assumption, $r$ and~$r'$ agree on initial states in $(j,0)$. 
    The induction step is proved based on the fact that 
    each local state which is not initial is determined by the previous
    local state of the same process and by ND events in that process
    in the last round. 
    Thus, $r_j(0)=r'_j(0)$ for all $j\in\Proc$. 
    The claim follows from the fact that $(i,t)\in\pas(r,(i,t))$.
\end{proofL}

Since the knowledge of a process in~$\Rmax$ is determined
by its local state,
Lemma~\ref{lem:past-local} implies that this knowledge
is determined by the past causal cone.
We can now state and prove, using Lemma~\ref{lem:past-local},
the following knowledge gain theorem for two
processes:

\begin{theorem}[Basic Knowledge Gain]
\label{thm:2proc}
Assume that
$e$ takes place at $(i_0,t)$ in~$r\in\Rmax=\Rrep(P,\gammax)$.
If ~$(\Rmax,r,t')\sat K_{i_1}\ndocc{e}$
then $(i_0,t)\spc (i_1,t')$.
\end{theorem}

\begin{proofT}
Let~$e$ be an ND event occuring at $(i_0,t)$ in the run $r\in\Rmax$.   
  We shall prove the contrapositive: If $(i_0,t)\not\spc
  (i_1,t')$ then ~$(\Rmax,r,t')\not\sat K_{i_1}\ndocc{e}$.  By
  assumption, $\Rmax=\Rrep(P,\gammax)$ for some protocol~$P$.
  Let $r'\in\Rmax$ be a run identical to $r$ until (but not
  including) time $t$, in which
    \begin{itemize}
    \item[(i)] the environment's actions at all nodes in
      $\pas(r,i_1,t')$ are identical to those in~$r$; and
      \item[(ii)] the environment's actions at $i_0$ in the interval
        $[t,t']$ are identical to those in~$r$ with the exception that
        $e$ does not occur in any of the related nodes. Thus if $t=0$
        then $i_0$'s initial state in $r'$ differs from that in $r$ by
        not including $e$, and similarly for other local
        states. Finally,
      \item[(iii)] all messages delivered to nodes not in
        $\pas(r,i_1,t')$ are delivered at the maximal possible
        transmission time according to the bounds~$\bij$.
    \end{itemize}
    To see that such a run $r'$ indeed exists in $\Rmax$, we note
    that clauses (ii) and (iii) relate to nodes outside
    $\pas(r,(i_1,t'))$ and thus by assumption do not contradict clause
    (i), and that by definition an external input event is entirely
    independent of the run's past history, so its non occurrence in an
    interval of time is possible. Since $\Rmax$ contains all runs
    of~$P$ in~$\gammax$, it must include~$r'$.

    Notice that $(\Rmax,r',t')\not\sat\ndocc{e}$: first, since events
    are distinct in a run and since $e$ occurs at time $t$ in~$r$, it does
    not occur in $r$ at any time previous to $t$. Since $r'$ is
    identical to $r$ until time $t$ the same applies for~$r'$. Next,
    if~$e$ is an external input, then the possibility for $e$ to occur
    at some $\bar t\in [t,t']$ is foiled by (ii), while if it is a
    message receive, then the following argument applies.  Since
    $(i_0,t)\not\spc(i_1,t')$ then also $(i_0,\bar t)\not\spc(i_1,t')$
    for all $\bar t\geq t$, as $(i_0,t)\spc(i_0,\bar t)$ is true of all
    $\bar t \geq t$. It follows that $(i_0,\bar t)\notin \pas(r,(i_1,t'))$
    for all $\bar t\geq t$. So either the message is postponed beyond
    time $t'$, or else it cannot be thus postponed, in which case $e$
    is not an ND event (an early receive) when it occurs in $r'$.  For
    all cases we get by definition of $\sat$ that
    ~$(\Rmax,r',t')\not\sat\ndocc{e}$.  By definition, $r'$ agrees
    with~$r$ on initial states, external inputs, and delivery times on
    nodes of $\pas(r,(i_1,t'))$.  Thus by Lemma~\ref{lem:past-local}
    we have that $(r,t')\sim_{i_1}(r',t')$, and therefore 
~$(\Rmax,r,t')\not\sat
    K_{i_1}\ndocc{e}$, as desired.  \renewcommand{\toto}{thm:2proc}
 \end{proofT}

The proof of Theorem~\ref{thm:2proc}  is obtained by constructing a
run $r'$ indistinguishable to $i_1$ at~$t'$ from~$r$ in which no ND
events occur outside\\
$\pas(r',(i_1,t'))=\pas(r,(i_1,t'))$.
Theorem~\ref{thm:2proc}  captures a natural sense in which
syncausality is a notion of potential
causality for the synchronous model.

The proof of Theorem~\ref{thm:1R2syncausality} can now be derived.
For protocols that recall responses, 
the theorem follows immediately from Theorems~\ref{thm:2proc}
and~\ref{thm:ORnestedK}. We prove the theorem for arbitrary
protocols by relating the general case to 
that of protocols that recall responses (Definition \ref{def:response_recall}). 

\renewcommand{\toto}{thm:1R2syncausality}
\begin{proofOfT}
	Let~$P$ be a protocol solving $\OR=\OR\angles{\trig,\alpha_1,..,\alpha_k}$.
	Define $P'$ to be a protocol that differs from~$P$ only in that every
	process~$i$ maintains a list called {\tt Responses}, to which it adds an
	item $(a_h,t_h)$ whenever~$i$ performs a response $\alpha_h$. (We assume
	w.l.o.g.~that no list with this name is used by~$P$.) Notice that this list
	is an auxiliary variable that does not affect the behavior of the protocol.
	Indeed, there is an isomorphism between the runs of~$P$ and those of~$P'$,
	where the same nondeterministic events and the same actions take place at
	all nodes of corresponding runs. In particular, since~$P$ solves~$\OR$, then
	so does~$P'$. By construction, $P'$ recalls responses. Denote
	$\Rmax'=\Rrep(P',\gammax)$. Assume that $\trig$ occurs in the run~$r$ of~$P$
	and let $r'$ be the corresponding run of~$P'$.

	Let $1\le h\le k$. Since $P'$ recalls responses, we have by
	Theorem~\ref{thm:ORnestedK} that   $(\Rmax',r',t')\sat  
	K_{i_h}\occurred{\trig}$. As every external input is, in particular, an ND
	event, this gives us $(\Rmax',r',t')\sat K_{i_h}\ndocc{\trig}$. We now use
	Theorem~\ref{thm:2proc} and the fact that $\trig$ occurs at $(i_0,t)$ to
	conclude that $(i_0,t)\spc(i_h,t')$ in $r'$.

	Since all actions and communication events in $r$ and in $r'$ are the same,
	it follows that  $(i_0,t)\spc(i_h,t')$ in $r$ too, as required.
\end{proofOfT}

\section{Nested Knowledge Requires Centipedes}\label{sec:NK2centipedes}
When we move beyond single response problems into the double and $k$
response variants, Theorem~\ref{thm:ORnestedK} provides us with nested
knowledge conditions. Showing that nested knowledge implies the
existence of a centipede requires a substantial formal theory. This
section develops the required theory. 

The first relevant notion is captured by the following definition:

\begin{definition}[Bridge nodes]
\label{def:bridge}
Fix~$r$ and let $\alpha\spc\alpha'$. We say that $\beta$ {\em
  bridges} $\alpha$ and $\alpha'$ if
\begin{enumerate}
    \item $\alpha\spc\beta\tspc\alpha'$ ~~and
    \item $\alpha\spc\beta'\tspc\beta$ ~implies $\beta'=\beta$, for all
      nodes $\beta'$.
\end{enumerate}
\end{definition}

Intuitively, a bridge is an earliest node that is syncausally affected
by~$\alpha$ and 
precedes~$\alpha'$ by way of a timing guarantee. 
Interestingly, 
bridges are guaranteed to exist:
\begin{lemma}
\label{lem:bridges-exist}
    If $\alpha\spc\alpha'$ then there is a node~$\beta$ bridging $\alpha$
    and~$\alpha'$.
\end{lemma}

\begin{proofL}
    Let $\alpha\spc\alpha'$, where $\alpha=(i,t)$ and $\alpha'=(j,t')$.
    By Lemma~\ref{lem:step-synch} we have that $t\le t'$.
    We prove the claim by induction on $d=t'-t$.
    The base case is $d=0$, in which case $t=t'$ and by
    Lemma~\ref{lem:step-synch} we have that $\alpha=\alpha'$. 
    Since $\alpha\spc\alpha\tspc\alpha$ holds, and
    $\alpha\spc\beta'\tspc\alpha$ holds only for $\beta'=\alpha$, 
    it follows that
    $\beta=\alpha$ is a bridge as required.
    For the inductive step, let $d>0$ and assume that the claim holds for
    all pairs of causally related nodes with time differences 
strictly smaller than~$d$. 
Since $\alpha'\tspc\alpha'$ by definition, the assumption that
$\alpha\spc\alpha'$ clearly implies that
$\alpha\spc\alpha'\tspc\alpha'$. We consider two cases. 
If there is no node $\beta'\ne\alpha'$ such that
$\alpha\spc\beta'\tspc\alpha'$ then $\beta=\alpha'$ is the desired
bridge node. Otherwise, such a $\beta'=(i_1,t_1)$ exists.  As before,
we obtain by Lemma~\ref{lem:step-synch} that $t_1< t'$.  In
particular, $d_1=t_1-t<t'-t=d$.  Thus, since $\alpha\spc\beta'$ we
have by the inductive assumption for~$d_1$ that there is a node
$\beta$ bridging $\alpha$ and~$\beta'$. It follows that
$\alpha\spc\beta\tspc\beta'\tspc\alpha'$ and $\beta$ satisfies the
minimality clause (2) of Definition~{\bf \ref{def:bridge}} with
respect to~$\alpha$. As $\tspc$ is a transitive relation, we obtain
that $\alpha\spc\beta\tspc\alpha'$ and the claim follows.
\end{proofL}

Bridges are closely related to early message receives: 
\begin{lemma}
\label{lem:bridges-early}
If $\alpha\ne\beta$ and $\beta$ bridges $\alpha$ and~$\alpha'$, then
there exists some~$\beta'$ such that $\alpha\spc\beta'\spc\beta$ and
the syncausal chain $\beta'\spc\beta$ consists of a single early
receive.
\end{lemma}
\begin{proofL}
    Denote $\alpha=(i_1,t_1)$ and $\beta=(i_2,t_2)$.
    If $\beta$ bridges $\alpha$ and~$\alpha'$ then, in particular,
    $\alpha\spc\beta$. If, in addition, $\alpha\ne\beta$
    then $t_1<t_2$ by Lemma~\ref{lem:step-synch}.
    It follows that $\alpha\spc\beta'\spc\beta$ where
    $\beta'\spc\beta$ is derivable by clause (1), (2), or (3),
    and $\beta'=(i',t')$ for some $t'<t_2$.
    If $\beta'\spc\beta$ is derivable by (1) or (3), then
    $\beta'\tspc\alpha'$ and $\beta$ does not bridge $\alpha$ and
    $\alpha'$, contradicting the assumption. The alternative is that
    $\beta'\spc\beta$ is derivable from (2) but {\em not} from (3), and
    hence $\beta'\spc\beta$ must be an early receive, as claimed.
\end{proofL}

The existence of bridge nodes as a special kind of node motivates an
alternative approach to defining centipedes, based on bridge nodes. 
We start with the notion of {\em centinodes}: 

\begin{definition}[Centinode]
\label{def:cnode}
We inductively define node~$\theta$ to be a \\
$\cnode{t}{i_0,\ldots,i_k}{t'}$ centinode in $\rmmp$ as follows.
\begin{description}
    \item[$k=0$:]\quad $\theta$ is a $\cnode{t}{i_0}{t'}$ centinode iff
      ~$\theta=(i_0,t)$; while
    \item[$k>0$:]\quad $\theta$ is a $\cnode{t}{i_0,\ldots,i_k}{t'}$
      centinode iff there exists a 
    $\cnode{t}{i_0,\ldots,i_{k-1}}{t'}$ centinode $\theta'$ in $(r,t..t')$, 
such that $\theta$ bridges between $\theta'$ and $(i_k,t')$ in~$r$.
\end{description}
\end{definition}

As a  straightforward conclusion of Lemma~\ref{lem:bridges-exist}
 we can show:
\begin{lemma}\label{lem:cent-equiv}
	The following three are equivalent:
	\begin{enumerate}
		\item A centipede for $\langle i_0,\ldots,i_k\rangle$ in $\rmmp$ exists;
		\item A $\cnode{t}{i_0,\ldots,i_k}{t'}$ centinode in $\rmmp$ exists; and
		\item A centipede $\langle\theta_0,\ldots,\theta_k\rangle$ for $\langle
		i_0,\ldots,i_k\rangle$ in $\rmmp$ exists, in which every node $\theta_j$ is
		a $\cnode{t}{i_0,\ldots,i_j}{t'}$ centinode in $\rmmp$, for $j=0,\ldots,k$.
	\end{enumerate}
\end{lemma}

\begin{proofL}
	The truth of $3\Rightarrow 2$ is immediate. A straightforward induction on~$k$ shows
	that $2\Rightarrow 1$, as the current centinode $\theta_{k}$ is syncausally
	related to $\theta_{k-1}$ and timing guarantee related to $(i_{k},t')$. 
	We now prove that $1\Rightarrow 3$. Assume that
	$\cC=\angles{\theta_0,..,\theta_k}$ is a centipede for $\langle
	i_0,\ldots,i_k\rangle$ in $\rmmp$. We define by induction on $h\le k$ centipedes
	$\cC_h=\angles{\theta'_0,..,\theta'_h,\theta_{h+1},..,\theta_k}$ in $\rmmp$, in
	which the nodes $\theta'_0$ to $\theta'_h$ are centinodes. The final
	centipede~$\cC_k$ in the construction satisfies the conditions of~$3$.
	\begin{description}
	    \item[$h=0:$] By definition, the initial node~$\theta_0$ in~$\cC$ is a
	    $\cnode{t}{i_0}{t'}$ centinode in $\rmmp$. Defining $\theta'_0=\theta_0$ we
	    have $\cC_0=\cC$.
	    \item[$h>0:$] Assume that a centipede
	    $\cC_{h-1}=\angles{\theta'_0,..,\theta'_{h-1},\theta_{h},..,\theta_k}$ as
	    described above has been constructed. By Lemma~\ref{lem:bridges-exist} there
	    exists a node $\theta'_h$ bridging $\theta'_{h-1}$ and $\theta_h$. Since
	    $\theta_h\tspc (i_h,t')$ we get that  $\theta'_h$ is a centinode for
	    $\langle i_0,\ldots,i_h\rangle$ in $\rmmp$. Define
	    $\cC_h=\angles{\theta'_0,..,\theta'_{h},\theta_{h+1},..,\theta_k}$. If $h=k$
	    then we are done. Otherwise, since
	    $\theta_{h'}\tspc\theta_h\spc\theta_{h+1}$ and $\cC_{h-1}=
	    \angles{\theta'_0,..,\theta'_{h-1},\theta_{h},..,\theta_k}$ is a centipede
	    for $\langle i_0,\ldots,i_k\rangle$ in $\rmmp$, we obtain that $\cC_h$ is
	    also such a centipede, as required.
	\end{description}
\end{proofL}

Lemma~\ref{lem:cent-equiv} allows using centinodes and centipedes
interchangeably.  Indeed, clause~3 suggests that we can without loss
of generality think of centipedes as consisting of a sequence of
centinodes. 
We are now ready to prove our main theorem, stating that 
the existence of a centipede is a necessary condition for attaining 
nested knowledge of an ND event: 

\begin{theorem}[Knowledge Gain]
\label{thm:cpede}
    Let~$P$ be a deterministic protocol, and let
    $r\in\Rmax=\Rrep(P,\gammax)$. Assume
	that   $e$ is an ND event at $(i_0,t)$ in~$r$.
    If $(\Rmax,r,t')\sat K_{i_k}K_{i_{k-1}}\cdots K_{i_1}\ndocc{e}$,
    then there is a centipede for $\langle i_0,\ldots,i_k\rangle$ in
    $\rmmp$.
\end{theorem}

\begin{proofT}
We shall prove the contrapositive form: if no centinode
$\cnode{t}{i_0,\ldots,i_{k-1},i_k}{t'}$ exists in $\rmmp$, then
$(\Rmax,r,t')\not\sat K_{i_k}K_{i_{k-1}}\cdots K_{i_1}\ndocc{e}$. We reason by
induction on $k\ge 1$:

\begin{description}
	\item[$k=1$]\quad This case is a rephrasing of Theorem~\ref{thm:2proc}: By
	assumption, there is no $\cnode{t}{i_0,i_1}{t'}$ centinode in~$\rmmp$. In
	other words, there is no node~$\theta$ bridging $(i_0,t)$ and $(i_1,t')$. By
	Lemma~\ref{lem:bridges-exist} it follows that $(i_0,t)\not\spc(i_1,t')$.
	Thus, by Theorem~\ref{thm:2proc} we have that ~$(\Rmax,r,t')\not\sat
	K_{i_1}\ndocc{e}$, as claimed.
	
	\item[$k\ge 2$]\quad Assume inductively that the claim holds for $k-1$.
	Moreover, assume that no $\cnode{t}{i_0,\ldots,i_k}{t'}$ centinode exists in
	$\rmmp$. For every $r'\in\Rmax$ let $C^{r'}=\{\theta^1,\ldots,\theta^d\}$ be
	the set of $\cnode{t}{i_0,\ldots,i_{k-1}}{t'}$ centinodes in $(r',t..t')$.
	Observe that $\theta'\not\spc (i_k,t')$ in~$r$ for all $\theta'\in C^r$,
	since otherwise we would have by Lemma~\ref{lem:bridges-exist} that there is
	a bridging node $\theta'\spc\beta\tspc(i_k,t')$. But $\beta$ would then be a
	$\cnode{t}{i_0,\ldots,i_{k-1},i_k}{t'}$ centinode in $\rmmp$, contradicting
	our assumption.
		
		We consider two cases. First suppose that $(i_0,t)\in C^r$. Given that
		$(i_0,t)\not\spc(i_k,t')$ we have that $(\Rmax,r,t')\not\sat
		K_{i_k}\ndocc{e}$ by Theorem~\ref{thm:2proc} above, and since
		$K_{i_k}K_{i_{k-1}}\cdots K_{i_1}\phi$ validly implies $K_{i_k}\phi$ is
		valid, we obtain that $(\Rmax,r,t')\not\sat K_{i_k}K_{i_{k-1}}\cdots
		K_{i_1}\ndocc{e}$, as claimed.
		
		Next suppose that $(i_0,t)\notin C^r$. Let~$r'\in\Rmax$ be a run such
		that $r'$ is identical to $r$ until (but not including) time $t$, and
		where
		\begin{itemize}
		    \item[(i)] the environment's actions at all nodes in
		    $\pas(r,(i_k,t'))$ are identical to those in~$r$; and
		    \item[(ii)] all messages delivered to nodes not in $\pas(r,(i_k,t'))$ are
		    delivered at the maximal possible transmission time according
		    to the bounds~$\bij$.
		\end{itemize}
		
		To see that such a run $r'$ indeed exists in $\Rmax$, we note that clause
		(ii) relates to nodes outside $\pas(r,(i_k,t'))$ and thus by assumption does not
		contradict clause (i), and that by definition all early message receives can
		be delayed, independent of the run's past or concurrent events. Since
		$\Rmax$ contains all runs of~$P$ in~$\gammax$, it must include~$r'$.

		Showing that such a run exists repeats the arguments in Theorem
		\ref{thm:2proc}. From $r'$ being identical to $r$ until time $t$, from
		clause (i) above and from Lemma~\ref{lem:past-local} it follows that
		$r'_{i_k}(t')=r_{i_k}(t')$. Notice that by construction of~$r'$ we have
		that $\alpha\spc\alpha'$ holds in $r'$ only if $\alpha\spc\alpha'$ in
		$r$, and that  every early receive in~$r'$ is an early receive in~$r$.
		Considering that bounds are universal in all runs, we obtain that every
		bridge node in $r'$ is also a bridge node in $r$, and hence that
		$C^{r'}\subseteq C^r$. By definition of~$r'$, none of the nodes in the
		set $C^r$, and hence also in  $C^{r'}$, experiences an early receive
		in~$r'$. Yet from Lemma~\ref{lem:bridges-early}  and from $(i_0,t)\notin
		C^{r'}$ it follows that every node $\theta'\in C^{r'}$ must be a
		nontrivial bridge node in~$r'$,  thus experiencing an early receive. We
		therefore conclude that $C^{r'}=\emptyset$. 
		
		Based on the inductive hypothesis and the definition of $C^{r'}$ we get
		$$(\Rmax,r',t')\not\sat K_{i_{k-1}}\cdots K_{i_1}\ndocc{e}.$$ As
		$r_{i_{k}}(t)=r'_{i_{k}}(t)$, we obtain using the definition of
		Knowledge operator that  \[(\Rmax,r,t')\not\sat K_{i_k}K_{i_{k-1}}\cdots
		K_{i_1}\ndocc{e},\] and we are done.
\end{description}
\renewcommand{\toto}{thm:cpede}
\end{proofT}

Based on the Knowledge Gain Theorem, we can proceed to prove
Theorem~\ref{thm:centi2OR}, the Centipede Theorem.
For protocols that recall responses, 
the Centipede theorem follows immediately from Theorems~\ref{thm:cpede}
and~\ref{thm:ORnestedK}. We prove the Centipede Theorem for arbitrary
protocols by relating the general case to 
that of protocols that recall responses, as we did in Theorem~\ref{thm:1R2syncausality}. 

\renewcommand{\toto}{thm:centi2OR}
\label{page:centi2OR}
\begin{proofOfT}
Let~$P$ be a protocol solving $\OR=\OR\angles{\trig,\alpha_1,..,\alpha_k}$.
Define $P'$ to be a protocol that differs from~$P$ only in that 
every process~$i$ maintains a list called {\tt Responses}, 
to which it adds an item $(a_h,t_h)$ whenever~$i$ perfoms a response
$\alpha_h$. (We assume w.l.o.g.~that no list with this name is used
by~$P$.) Notice that this list is an auxiliary variable that does not 
affect the behavior of the protocol. Indeed, there is an isomorphism
between the runs of~$P$ and those of~$P'$, where the same
nondeterministic events and the same actions take place 
at all nodes of corresponding runs. In particular, since~$P$
solves~$\OR$, then so does~$P'$. 
By construction, $P'$ recalls responses. 
Denote $\Rrep'=\Rrep(P',\gammax)$. 
Assume that $\trig$ occurs in the run~$r$ of~$P$ and let $r'$ be the 
corresponding run of~$P'$. 
Since $P'$ recalls responses, we have by Theorem~\ref{thm:ORnestedK} that 
\[(\Rrep',r,t_k)\sat~K_{i_k}K_{i_{k-1}}\cdots K_{i_1}\occurred{\trig}.\] 
Since $\trig$ is an external input in all runs of $\Rrep'$, we have
that $\Rrep'\sat \occurred{\trig}\Rightarrow\ndocc{\trig}$. 
Hence,  
\[(\Rrep',r,t_k)\sat~K_{i_k}K_{i_{k-1}}\cdots
K_{i_1}\ndocc{\trig}.\] 
By Theorem~\ref{thm:cpede} we thus obtain that 
there is a centipede for $\langle i_0,\ldots,i_k\rangle$ in
    $(r',t..t')$. 
Since all actions and communication events in~$r$ and in~$r'$ are the
same, it follows that there is a centipede for $\langle i_0,\ldots,i_k\rangle$ in
    $\rmmp$, and we are done. 
\end{proofOfT}

\section{Varying Nondeterminism in Message Transmission}\label{sec:nd_variations}
A better grasp of the dynamics and flexibility of the 
centipede structure, and of the scope of the related Knowledge Gain
Theorem, is afforded by considering two particular models, with very
different characteristics. On one extreme, we define the 
\emph{Asynch-delivery} model to be one in which $\bij=\infty$ for {\em
  all} channels $i\mapsto j$. 
On the other extreme, we consider the
\emph{Fixed-delivery} model to be one in which every message on a
channel $i\mapsto j$ spends exactly $\bij<\infty$ time units in transit. 

We define the Asynch-delivery model as a context $\gamma^{\mathsf{as}}$ which is 
a $\gammab$ context where $\bij=\infty$
for all existing channels. This is a model where processes share a global
clock but communication is asynchronous.  In this case, clause (3) in the
definition of syncausality cannot be used to infer syncausality of any
pair of nodes, and thus $\spc$ coincides with~$\Lam$.

In this model, the Knowledge Gain then reduces to a theorem equivalent
to Chandy and Misra's Knowledge Gain Theorem for totally asynchronous
contexts \cite{ChM}.

\begin{lemma}
	\label{lem:async_cpede}
	Let~$P$ be an arbitrary protocol, let
        $\Rrep^{\mathsf{as}}=\Rrep(P,\gamma^{\mathsf{as}})$.
        Assume
	$e$ is an ND event occurring at $(i_0,t)$ in~$r$.\\
	If $(\Rrep^{\mathsf{as}},r,t')\sat K_{i_k}K_{i_{k-1}}\cdots K_{i_1}\ndocc{e}$,\\
	then there is a chain
        $(i_0,t)\Lam(i_1,t_1)\Lam\cdots\Lam(i_k,t_k)$ in $\rmmp$.
\end{lemma}

\begin{proofL}
  By the theorem's assumptions and by applying the Knowledge Gain
  Theorem, we obtain that there must exist a centipede\\
  $\angles{(i_{0},t),\theta_{1},..,\theta_{k-1},(i_{k},t')}$ for
  $\angles{i_{0},i_{1},..,i_{k}}$ in $\rmmp$.

  Yet when $\dist=\infty$ for all channels, we get that if
  $(i,t)\spc(j,t')$ and $t\leq t'<\infty$ then the syncausal relation
  cannot be based on applications of clause (3) of the definition of
  syncausality on page \pageref{def:syncau}. Thus, it must be that
  $(i,t)\Lam(j,t')$. Moreover, if $t\leq t'<\infty$ and
  $(i,t)\tspc(j,t')$ then it must be that $i=j$.
	
	Thus in the existing centipede it must be that for all $h< k$,
	$\theta_{h}=(i_{h},t_{h})$ and
        $\theta_{h}\Lam\theta_{h+1}$. Thus providing us with  
	a message chain linking $i_{0},i_{1},..,i_{k}$ in $\rmmp$.
\end{proofL}

As seen in Figure~\ref{fig:nd_variations}, under the Asynch-delivery
model the centipede's legs are shortened to length $0$, 
and the syncausal relations between it's body nodes collapse into
Lamport's happened-before. 

The Fixed-delivery model runs opposite to the Asynch-delivery one in removing
the nondeterministic aspect in message delivery. Thus, not only do
processes share a global clock, but they also share knowledge of the
\emph{exact} time it takes each message to be delivered.

We define this model as a $\gammab$ context where $min_{ij}=max_{ij}$ for all
channels, and denote it $\gamma^{\mathsf{f}}$.
Under this model, every message sent arrives exactly at its related
channel's bound guarantee. The syncausal relation then acquires the
same extension as that of the timing guarantee, and centipede body
nodes are collapsed into a single node, as shown in
Figure~\ref{fig:nd_variations}.

\begin{lemma}
	\label{lem:ddmodel_cpede}
	Let~$P$ be an arbitrary protocol, and let $\Rrep^{\mathsf{f}}=\Rrep(P,
       \gamma^{\mathsf{f}} )$. \\ 
	Fix $r$ and assume	$e$ is an ND event occurring at
        $(i_0,t)$ in~$r$.\\ 
	If $(\Rrep^{\mathsf{f}},r,t')\sat K_{i_k}K_{i_{k-1}}\cdots K_{i_1}\ndocc{e}$,\\
	then $(i_{0},t)\tspc(i_{h},t')$ for all $h\leq k$.
\end{lemma} 

\begin{proofL}
	Note that in the context $\gamma^{\mathsf{f}}$ we have that
        $(i,s)\spc(j,s')$ iff 
	$(i,s)\tspc(j,s')$, for all processes $i,j$ and times $s,s'$. 
	By the Knowledge Gain Theorem there must exist a centipede for 
	$\angles{i_{0},i_{1},..,i_{k}}$ in $\rmmp$. In other words,
        for each $h\leq k$ 	there exists some $\theta_{h}$ such
        that $(i_{0},t)\spc\theta_{h}\tspc(i_{h},t')$. 
	Thus we get that $(i_{0},t)\tspc\theta_{h}\tspc(i_{h},t')$ and
        hence $(i_{0},t)\tspc(i_{h},t')$ as required.
\end{proofL}

\begin{figure}
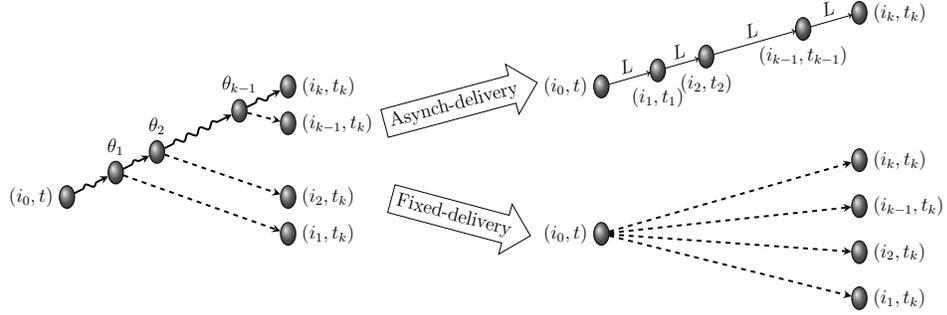

	\centering
		\newcommand{\tikzfname}{tikz_nd_variations}
		\tikzsetnextfilename{\cfcurrentfolder Figures/\tikzfname}
		\resizebox{1\textwidth}{!}{\cfinput{Tikz/\tikzfname}}	
	\caption{Collapsed centipede variations}
	\label{fig:nd_variations}
\end{figure}

\section{Sufficiency of Centipedes for Knowledge Gain}\label{sec:suff_NKG}
The Knowledge Gain and Centipede Theorems (Theorems ~\ref{thm:cpede}
and~\ref{thm:centi2OR} respectively) state that the centipede  structure is
{\em necessary} for gaining nested knowledge occurrence of nondeterministic
events and for solving the \OR problem. These results hold in a strong sense,
{\em regardless} of the protocol used by the processes. Our goal in this section
is to show that these results are tight. 

We cannot prove that centipedes are sufficient means to achieve these ends {\em
for all protocols}, because the knowledge actually transferred by messages
depends on the protocol, and may be insufficient.\footnote{See \cite{Parikh03}
and \cite{Mo92} for some interesting observations on the connections between
protocols and message meanings.} The most we can do in order to prove the
tightness of our definitions is to show that there exist \emph{specific}
protocols under which centipedes are sufficient for knowledge gain. We will do
so for the following version of the \emph{full information protocol}.

\begin{definition}[Full-information Protocol] 
In the \emph{full information protocol} for synchronous systems,
denoted~$\Pfip$, every process $i\in   \Proc$ sends its local state on each of
its outgoing channels at every time step. Moreover, each process retains
a history of every event that has taken place locally, and every
message received, along with their times of occurrence.  
\end{definition}

We will denote with $\Rfip$ the system $\Rrep(\fip,\gammax)$.
In $\fip$ the processes convey all of their knowledge as fast as they
can. Roughly speaking, knowledge is spread in the system as fast as
possible, given the transmission times allowed by the environment
in the given run. While our stated goal is to prove that under $\fip$
the Knowledge Gain and Centipede Theorems  are tight, our
results will be somewhat stronger. These theorems 
show that centipedes  are necessary for knowledge gain 
regarding nondeterministic events. 
As we shall see, in the context of $\fip$, there is no need to
restrict attention to nondeterministic events. The causal structures in question 
are sufficient for knowledge gain regarding general events (and more
general facts).

Three simple but very useful properties of the
timestamping operator $\atn{t}{}$ are captured by the following immediate lemma:

\begin{lemma}\label{lem:time_knowledge}
	For every formula $\vphi$ and times $t,t'$, the following
        formulas are valid in $\Rmax$ 
		\begin{description}
			\item[TS1] \quad$\sat ~\:\atn{t}{(\vphi \leftrightarrow
                            \atn{t}{\vphi})}$  
			\item[TS2] \quad$\sat ~\:\atn{t'}{\atn{t}{\vphi}}
                          \leftrightarrow \atn{t}{\vphi}$ 
			\item[TS3] \quad $\sat ~\:\atn{t}{(K_{i}\vphi
                            \leftrightarrow K_{i}\atn{t}{\vphi})}$ 
		\end{description}	
\end{lemma}

Notice that, as described, processes following $\fip$ have perfect recall. Since
they maintain their local histories, they do not forget what they knew. Of
course, the truth of a transient fact can change over time. Knowing that the
time is~3 is possible at time~3 but not at time~4. But whether $\varphi$ held at
time~$t$ does not change. More formally, perfect recall and the presence of
clocks give us the following knowledge-preservation property, which states that
if at time $t$ process $i$ knows that $\vphi$, then at every future time point
the process will know (or remember) that it knew $\vphi$ at $t$.
  
\begin{lemma}
	If $t'\geq t$, then the  following
        formula is valid in $\Rfip$:
		\begin{description}
			\item[TS4] \quad$\sat ~\:\atn{t}{K_{i}\vphi} \imp
                          \atn{t'}{K_{i}\atn{t}{K_{i}\vphi}}$.
		\end{description}	
\end{lemma}

The proof of the following lemma will make use of the $\spcnt$
relation, which results from a single application of one of the
clauses (1), (2), or (3) of syncausality.

\begin{figure}[!h]
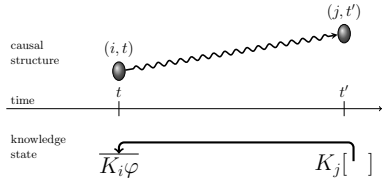

	\centering
	\newcommand{\tikzfname}{tikz_fip1b}
	\tikzsetnextfilename{\cfcurrentfolder Figures/\tikzfname}
	\resizebox{0.4\linewidth}{!}{\cfinput{Tikz/\tikzfname}}
	\caption{The syncausal relation and reflected knowledge state}
	\label{fig:fip1}
\end{figure}

We are now ready to show that syncausality alone is sufficient to
ensure knowledge transfer under $\Pfip$.  Figure~\ref{fig:fip1} below provides
a graphical visualization of what Lemma~\ref{lem:fip1} shows. Namely,
that if $K_{i}\vphi$ holds at time $t$ and  $(i,t)\spc(j,t')$, then 
at $t'$ process $j$ knows that at time $t$ process $i$ knew that $\vphi$.

\begin{lemma}
\label{lem:fip1}
If ~$(\Rfip,r,t)\sat K_i\varphi$ and~$(i,t)\spc(j,t')$ in~$r$, 
then\\ $(\Rfip,r,t')\sat K_j(\atn{t}{K_{i}\varphi})$~.
\end{lemma}

\begin{proofL}
Given that $(i,t)\spc(j,t')$, we have that 
\[(i,t)=(j_0,s_0)\spcnt(j_1,s_1)\spcnt\cdots\spcnt(j_n,s_n)=(j,t')\]
We prove the claim by induction on~$n$. 
\begin{description}
\item[$\mathbf{n=0:}$] In this case, $j=i$ and $t=t'$. Positive
introspection gives us $(\Rfip,r,t)\sat K_{i}K_i\varphi$. Let 
$\vphi'=K_{i}\vphi$, and apply $TS3$ based on $(\Rfip,r,t)\sat K_i\varphi'$.
This gives us  $(\Rfip,r,t')\sat K_i\atn{t}{K_{i}\varphi}$ as required.
\item[$\mathbf{n>0:}$] Assume inductively that $(\Rfip,r,s_{n-1})\sat
  K_{j_{n-1}}(\atn{t}{K_{i}\varphi})$.  By definition of the sequence we
  have that $(j_{n-1},s_{n-1})\spcnt(j_{n},s_{n})$.  By definition of
  $\spcnt$ there are three options to consider, corresponding to
  clauses (1)--(3) of syncausality:
	\begin{enumerate}
        \item $j_{n-1}=j_n$ and $s_{n-1}\leq s_n$:  In this case by
          $TS4$ we have that $$(\Rfip,r,s_{n-1})\sat
          \atn{s_{n}}{K_{j_{n}}\atn{s_{n-1}}{K_{j_{n-1}}(\atn{t}{K_{i}\varphi})}},$$  
          which is reduced to
          $$(\Rfip,r,s_{n})\sat
          K_{j_{n}}\atn{s_{n-1}}{K_{j_{n-1}}(\atn{t}{K_{i}\varphi})}.$$ 
          Applying the Knowledge Axiom and $TS2$ to every
          $(r',s_{n})\sim_{j_{n}}(r,s_{n})$ we obtain
          $(\Rfip,r,s_{n})\sat K_{j_{n}}(\atn{t}{K_{i}\varphi})$.	
		
        \item process $j_{n-1}$ sends a message in round $s_{n-1}$,
          which is received by $j_n$ in round $s_n$: Since the
          protocol used is $\Pfip$, message contents consist of the local
          state of sender. Based on the inductive assumption we get
          that $(\Rfip,r,s_{n})\sat
          K_{j_{n}}\atn{s_{n-1}}{K_{j_{n-1}}(\atn{t}{K_{i}\varphi})}$.
          This is implies $(\Rfip,r,s_{n})\sat
          K_{j_{n}}(\atn{t}{K_{i}\varphi})$ as in case (1).
		
        \item $(j_{n-1},j_n)$ is a network channel and no message is
          sent by $j_{n-1}$ to $j_{n}$ at time $s_{n}$: Since in
          $\Pfip$ every process sends its local state to all neighbors
          in every round, this option is not viable in $r$.
	\end{enumerate}
\end{description}
\end{proofL}

The lemma is proved based on the fact  that in $\fip$ processes constantly send
explicit messages on all outgoing channels, so a syncausal chain in $\fip$ never
contains a link that is based on clause (3) of syncausality.  Hence, if
$(i,t)\spc(j,t')$ in $r$, then there must exist a chain of ``real'' messages
linking the two nodes. Messages in~$\fip$  contain the local state of the
sender. Hence, $i$'s local state at time $t$ is propagated through the message
chain until it reaches $j$.

\begin{figure}[!h]
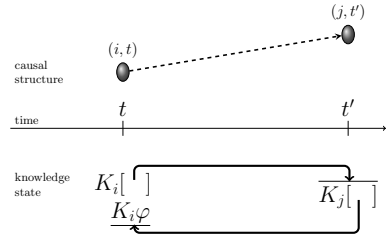

	\centering
	\newcommand{\tikzfname}{tikz_fip2b}
	\tikzsetnextfilename{\cfcurrentfolder Figures/\tikzfname}
	\resizebox{0.4\linewidth}{!}{\cfinput{Tikz/\tikzfname}}
	\caption{The timing guarantee and reflected knowledge state}
	\label{fig:fip2}
\end{figure}

Lemma~\ref{lem:fip2} below makes use of the further guarantees made by the
$\tspc$ relation.  Recall that $(i,t)\tspc(j,t')$ implies that
$(i,t)\spc(j,t')$,  by Lemma~\ref{lem:guarantee2sync}. Moreover,
The $\tspc$ relation  is
determined by the context $\gammax$ alone. So that if
$(i,t)\tspc(j,t')$ holds in a run $r$ of the system, it will do so in
\emph{all} runs of the system. Thus, process $i$ knows already at time~$t$ that 
its current knowledge will be available to~$j$ at~$t+ \sD(i,j)$. This
situation is depicted in Figure~\ref{fig:fip2}. 
Translated into English, the figure shows that if at time $t$ process
$i$ knows that $\vphi$, and if $(i,t)\tspc(j,t')$, then 
at time $t$ process $i$ also knows that at time $t'$ process $j$ will know
that at time $t$ process $i$ knew that~$\vphi$.

\begin{lemma}
\label{lem:fip2}
If ~$(\Rfip,r,t)\sat K_i\varphi$ ~and~ $(i,t)\tspc(j,t')$, then\\
$(\Rfip,r,t)\sat K_i(\atn{t'}{K_j(\atn{t}{K_{i}\varphi})})$~.
\end{lemma}

\begin{proofL}
	Since $(i,t)\tspc(j,t')$, and since this property is determined 
	by the network independently of the particular run $r$, 
	we have that $(i,t)\tspc(j,t')$ in every run $r'\in \Rfip$. 
	Moreover,  by Lemma~\ref{lem:guarantee2sync} we have that 
	$(i,t)\spc(j,t')$ in every such run. 
	Applying Lemma \ref{lem:fip1} to every run $r'$ such that
        $(r',t)\sim_{i}(r,t)$  
	we obtain that  $(\Rfip,r',t')\sat K_j(\atn{t}{K_{i}\varphi})$. 
	By $TS1$ we obtain $(\Rfip,r',t')\sat \atn{t'}{K_j(\atn{t}{K_{i}\varphi})}$. 
	By choice of runs $r'$ we now conclude that  
	$(\Rfip,r,t)\sat K_{i}(\atn{t'}{K_i(\atn{t}{K_{i}\varphi})})$
\end{proofL}

\begin{figure}[!h]
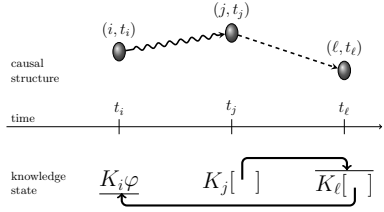

	\centering
	\newcommand{\tikzfname}{tikz_fip3b}
	\tikzsetnextfilename{\cfcurrentfolder Figures/\tikzfname}
	\resizebox{0.4\linewidth}{!}{\cfinput{Tikz/\tikzfname}}
	\caption{Syncausal relation and timing guarantee, with induced knowledge state}
	\label{fig:fip3}
\end{figure}

Lemmas~\ref{lem:fip1} and ~\ref{lem:fip2}
capture essential epistemic aspects of the $\fip$
in the synchronous context~$\gammax$, based in part on perfect recall. 
Composing them gives us Lemma~\ref{lem:fip3}, 
which is at the heart of the proof of the sufficiency
Theorem~\ref{thm:centipede_suff} below. The causal and epistemic
states described by the lemma are shown in Figure~\ref{fig:fip3}.

\begin{lemma}\label{lem:fip3}
If ~$(\Rfip,r,t_{i})\sat K_i\varphi$ ~and~ $(i,t_{i})\spc(j,t_{j})\tspc(\ell,t_{\ell})$ \\
then~~$(\Rfip,r,t_{j})\sat K_{j}\atn{t_{\ell}}{K_{\ell}\atn{t_{i}}{K_{i}\vphi}}$~.
\end{lemma}

\begin{proofL}
	Since $(\Rfip,r,t_{i})\sat K_i\varphi$ and
        $(i,t_{i})\spc(j,t_{j})$, Lemma~\ref{lem:fip1} 
	gives us $(\Rfip,r,t_{j})\sat
        K_{j}\atn{t_{i}}{K_i\varphi}$. Now as
        $(j,t_{j})\tspc(\ell,t_{\ell})$, 
	using Lemma~\ref{lem:fip2} we get 
	$(\Rfip,r,t_{j})\sat K_{j}\atn{t_{\ell}}{K_{\ell}\atn{t_{j}}{\atn{t_{i}}{K_i\varphi}}}$.
	Finally, applying validity $TS2$ reduces the result to
	$(\Rfip,r,t_{j})\sat K_{j}\atn{t_{\ell}}{K_{\ell}\atn{t_{i}}{K_i\varphi}}$
	 as required.
\end{proofL}

We are now ready to prove that in $\fip$, the existence of a centipede
is sufficient for nested knowledge gain.

\begin{theorem}\label{thm:centipede_suff}
If ~$(\Rfip,r,t)\sat K_{i_0}\varphi$ and there is a centipede
    for $\angles{ i_0,\ldots,i_k}$ in $(r,t..t')$, then
    ~$(\Rfip,r,t')\sat K_{i_k}K_{i_{k-1}}\cdots
    K_{i_1}K_{i_0}(\atn{t}{K_{i_0}\varphi})$~.
\end{theorem}

\begin{proofT}
	Let $\angles{(j_0,t_0),..,(j_{k},t_{k})}$ be
	a centipede for $\langle i_0,\ldots,i_k\rangle$ in $(r,t..t')$, 
	such that $(i_{0},t)=(j_0,t_0)$ and
        $(j_{k},t_{k})=(i_{k},t')$, as seen in
        Figure~\ref{fig:centi_suff}.   
	
	\begin{figure}[!h]
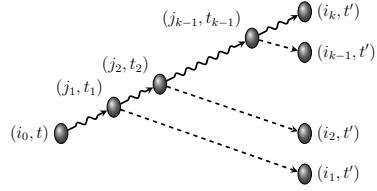

		\centering
		\newcommand{\tikzfname}{tikz_centi_suff}
		\tikzsetnextfilename{\cfcurrentfolder Figures/\tikzfname}		
		\resizebox{0.4\linewidth}{!}{\cfinput{Tikz/\tikzfname}}
		\caption{Centipede for Theorem~\ref{thm:centipede_suff}}
		\label{fig:centi_suff}
	\end{figure}
	
	We show by proceeding inductively on each ``body'' node $(j_{h},t_{h})$ for $0\leq h \leq k$ that 
	$(\Rfip,r,t_h)\sat K_{j_h}(\atn{t'}{K_{i_h}K_{i_{h-1}}\ldots K_{i_0}(\atn{t}{K_{i_{0}}\varphi})})$.
	Recall that the global time~$t$ appears as a component of all local
	states. Thus, $(r^1,t^1)\sim_i(r^2,t^2)$ is possible only if $t^2=t^1$.
	
	\begin{description}
		\item[$h=0$:] As $(i_{0},t)\tspc (i_{0},t')$, applying
                  Lemma~\ref{lem:fip2} to assumption  
		$(\Rfip,r,t)\sat K_{i_0}\varphi$ gives us
                $(\Rfip,r,t)\sat
                K_{i_{0}}\atn{t'}{K_{i_0}\atn{t}{K_{i_{0}}\varphi}}$. 
		Since $i_{0}=j_{0}$ we get
		$(\Rfip,r,t)\sat
                K_{j_{0}}\atn{t'}{K_{i_0}\atn{t}{K_{i_{0}}\varphi}}$. 
				
		\item[$h>0$:] Assume for $h-1$ and show for $h$. For
                  clarity, define 
		$$\Psi_{h} = K_{i_{h}}\ldots K_{i_0}(\atn{t}{K_{i_{0}}\varphi}).$$
		The inductive
		assumption gives us that 
		$$(\Rfip,r,t_{h-1})\sat K_{j_{h-1}}(\atn{t'}{\Psi_{h-1}}).$$
		By definition of the centipede we
		have that $(j_{h-1},t_{h-1})\spc(j_h,t_h)$ and that 
		$(j_h,t_h)\tspc(i_{h},t')$. By Lemma~\ref{lem:fip3}
		we have that 
		$$(\Rfip,r,t_{h})\sat K_{j_{h}}(\atn{t'}{K_{i_{h}}\atn{t'}{\Psi_{h-1}}}).$$
		Using validity $TS3$ we reduce this to
		$$(\Rfip,r,t_{h})\sat K_{j_{h}}(\atn{t'}{K_{i_{h}}\Psi_{h-1}}),$$
		which is the required
		$(\Rfip,r,t_{h})\sat K_{j_{h}}(\atn{t'}{\Psi_{h}})$.
		This concludes the inductive proof. 		
	\end{description}
													
	In particular, we obtain
	$(\Rfip,r,t')\sat K_{i_k}(\atn{t'}{K_{i_k}K_{i_{k-1}}\ldots
	K_{i_0}(\atn{t}{K_{i_{0}}\varphi})})$ since $j_k=i_k$ and $t_k=t'$. 
	Using the Knowledge Axiom we obtain that \\
	$(\Rfip,r,t')\sat \atn{t'}{K_{i_k}K_{i_{k-1}}\ldots
	K_{i_0}(\atn{t}{K_{i_{0}}\varphi})}$. Finally, using TS1, 
	$(\Rfip,r,t')\sat K_{i_k}K_{i_{k-1}}\ldots
	K_{i_0}(\atn{t}{K_{i_{0}}\varphi})$ as desired.
	\renewcommand{\toto}{thm:centipede_suff}
\end{proofT}
														
Theorem~\ref{thm:centipede_suff} proceeds by tracing the knowledge states of the
centipede's ``body'' nodes.\footnote{The knowledge state of a node is a
convenient abuse of language, that refers to the knowledge state of the process
related to the node, at the time related to the node.} These nodes provide a
communication path that ``feeds'' the endpoints $i_{1},i_{2}$, etc. A subtle
point is that each body node $\theta_{h}$ already knows that its related
endpoint $i_{h}$ will know by $t'$ what it (i.e. $\theta_{h}$) knows. This is
information that $\theta_{h}$ can also pass on to the next body node
$\theta_{h+1}$.

We obtain the following result by immediate application of 
Lemma~\ref{lem:suffOR} to Theorem~\ref{thm:centipede_suff}.

\begin{theorem}[Nested Knowledge Sufficiency]\label{thm:NKsuffRP}
    Let $P$ be an $\fip$ protocol that is also non-hesitant for
    $\OR=\angles{\trig,\alpha_1,\ldots,\alpha_k}$. If for every
    $r\in\Rfip=\Rrep(P,\gammax)$ in which $e$ is an ND event at $(i_0,t)$
    there exists time $t'$ such that  a centipede for $\langle
    i_0,\ldots,i_k\rangle$ exists in $(r,t..t')$, then $P$ solves $\OR$.
\end{theorem}

\section{Conclusions}\label{sec:NGK_conclusions}
This chapter starts out by introducing and discussing several new concepts 
related to causality in synchronous systems. Thus, the \emph{bound guarantee}
and \emph{syncausality} relations lead up to the \emph{centipede} structure. Then
the formal theory is developed that results with the Knowledge Gain Theorem, and thence
the Centipede Theorem. Finally, it is shown that the causal structures defined are \emph{tight},
in the sense that there exists a protocol where the existence of a centipede is sufficient
for knowledge gain and for solving the \OR problem.

Our results all hold in particular in the case in which $\bij=\infty$
for all channels, so that communication is asynchronous (although
processes share the global clock and can move at each step). Because
communication is asynchronous, bound guarantees are useless in this
setting. Syncausality reduces to Lamport's happened-before, and all
possible centipedes collapse to message chains. Thus, our results also
apply to such contexts, reproving Chandy and Misra's Knowledge-gain
theorem in a slightly more general setting. Asynchrony of
communication alone suffices for this type of implosion.

How knowledge actually evolves in a system will
depend on the particular protocol used.   As a first study of the role that
protocols play in determining information flow in the synchronous
contexts $\gammax$, we have analyzed the full-information protocol
and have shown that the definitions for syncausality and centipede are
not only necessary but also sufficient for nested knowledge under such
protocols. If one adds the non-hesitance property then the causal
structures also suffice for solving the \OR problem. It follows that
our characterization of coordination in terms syncausality and
centipedes is, in a precise sense, tight.

%
\ifx \isEmbedded \undefined


\maketitle

\else
\fi
\label{ch:CKG}

\section{Introduction}\label{CKG_intro}
This chapter  analyzes the causal
relations that lead to common knowledge gain and to simultaneous coordination. A well-known result
\cite{HM1} shows that common knowledge cannot be gained in asynchronous systems.
Common knowledge can, however, be gained in synchronous ones. As such, the
results in this chapter have no counterpart in asynchronous systems. The state
of common knowledge has been shown to play an important role in agreements and
in coordinating simultaneous actions \cite{HM1,FHMV96,FHMV}.

As before, we provide a more concrete motivation for our investigation by
considering the Simultaneous Response problem, defined in
Section~\ref{sec:coord_via_response}. Consider the scenario depicted in the
following example.

\begin{figure}[h!]
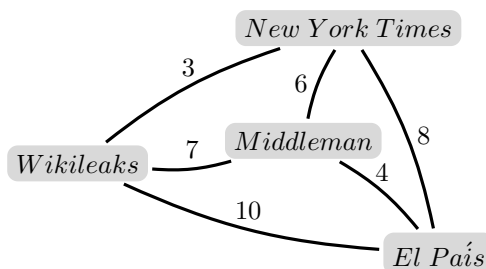

	 \begin{center}
		\newcommand{\tikzfname}{tikz_ck_net}
		\tikzsetnextfilename{\cfcurrentfolder Figures/\tikzfname}
		\cfinput{Tikz/\tikzfname}
	\caption{The network of Example~\ref{ex:ck1}}
	\label{fig:ck_net1}
	  \end{center}
\end{figure}

\begin{example}\label{ex:ck1}

The Wikileaks whistle blowing site is about to uncover yet another state secret. It strikes a
bargain with \emph{El Pa\'is} and~\emph{The New York Times}. As soon  the
secret becomes available to Wikileaks (the exact timing depends upon an external
source and is thus unknown), it will pass on the information to the papers
using time stamped messages. The contract with Wikileaks
states that both papers are to publish the scoop \emph{simultaneously}, or not
at all. The parties involved communicate over the network shown in
Figure~\ref{fig:ck_net1}. Note that the scenario sketches out an instance of
\SR\ where a spontaneous event at Wikileaks is to be followed by a pair
of simultaneous publication events.

\begin{figure}[!h]
	\centering
	\subfloat[Publishing simultaneously at $t+10$]{
		\label{fig:ck_run1}
		\newcommand{\tikzfname}{tikz_ck_case1}
		\tikzsetnextfilename{\cfcurrentfolder Figures/\tikzfname}
		\scalebox{0.75}{\cfinput{Tikz/\tikzfname}}
	}
	\subfloat[Publishing simultaneously at $t+9$]{
		\label{fig:ck_run2}
		\newcommand{\tikzfname}{tikz_ck_case2}
		\tikzsetnextfilename{\cfcurrentfolder Figures/\tikzfname}
		\scalebox{0.75}{\cfinput{Tikz/\tikzfname}}
	}
	
	\caption{Example~\ref{ex:ck1}}
	\label{fig:ex_ck1}
\end{figure}

Suppose that the secret becomes available at time $t$ and that Wikileaks
sends messages to the \emph{NYT} and \emph{El Pa\'is} right after (let's
keep the Middleman out of it for now).

In Figure~\ref{fig:ck_run1} Wikileak's messages to the \emph{NYT} and \emph{El Pa\'is} 
arrive at times $t+3$ and $t+5$ respectively. The editors both wait until $t+10$
before simultaneously publishing the secret.

Figure \ref{fig:ck_run2} offers an alternative scenario. Here the Middleman 
is also notified by Wikileaks, and it sends on messages to both papers. 
Despite the fact that the messages sent by Wikileaks to the papers both arrive
by $t+3$, and that the Middleman's messages arrive by $t+8$,
the papers must wait until $t+9$ in order to ensure simultaneous publication
of the scoop.

\hfill$\Box$
\end{example}

Recall that, given Theorem~\ref{thm:SRck}, the simultaneous response requirement
is reduced to a requirement for common knowledge of the occurrence of the ND
event. Example~\ref{ex:ck1} is thus best analyzed in terms of knowledge gain. In
Figure~\ref{fig:ck_run1}, as soon as the message to \emph{El Pa\'is} arrives, we
have  that $K_{E}\mathsf{Secret} \wedge K_{N}\mathsf{Secret}$. But we also have
$K_{E}K_{N}\mathsf{Secret}$, since the Spanish editor can work out that send time
was $t$ and that a message to the \emph{NYT} will have arrived by $t+3$ at the
latest. $K_{N}K_{E}\mathsf{Secret}$ does not hold however, because the message 
to  \emph{El Pa\'is} may take longer than $5$. By waiting until $t+10$ we also
have $K_{N}K_{E}\mathsf{Secret}$. Since both bounds have been reached, 
and since the bounds are common knowledge, we also
get that $K_{E}K_{N}K_{E}\mathsf{Secret}$, $K_{N}K_{E}K_{N}K_{E}\mathsf{Secret}$, etc.
As this ever lengthening nesting of knowledge  points out, at
$t+10$ the group of papers \{\emph{NYT}, \emph{El Pa\'is}\} has gained common
knowledge of the secret, $C_{\{E,N\}}\mathsf{Secret}$.

In Figure~\ref{fig:ck_run2} similar calculations will convince the reader that,
based on the Middleman's messages, common knowledge arises already at $t+9$.
Note that mutual knowledge ($K_{E}\mathsf{Secret} \wedge K_{N}\mathsf{Secret}$,
established at time $t+3$) and even mutual nested knowledge 
($K_{N}K_{E}\mathsf{Secret} \wedge K_{E}K_{N}\mathsf{Secret}$, established at
time $t+7$) do not necessarily lead to common knowledge. For example, at $t+7$
$K_{E}K_{N}K_{E}\mathsf{Secret}$ does \emph{not} hold: the Spanish editor
is thinking that as far as the editors in New-York are concerned, 
a message from Wikileaks  to \emph{El Pa\'is} could arrive as late as $t+10$, and 
that the message from the Middleman may not have arrived in New-York as yet.

\section{Centibrooms}\label{sec:centibroom}
As illustrated above the existence of a centipede, even under the best of terms where 
messages contain all relevant information, may not suffice for ensuring common knowledge gain.
The analysis suggests that it is only when a node exists from which messages are 
\emph{guaranteed} to have arrived at the sites of all parties concerned, that common knowledge
may arise. 

We now define a communication structure that echoes this intuition.

\begin{figure}
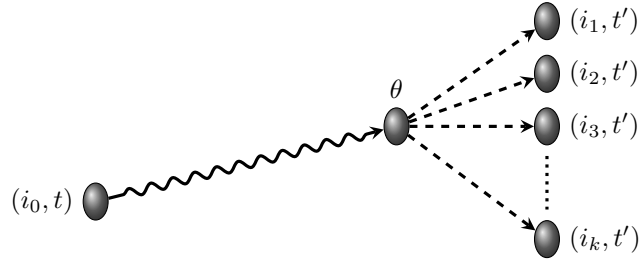

	\centering
		\newcommand{\tikzfname}{tikz_centibroom}
		\tikzsetnextfilename{\cfcurrentfolder Figures/\tikzfname}
		\cfinput{Tikz/\tikzfname}
	\caption{A centibroom}
	\label{fig:centibroom}
\end{figure}

\begin{definition}[Centibroom]
	\label{def:centibroom}
	Let $t\le t'$	and $G\subseteq\Proc$.
	Node $\theta$ is a {\em centibroom} for  $\angles{i_0,G}$ in $\rmmp$ if
	~$(i_0,t)\spc\theta$ ~and ~$\theta\tspc (i_h,t')$ holds for
        all ~$i_h\in G$. 
\end{definition}

The centibroom node $\theta$ is syncausally connected to the
originating node of the nondeterministic event, which enables it to be
informed of the event's occurrence. Node~$\theta$ is also connected 
by bound guarantees to the time~$t'$ nodes of all processes in~$G$.
Intuitively, this makes it possible for $\theta$ to guarantee that a
message sent to any $i_h\in G$ will have arrived by $t'$. Note that,
once again,  Figures~\ref{fig:ck_run1} and~\ref{fig:ck_run2} contain
centibroom structures (in both figures all communication that 
is not a part of the centibroom is dimmed out).

The Centibroom Theorem, formulated below and proved in the next section, shows that
indeed in order to coordinate a simultaneous response, a centibroom
must exist that connects the responding sites to the triggering one. The 
Centibroom Theorem can be seen as an extension of the Centipede
Theorem that applies to the \SR\ problem.

\begin{theorem}[Centibroom Theorem]\label{thm:centibroom}
	Let $P$ be a protocol solving $\SR=\angles{\trig,\alpha_1,\ldots,\alpha_k}$
	in $\gammax$, and assume that $\trig$ occurs at $(i_0,t)$ in
	$r\in\Rmax$.
	If the response actions are performed at time $t'$ in~$r$, then
	there is a centibroom $\langle i_0,G\rangle$ in
	$\rmmp$.
\end{theorem}

\section{Common Knowledge Requires Centibrooms}\label{sec:CKreqCentibroom}
Clearly, centibrooms are simpler structures than general centipedes. Notice,
however, that a  centibroom for $G=\{j_1,\ldots,j_\ell\}$ can be considered as a
condensed representation of infinitely many centipedes, each of which can
support knowledge gain of a particular formula. More concretely, we have the following.

\begin{lemma}\label{lem:broom-is-cent}
	Let $G\subseteq\Proc$, and let $\theta$ be a centibroom for $\langle
	i_0,G\rangle$ in $\rmmp$. Then for every sequence $\langle
	i_1,\ldots,i_k\rangle\in G^k$ of processes in~$G$, the sequence
	$(i_0,t)\cdot\theta^k$ (where $\theta$ repeats $k$ times) is a centipede for
	$\langle i_0,\ldots,i_k\rangle$ in $\rmmp$.
\end{lemma}

\begin{proofL}
	Fix a sequence $\cC=\angles{i_{1},\ldots,i_{k}}\in G^{k}$. The sequence\\
	$\angles{(i_{0},t),\overbrace{\theta\cdots\theta}^{k-1},(i_{k},t')}$, with
	$k-1$ repetitions of $\theta$, is a
	centipede for $\cC$, since: 
	\begin{itemize} 
		\item•$\theta$ is a centibroom for $\langle i_0,G\rangle$, so $(i_{0},t)\spc
		\theta$, and 
		\item•$\theta \tspc (i_{k},t')$ implies $\theta \spc (i_{k},t')$, and
		\item $\theta\spc \theta$ due to reflexivity of $\spc$, finally 
		\item $\theta$ is a centibroom for $\langle i_0,G\rangle$ so
		$\theta\tspc (i_{h},t')$ for all $1\leq h \leq k-1$.
	\end{itemize}•
\end{proofL}

Notice that Lemma~\ref{lem:broom-is-cent} does not bound the value
of~$k$, nor does it restrict the possibility of repetitions in the
sequence $\langle i_1,\ldots,i_k\rangle$ in question.
We are now ready to show that the centibroom  serves as the structure underlying
common knowledge. 

\begin{theorem}[Common Knowledge Gain]
	\label{thm:CKgain}
	Let~$P$ be an arbitrary protocol, let $G\subseteq\Proc$,
	and let $r\in\Rmax$. Assume
	that   $e$ is an ND event at $(i_0,t)$ in~$r$. 
	If $t'>t$ and $(\Rmax,r,t')\sat C_G(\ndocc{e})$,
	then there is a  centibroom $\hat \theta$ for $\angles{ i_0,G}$ in
	$\rmmp$.
\end{theorem}

\begin{proofT}
	Assume the notations and conditions of the theorem. Denote
	$G=\{i_1,\ldots,i_k\}$ and $d=t'-t$. Since $(\Rmax,r,t')\sat C_G \ndocc{e})$
	we have by definition of common knowledge that\\
	$(\Rmax,r,t')\sat E^{k(d+1)}_G\ndocc{e}$. In particular, this implies that \[(\Rmax,r,t')\sat
	~(K_{i_k}\cdots K_{i_1})^{d+1}\ndocc{e},\] where $(K_{i_k}\cdots
	K_{i_1})^{d+1}$ stands for~$d+1$ consecutive copies of $K_{i_k}\cdots
	K_{i_1}$. By the Knowledge Gain Theorem~\ref{thm:cpede}, there is a
	corresponding centipede
	$\sigma=\langle\theta_0,\theta_1,\ldots,\theta_{k(d+1)}\rangle$ in $\rmmp$.
	Denote $\theta_h=(i_h,t_h)$ for all $0\le h\le k\!\cdot\!(d+1)$. Recall
	that, by definition, $\theta_h\spc\theta_{h+1}$ holds for all
	$h<k\!\cdot\!(d+1)$. By Lemma~\ref{lem:step-synch} we obtain that if
	$\theta_h\ne\theta_{h+1}$ then $t_h<t_{h+1}$. It follows that there can be
	at most~$d+1$ distinct nodes $\alpha_1\spc\alpha_2\spc\cdots\spc\alpha_\ell$
	in~$\sigma$. Every $\alpha_h$ represents a segment
	$\theta_x,\ldots,\theta_{x+s}$ of the nodes in~$\sigma$. By the pigeonhole
	principle, one of the $\alpha$'s must represent a segment consisting of at
	least~$k$ of the $\theta$'s in~$\sigma$. Denoting this node by~$\hat\alpha$,
	we obtain that $\hat\alpha\tspc (i_h,t')$ for every $i_h\in G$. Moreover, by
	definition of the centipede and transitivity of~$\spc$ we have that
	$(i_0,t)\spc\hat\alpha$. It follows that $\hat\alpha$ is a centibroom 
	for $\langle i_0,G\rangle$ in $\rmmp$.	
	\renewcommand{\toto}{thm:CKgain}
\end{proofT}

The proof of Theorem~\ref{thm:CKgain} is based on the
Knowledge Gain Theorem~\ref{thm:cpede}. Recall that  $C_G\varphi$ implies
arbitrarily deeply nested knowledge of~$\varphi$.
Every such nested knowledge formula implies the existence of a centipede.
A nested knowledge formula is constructed whose centipede has
sufficiently many nodes that at least one of them
must be a centibroom for $G$ at~$t'$. 

In Chapter~\ref{ch:NKG} we defined the centinode, which is an instance of the
centipede whose every ``body'' node is a bridge to the related ``leg'' node. 
We now similarly identify and prove the existence of a \emph{bridging centibroom}.

\begin{definition}[Bridging centibroom]\label{def:broom_bridge}
	Let $t\le t'$	and $G\subseteq\Proc$.
	Node $\theta$ is a {\em bridging centibroom} for  $\angles{i_0,G}$ in $\rmmp$ if
	\begin{itemize}
		\item•~$\theta$ is a centibroom for $\angles{i_0,G}$ in $\rmmp$; and
		\item $\theta$ bridges $(i_{0},t)$ and $g$ for every $g\in G$.
	\end{itemize}•
\end{definition}

\begin{lemma}\label{lem:bribroom_exists}
	Fix $r\in \Rmax$ and assume that $\theta$ is a  centibroom for $\angles{
	i_0,G}$ in $\rmmp$. Then there exists a node $\theta'$ that is a bridging
	centibroom for $\angles{ i_0,G}$ in $\rmmp$.
\end{lemma}•

\begin{proofL}
	By Lemma~\ref{lem:bridges-exist} there exists a node $\psi$ such 
	that $\psi$ bridges $(i_{0},t)$ and $\theta$. Node $\psi$ is a bridging
	centibroom since
	\begin{itemize}
		\item•$(i_{0},t)\spc\psi\tspc\theta\tspc g$ implies 
		$(i_{0},t)\spc\psi\tspc g$ for all $g\in G$
		\item $\alpha\spc\psi'\tspc\psi$ ~implies $\psi'=\psi$ by definition of bridge.
	\end{itemize}•
\end{proofL}

Theorem~\ref{thm:CKgain} shows that 
common knowledge can arise in synchronous systems 
\emph{only} when there exists a centibroom structure, centered about 
 the centibroom node. The above Lemma~\ref{lem:bribroom_exists}, 
 together with Lemma~\ref{lem:bridges-early},
 points out that there must exist a bridging centibroom for the group, in which
a nondeterministic \emph{pivotal event}, either an early receive or possibly
an external input when $\psi=(i_{0},t)$, occurs.
This
demonstrates that the nature of common knowledge is finitistic,
despite its familiar definition
being based on an infinite conjunction of facts.
This phenomenon is consistent with the
analysis of common knowledge in the work on fault-tolerance
\cite{DM,MT,MM08}. There, too, common knowledge arises at some
time~$t'$ exactly if there is some property~$S$ of the correct nodes that
ensures that all processes will know by time~$t'$ that the property~$S$
held in the run.

We remark that Theorem~\ref{thm:CKgain} relates to a familiar situation
involving the evolution of knowledge in broadcasts. In a flooding protocol or a
radio broadcast, for example, the contents being broadcast become common
knowledge to a growing set of participants with time. Typically, after a time
interval equivalent to the diameter of the system, the contents can become
common knowledge to {\em all} processes in the system.

The proof of Theorem~\ref{thm:centibroom} is now immediate: we show that the
existence of a centibroom is a necessary condition for solving the Simultaneous
Response problem by applying the Common Knowledge Theorem~\ref{thm:CKgain} to
Theorem~\ref{thm:SRck}.
 
\section{The Simultaneous Global Snapshot Protocol}\label{sec:snapshot}
Before exploring further the theoretical implications for the centibroom
structure, we pause to consider a possible application.%
\footnote{We thank Gadi Taubenfeld for suggesting this application to us.} 

A well known application for Lamport's causal relation is the \emph{global
snapshot algorithm}, proposed by Chandy and Lamport in \cite{ChL}. This
algorithm is used to record a consistent global state in  asynchronous systems.
A \emph{global snapshot} of the system at a given run $r$ and time $t$,
which we will denote with $Snap(r,t)$, consists of
records of the local states of all processes in the system, and of the
communication channels, at that point in the run. 
Technically, communication channels do not posses a memory,
component so their state must be reconstructed by the processes. 
Interestingly, the Chandy-Lamport algorithm cannot ensure that the global snapshot that it
actually records is in fact a global state in the current run. No protocol can grant
such assurances in an asynchronous system. Rather, the algorithm ensures that the recorded
snapshot is \emph{consistent} with the current run in the following sense:

\begin{definition}[Snapshot consistency]\label{def:snapshot_cons}
	Fix $r\in \Rrep(P,\gamma)$ for arbitrary protocol $P$ and context $\gamma$.
	Snapshot $S^{*}$ is consistent with the interval $[t_{s},t_{e}]$ of $r$
	if there exists $r'\in \Rrep(P,\gamma)$ and  times $t'_{s}\leq t'_{*} \leq t'_{e}$ such that

	\begin{enumerate}
		\item $Snap(r,t_{s})=Snap(r',t'_{s})$,
		\item $S^{*}=Snap(r',t'_{*})$, and
		\item $Snap(r,t_{e})=Snap(r',t'_{e})$.
	\end{enumerate}• 
\end{definition}•

Mechanisms for recording global states come in useful, for example,  in
association with  recovery from system failure. In fact, many applications use
such algorithms in order to retain ``checkpoints'': global states that can be
``rolled back'' into, when failure occurs \cite{Neves1996}. The Centibroom Theorem
suggests a synchronous variation for Chandy and Lamport's original algorithm.
We will
actually consider two variants: the first being message optimal, and the second 
providing time optimization.

When activated, the \emph{Simultaneous Global Snapshot Protocol} results with
all processes simultaneously recording their local states at a time $t$, and all
messages that are in transit on inbound communication channels at that time.
Observe that given the synchronous nature of the system, simultaneity is a
necessary requirement for achieving a consistent global state. Allowing two
processes $i$ and $j$ to record their local states at $t_{i}$ and $t_{j}$
respectively, where $|t_{i}-t_{j}|>0$ may, in the general case, result in an
inconsistency: it may be the case that there are no possible global states
that includes the local states defined by $(i,t_{i})$ and $(j,t_{j})$ both,
 due to simultaneous actions that are always
performed by $i$ and $j$ together at some time $t_{i}< t'\leq t_{j}$.
Summing up, if snapshot $S^{*}$ is consistent with 
 the interval $[t_{s},t_{e}]$ of run $r$, then there exists some time $t_{*}\in [t_{s},t_{e}]$
 such that $S^{*}=Snap(r,t_{*})$.

The algorithm is quite simple. We mark with $Diameter_{i}$ the distance of the
process $j$ furthest from $i$, when measuring based on $\sD{ij}$.  
We assume that the protocol may be
initiated (from the outside) at any process in the system, 
or even in several places in the system. Algorithm~\ref{alg:snap-1}
shows the protocol's pseudo code. The (arbitrary) initiator node $(i,t)$ floods
the system with $initiate$ messages, that indicate time $t'=t+Diameter_{i}$ as the
time at which the ``snapshot'' must be taken. By definition of $Diameter_{i}$, 
these messages arrive at all sites by time $t'$. At $t'$
every process $j$ records its own local state, and starts recording incoming
communications on each of its inbound channels. 
Recording the channel $h \mapsto j$ takes place from time $t'$,
until $t'+\bound{hj}$, but only messages that are not marked
with an extra ``ignore'' bit are recorded.  Apart from carrying on these
recordings, the processes are free to carry on with their (non snapshot related)
tasks. However, if these tasks demand that a process $j$ send a message on some
\emph{outbound} channel $j \mapsto h$ prior to time $t'+\bound{jh}$, 
then this message is marked ``ignore'' by appending an extra bit set to 1 to the message. 

Note that a different mechanism could be employed for the purpose of recording the contents
of communication channels. Rather than starting to record upon snapshot, the alternative
mechanism would have each process constantly keeping a long-enough tail
on its history so that when snapshot occurs at time $t$, for each channel $i\mapsto j$,
process $i$ can recount all messages sent on the channel which may, potentially still be
en route. Those would be all messages sent after $t-\bound{ij}$. At the price of greater
stress on memory resources,  the algorithm would complete the snapshot recording
faster. Although in order to gain a complete picture of the state of the channel $i\mapsto j$,
we would have to further compare the local states of $i$ and $j$ at the snapshot time.
For this reason we opt for the version  presented below, its simplicity being better suited for our
explanatory purposes.

\begin{algorithm} 
\caption{Simultaneous Global Snapshot~~-~~$P^{snap-1}$}\label{alg:snap-1} 
\begin{algorithmic}[1] 
\Procedure{Initiator node $(i,t)$:}{}
\State $snapshot \gets t+Diameter_{i}$
\ForAll {outgoing channels $i\mapsto h$}
	\State $send_{h}(initiate(snapshot))$ 
\EndFor
\EndProcedure 

\Procedure{Arbitrary node $(j,t')$:}{}
\If {receive initiate(S)} 
\State $snapshot\gets S$ 
\ForAll {outgoing channels $i\mapsto h$}
	\State $send_{h}(initiate(snapshot))$
\EndFor
\EndIf
\If {$t'=snapshot$} record local state \EndIf
\ForAll {incoming channels $h\mapsto j$}
	\State receive $msg$ on channel 
	\If {$snapshot \leq t'< snapshot+\sD(h,j) \wedge msg.transparent\neq 0$}
		\State record $msg$
	\EndIf
\EndFor
\EndProcedure 
\end{algorithmic} 
\end{algorithm}

The following lemma proves the
protocol's correctness.

\begin{lemma}\label{lem:snapshot_correct}
	Choose $r\in\Rrep(P^{snap-1},\gammax)$ where snapshot initiation
	occurs at $(i,t)$. Then there exists a time $t'\geq t+Diameter_{i}$ where
	each process $j$ contains 
	\begin{enumerate}
		\item record of its local state at time $t+Diameter_{i}$, and
		\item record of incoming messages en route at time $t+Diameter_{i}$.
	\end{enumerate}•
\end{lemma}

\begin{proofL}
	That all process local states are simultaneously recorded at $t'$
	is straightforward from the definitions. That exactly those messages that were 
	in transit at time $t'$ are recorded can be seen by noting first that all
	messages in transit on channel $h\mapsto j$ at $t'$ are guaranteed
	to arrive by time $t'+\bound{hj}$, at which point recording on that channel 
	stops. Moreover, messages sent after $t'$ but which arrive at $j$ before 
	$t'+\bound{hj}$ will be marked transparent and will not be recorded. Thus, the 
	algorithm is correct in recording the global state at time $t'$.  
\end{proofL}


The algorithm is straightforward. A revised version of the algorithm can ensure time
optimality. The protocol starts the same, with the initiating node $(i,t)$
flooding the system with $initiate$ messages bearing the value $t+Diameter_{i}$.
However, in this version, every process $j$ that gets such a message at time
$t'$ checks to see whether it can ensure an even quicker simultaneous recording
response, i.e. whether $t'+Diameter_{j}<t+Diameter_{i}$. If so, it will start to
flood the system with $initiate$ messages bearing  $t'+Diameter_{j}$. 

\begin{algorithm} 
\caption{Simultaneous Global Snapshot~~-~~$P^{snap-2}$}\label{alg:snap-2} 
\begin{algorithmic}[1] 
\Procedure{Initiator node $(i,t)$:}{}
\State $snapshot \gets t+Diameter_{i}$
\ForAll {outgoing channels $i\mapsto h$}
	\State $send_{h}(initiate(snapshot))$ 
\EndFor
\EndProcedure 

\Procedure{Arbitrary node $(j,t')$:}{}
\If {receive initiate(S)} 
	\If {$S\leq t'+Diameter_{j}$}
		\State $snapshot\gets S$ 
	\Else
		\State $snapshot\gets t'+Diameter_{j}$ 
	\EndIf	
\ForAll {outgoing channels $i\mapsto h$}
	\State $send_{h}(initiate(snapshot))$
\EndFor
\EndIf
\If {$t'=snapshot$} record local state \EndIf
\ForAll {incoming channels $h\mapsto j$}
	\State receive $msg$ on channel 
	\If {$snapshot \leq t'< snapshot+\sD(h,j) \wedge msg.transparent\neq 0$}
		\State record $msg$
	\EndIf
\EndFor
\EndProcedure 
\end{algorithmic} 
\end{algorithm}

\begin{lemma}\label{lem:snap2}
	Protocol $P^{snap-2}$ has the following two properties:
	\begin{description}
		\item[\textbf{Correctness:}] It is correct.
		\item[\textbf{Optimality:}] No other protocol can ensure a shorter delay between initiation
		and time of snapshot.
	\end{description}•
\end{lemma}

\begin{proofL}
	\begin{description}
		\item[\textbf{Correctness:}] Fix a run $r\in \Rrep(P^{snap-2}, \gammax)$ where
		initiation of snapshot algorithm occurs at $\theta_{0}=(i_{0},t_{0})$, setting snapshot
		time for $t'_{0}=t_{0}+Diameter_{i_{0}}$. If no shorter term initiate messages are issued within
		the interval $[t_{0},t'_{0}]$ then $r$ is also a $P^{snap-1}$ run, and is thus correct by
		Lemma~\ref{lem:snapshot_correct}.
		
		Otherwise, let $t'_{1}<t'_{0}$ be the earliest snapshot time suggested after initiation, and
		let $\theta_{1}=(i_{1},t_{1})$ be the issuing node. As $t'_{1}=t_{1}+Diameter_{i_{1}}$ and
		no process issues a shorter term $initiate$ message, $initiate(t'_{1})$ is guaranteed
		to arrive at all processes no later than $t'_{1}$. Again, as no process issues a shorter
		term snapshot suggestion, the local variable $snapshot$ is equal to $t'_{1}$ at time $t'_{1}$
		in all processes. Now, based on Lemma~\ref{lem:snapshot_correct}, the run is 
		shown to be correct.
		
		\item[\textbf{Optimality:}] By the Centibroom Theorem, any protocol in which a 
		simultaneous action on the part of all processes   
		is dependent upon snapshot initiation must contain a centibroom  for $\angles{\theta_{0},\Proc}$
		where $\theta_{0}=(i_{0},t_{0})$ is the initiation node. Choose a run $r\in \Rrep= \Rrep(P^{snap-2}, \gammax)$
		initiation occurs at $theta_{0}$ and snapshot at $t'_{0}$. 
		
		Suppose that there exists a centibroom
		node $(i_{1},t_{1})$ for  $\angles{\theta,\Proc}$ in $(r,t..t'_{1})$, where $t'_{1}<t'_{0}$. Assume
		without loss of generality that
		for every $t''<t'_{1}$ there are no centibrooms for  $\angles{\theta,\Proc}$ in $(r,t..t'')$.
		By definition of centibroom, $\theta_{0}\spc\theta_{1}$ and $\theta_{1}\tspc (h,t'_{1})$ for all $h\in \Proc$.
		At $t_{1}$ or sooner, $i_{1}$ receives an $initiate(S)$ message with
		some suggested snapshot time $S$. Since $t_{1}+Diameter_{i_{1}}=t'_{1}<t'_{0}\leq S$,  
		and as $i_{1}$ is following $P^{snap-2}$, it immediately starts to flood the system with  
		$initiate(t'_{1})$ messages. As no shorter term suggestion is made, by the above proof of
		the correctness of $P^{snap-2}$, snapshot occurs at $t'_{1}<t'_{0}$, in contradiction to the assumption
		that snapshot occurs at  $t'_{0}$. 
		
		We thus obtain that for every run $r\in \Rrep$ in which initiation occurs at $\theta_{0}$, the shortest interval
		within which a centibroom can be established is $[t_{0},t'_{0}]$, where $t'_{0}$ is the time at which
		snapshot actually occurs. As all nodes in $\fut(\theta_{0})$ flood the $initiate$ messages, there cannot
		be  a protocol $P'$ where information about initiation decimates any faster than in $P^{snap-2}$, and hence
		in particular a centibroom cannot be established any faster than in $P^{snap-2}$, and so delay between 
		initiation and snapshot is at least as long as it is in $P^{snap-2}$.		
	\end{description}•	
\end{proofL}

\section{Sufficiency of Centibrooms for Common Knowledge Gain}\label{sec:suff_CKG}
We proceed to show that the centibroom indeed characterizes common knowledge gain 
in synchronous systems, in the same way nested knowledge gain is characterized
by centipedes. We will show that the existence of a centibroom is sufficient for common
knowledge gain in every $\Rfip$ system by using 
the \emph{Induction Rule for Common Knowledge}, which states that
 from $\Rfip \sat \alpha \imp E_G(\alpha \wedge \beta)$ we can infer
$\Rfip \sat \alpha \imp C_G\beta$.  Importantly,
processes must now make explicit use of their capability to discern global time
in order to gain common knowledge, due to the essential
part played by bound guarantees.

\begin{theorem}\label{thm:centibroom_suff}
	If   ~$(\Rfip,r,t)\sat K_{i_0}\varphi$ and there is a centibroom  node
	$\theta$ for $\angles{i_0,G}$ in ~$(r,t..t')$, then $(\Rfip,r,t')\sat
	C_G(\at{\varphi}{t})$.
\end{theorem}

\begin{proofT}
	Assume that the conditions of the theorem hold, and let $\theta=(j,t_j)$. In
	particular, $(j,t_j)\tspc (i,t')$ for every $i\in G$. From $(\Rfip,r,t)\sat
	K_{i_0}\varphi$ and $(i_0,t)\spc(j,t_j)$ in $r$ we have by
	Lemma~\ref{lem:fip1} that $(\Rfip,r,t_j)\sat K_j(\at{\varphi}{t})$. We now
	use the induction rule with $\alpha$ set to
	$(\ttime=t')\wedge\at{(K_j\,(\at{\varphi}{t}))}{t_j}$, and $\beta$ being
	$\at{\varphi}{t}$. Since $\Rfip\sat\alpha\imp\beta$ in this case, it
	suffices to show that $\Rfip\sat\alpha\imp E_G\alpha$. Thus, let
	$r'\in\Rfip$ and fix time $\hat{t}$. If $(\Rfip,r',\hat{t})\nsat \alpha$
	then $\alpha\imp E_G\alpha$ is trivially satisfied in $(r,\hat{t})$. Now
	suppose that $(\Rfip,r',\hat{t})\sat \alpha$, giving us that $\hat{t}=t'$
	and thus $(\Rfip,r',t')\sat\atn{t_{j}}{K_j\,(\atn{t}{\varphi})}$. This, in
	turn, gives us $(\Rfip,r',t_{j})\sat K_j\,(\atn{t}{\varphi})$ by application
	of $TS1$ (Lemma~\ref{lem:time_knowledge}). Fix $i\in G$. Since $(j,t_{j})$
	is a centibroom node, we have $(j,t_{j})\tspc(i,t')$. By
	Lemma~\ref{lem:guarantee2sync} it is also the case that
	$(j,t_{j})\spc(i,t')$ in $r'$. Using Lemma~\ref{lem:fip1} we now obtain
	$(\Rfip,r',t')\sat K_{i}\atn{t_{j}}{K_j\,(\atn{t}{\varphi})}$. Moreover, the
	fact that the time is part of the local state in~$\gammax$ implies that
	$(\Rfip,r',t')\sat K_i(\ttime=t')$. It follows that $(\Rfip,r',t')\sat
	K_i\alpha$, and since~$i$ was an arbitrarily chosen member of~$G$ then
	$(\Rfip,r',t')\sat E_G\alpha$. It follows that $\Rfip\sat\alpha\imp
	E_G\alpha$. Since $\beta=\at{\varphi}{t}$ we obtain by the Induction Rule
	that $\Rfip\sat\alpha\imp C_G(\at{\varphi}{t})$. Finally, since
	$(\Rfip,r,t')\sat\alpha$ we obtain that $(\Rfip,r,t')\sat
	C_G(\at{\varphi}{t})$, as desired.
	\renewcommand{\toto}{thm:centibroom_suff}
\end{proofT}

In order to relate the centibroom in a $\fip$ system to a solution
to the simultaneous response problem $\SR$, we must tie in common knowledge
to action. Such a connection is established if we assume that the protocol
is also considerate with respect to $\SR$ (see  Definition~\ref{def:considerate}).	
We obtain the following result by immediate application of 
Lemma~\ref{lem:suffSR} to Theorem~\ref{thm:centibroom_suff}.

\begin{theorem}[Nested Knowledge Sufficiency]\label{thm:CKsuffRP}
    Let $P$ be an $\fip$  that is also considerate with respect to
    $\SR=\angles{\trig,\alpha_1,\ldots,\alpha_k}$. If for every
    $r\in\Rfip=\Rrep(P,\gammax)$ in which $e$ is an ND event at $(i_0,t)$
    there exists time $t'$ such that  a centibroom for $\langle
    i_0,\ldots,i_k\rangle$ exists in $(r,t..t')$, then $P$ solves $\SR$.
\end{theorem}
	
\section{Common Knowledge as a Finite Conjunction}\label{sec:finite_conjunction}
Common knowledge is typically perceived in terms of an infinite conjunction of
$E^k$, for $k>0$. There are also definitions of common knowledge in terms of a
fixed point (see, e.g., \cite{Lew,FHMV,Bar1}). The centibroom structure and the
necessity of centibrooms for common knowledge supports the fixed-point view: the
only way in which a new fact can become common knowledge is if there is a {\em
singular point}, represented by the centibroom node~$\theta$, which carries the
information that $\theta$ is a centibroom for all processes in $G$ at time~$t'$.
At time~$t'$, everyone can become aware of its existence, and the fixed-point
yields common knowledge. This is also consistent with the view advocated by
\cite{CM81,Lew}, that a \emph{shared environment} is required for common
knowledge to arise.

Even though the fixed point definition implies the infinite conjunction, Fischer
and Immerman \cite{FI1} showed that in {\em finite-state} systems, where the set
of all global states in a system~$R$ is finite, there is a power~$m$ such that
$C_G\varphi$ is equivalent to $E_G^m\varphi$.  The $\fip$ protocol, with its
perfect recall property in the synchronous context~$\gammax$, produces a state
space whose size is unbounded.  Nevertheless, given the role of the centipede
and centibroom structures in~$\gammax$, we now show that there are cases in
which common knowledge is a finite conjunction under $\fip$ in $\gammax$ as
well.

Roughly speaking, when running $\fip$ it takes time to obtain deep knowledge
{\em without} having common knowledge. Indeed, we obtain a sharp bound on the
depth of $E_G^k$ that can be obtained~$d$ time units after the occurrence of a
nondeterministic event. Given a group of size $|G|=g$ and natural number $d>0$, 
we denote by $\Mdg=(d-1)(g-1)+2$. We prove

\begin{theorem}\label{thm:CK_fip}
    Let~$e$ be an ND event occurring at $(i_0,t)$ in~$r\in \Rfip$, let $d>0$,
    and~$|G|=g$. If  ~$(\Rfip,r,t+d)\sat E_G^{\Mdg}\occurred{e}$ ~then
    ~$(\Rfip,r,t+d)\sat C_G\ndocc{e}$.
\end{theorem}

Note that although a centipede's ``body'' nodes
$\angles{\theta_{0},\ldots,\theta_{k}}$  are naturally conceived of as distinct,
they need not be such. Yet recall that by  Lemma~\ref{lem:step-synch}, when two
body nodes \emph{are} distinct, their time components must also be distinct.

Theorem~\ref{thm:CK_fip} follows directly by Theorem~\ref{thm:centibroom_suff}
from the following lemma:

\begin{lemma}\label{lem:CK_bound_exists}
   Let $r\in\Rmax$,  $d>0$, $G\subseteq \Proc$ with $g=|G|$, and assume that $e$ is
   an ND event at $(i_0,t)$ in~$r$. If  ~$(\Rmax,r,t+d)\sat
   E_G^{\Mdg}\!\occurred{e}$ ~then there exists a centibroom node for
   $\angles{i_0,G}$ in $(r,t\ldots t+d)$.
\end{lemma}

\begin{proofL}
	Assume that $(\Rmax,r,t+d)\sat E_G^{\Mdg}\!\occurred{e}$. If $(i_0,t)$ is a
	centibroom for $\angles{i_0,G}$ in~$(r,t\ldots t+d)$ then we are done. Otherwise $|G|>1$,
	and moreover there is some $j\in G$ such that $(i_0,t)\not\tspc(j,t+d)$. For
	notational convenience, let us denote the processes of $G$ by $\{j_0,\ldots,j_{g-1}\}$,
	where $(i_0,t)\not\tspc(j_0,t+d)$. Denote $M=\Mdg-1$ and let
	$f(h)=j_{(h\,\mathsf{mod}\,g)}$ for all $h\le M$. Thus, $f$ maps natural
	numbers into members of~$G$,  every interval of $g$ adjacent numbers are
	mapped to the full set $\{j_0,\ldots,j_{g-1}\}=G$, and $f(0)=j_0$. We focus on a
	knowledge formula of the form \[\Psi(e)=K_{f(M)}\,K_{f(M-1)}\cdots
	K_{f(1)}K_{f(0)}\,\occurred{e}~.\] Observe that there are
	$M+1=(|G|-1)\cdot(d-1)+2$ knowledge operators in~$\Psi(e)$, all of which belong
	to processes in~$G$. By assumption, $(\Rmax,r,t+d)\sat E_G^{M+1}\occurred{e}$, and
	hence in particular $(\Rmax,r,t+d)\sat\Psi(e)$. The Knowledge Gain Theorem implies
	that there exists a centipede for $\angles{i_0,f(0),f(1),..,f(M)}$ in $(r,t\ldots t+d)$.
	Let $$\angles{(i_0,t),\Omega^0,\Omega^1,\ldots,\Omega^{M-1},(f(M),t+d)}$$ be
	such a centipede. By definition of a centipede we have that
	$(i_0,t)\spc\Omega^0$ and $\Omega^0\tspc (j_0,t+d)$. Since `$\tspc$' is
	transitive, the fact that $(i_0,t)\not\tspc(j_0,t+d)$ implies that
	$(i_0,t)\not\tspc\Omega^0$. Since `$\tspc$' is reflexive we have that
	$(i_0,t)\ne\Omega^0$. Recall by definition of~$f$ that
        $f(M)=f(M-1)+1\mathsf{mod}\,g$. Since $g>1$, clearly
	$f(M)\ne f(M-1)$. Hence, by Lemma~\ref{lem:step-synch} we have that
	$(f(M),t+d)\not\tspc(f(M-1),t+d)$. It follows that $\Omega^{M-1}\ne(f(M),t+d)$.
	
	By Lemma~\ref{lem:step-synch}, if $\Omega^h=\Omega^{h'}$ then
	$\Omega^h=\Omega^{h''}=\Omega^{h'}$ for every $h''$ in the range $h\le h''\le
	h'$. Let $\Phi_1,\ldots,\Phi_D$ denote the maximal sub-sequence of distinct
	nodes in the sequence $\Omega^0,\ldots,\Omega^{M-1}$. Lemma~\ref{lem:step-synch}
	implies that the times at which the nodes
	$(i_0,t),\Phi_1,\ldots,\Phi_D,(f(M),t+d)$ occur form a strictly increasing
	sequence, and so $D\le d-1$. For all $b$ in the range $1\le b\le D$ define
	$s(b)=\{k:\,\Omega^k=\Phi_b\}$. Since $M=(|G|-1)\cdot(d-1)+1$ and $D\le d-1$, we
	have by the pigeonhole principle that $|s(\hat{b})|\ge |G|$ for at least one
	such index~$b'$. Since the set $s(b')$ consists of at least $|G|=g$
	consecutive natural numbers, we have that $\{f(k):\,k\in
	s(b')\}=\{j_0,\ldots,j_{g-1}\}=G$. By definition of the centipede it follows
	that $\Omega_{b'}\tspc (j,t+d)$ for all $j\in s(b')=G$, and so
	$\Omega_{b'}$ is a centibroom node for $\angles{i_0,G}$ in~$(r,t\ldots t+d)$, as
	required.
\renewcommand{\toto}{lem:CK_bound_exists}
\end{proofL}

As the next lemma shows, the bound of $\Mdg=(d-1)(g-1)+2$ of
Lemma~\ref{lem:CK_bound_exists} is tight.

\begin{lemma}\label{lem:CK_bound_tight}
    For every $t\ge 0$, $d>0$ and $g>1$ there exists a run $r\in\Rfip$, an ND
    event $e$ at $(i_0,t)$ in~$r$ and a set of processes $G\subseteq \Proc$ of
    size $|G|=g$, such that \[(\Rfip,r,t+d)\sat ~~~E_G^{\Mdg-1}
    \occurred{e}~~\wedge~~ \neg C_G\occurred{e}.\]
\end{lemma}

\begin{figure}[!h]
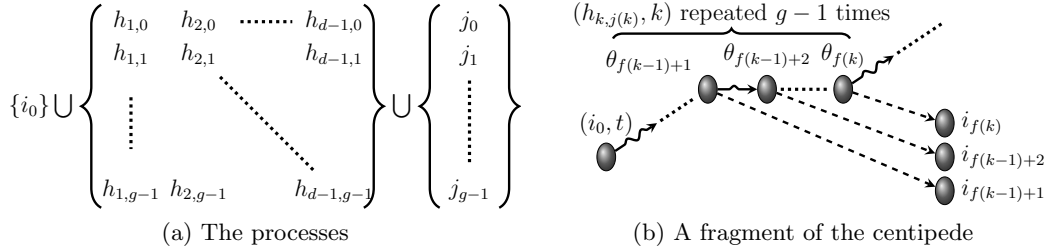

	\centering
	\advance\leftskip-1.2cm
	\subfloat[The processes]{
		\label{fig:bounds1}
		\newcommand{\tikzfname}{tikz_bound_tight1}
		\tikzsetnextfilename{\cfcurrentfolder Figures/\tikzfname}
		\scalebox{0.75}{\cfinput{Tikz/\tikzfname}}
	}
	\subfloat[A fragment of the centipede]{
		\label{fig:bounds2}
		\newcommand{\tikzfname}{tikz_bound_tight2}
		\tikzsetnextfilename{\cfcurrentfolder Figures/\tikzfname}
		\scalebox{0.75}{\cfinput{Tikz/\tikzfname}}
	}
	
	\caption{The setup for Lemma~\ref{lem:CK_bound_tight}}
	\label{fig:ebounds_tight}
\end{figure}

\begin{proofL}
  Fix $d,g$. Define $\gammax_{d,g}$ to be a synchronous context
  with the following  properties:
    \begin{itemize}
        \item Let $G=\{j_0,..,j_{g-1}\}$. For every $m<g$, denote by
          $G_{-m}$ the set $G\setminus\{j_m\}$.
        \item Let $\Proc=G \bigcup\{i_0\}\bigcup
      \{h_{k,m}\}_{1\leq k< d, 0\leq m< g}$. The set of processes is seen in Figure~\ref{fig:bounds1}.
        \item The network graph is complete, and the bounds on
          transmission times are as follows
        \begin{enumerate}
        \item for every $k<d$ and $m< g$, $\dist(h_{k,m},j)=1$ for all
          $j\in G_{-m}$
            \item for every other $i,j\in \Proc$, $\dist(i,j)=d+1$
        \end{enumerate}
    \end{itemize}

    For every $1\leq k < d$, use $H_k$ to denote the set
    $\{h_{k,m}\}_{0\leq m < g}$. Note that for every $r\in
    \Rfip_{d,g}=\Rrep(\fip,\gammax_{d,g})$, as the processes are running the $\fip$, every process
    sends every other process a message at every time unit. Note also
    that there can be no centibroom node for $\angles{i_0,G}$ in
    $(r,0\ldots d)$, because for every process $i$ there exists at least one
    $j\in G$ such that $\dist(i,j)>d$. Hence, by Theorem~\ref{thm:CKgain},
    $(\Rfip,r,t+d)\sat \neg C_G\occurred{e}$.

    Choose $r\in \Rfip_{d,g}$ such that an ND event $e$ occurs at
    $(i_0,t)$ and such that all sent messages arrive at the maximally allowed
    transmission time, except for the following ones:
    \begin{enumerate}
        \item For every $h\in H_1$, the message sent from $i_0$ to $h$ at time
        $0$ arrives at time $1$.
        \item For every $1\leq k < d-1$, for every pair of processes $h_1\in H_k$
       and $h_2\in H_{k+1}$, the message sent by $h_1$ to $h_2$ at time $k$
       arrives at $k+1$.
        \item For every $h\in H_{d-1}$ and $j\in G$, the message sent from $h$
        to $j$ at time $d-1$ arrives at time $d$
    \end{enumerate}

    The existence of $r$ is guaranteed by definition of
    $\Rfip_{d,g}$: the run is a legal possible execution of $\fip$ in
    the defined context.

    Use $f(k)$ to denote the value $(g-1)\cdot k$ for every $k>0$. Fix
    a sequence $S=\angles{i_0,i_1,..,i_{f(d-1)+1}}$ such that
    $\{i_1,..,i_{f(d-1)+1}\}\subseteq G$.  Observe that for every
    $1\leq k < d$, the subsequence
    $s_k=\angles{i_{f(k-1)+1},..,i_{f(k)}}$ contains exactly $g-1$
    elements, and so there must exist some $j(k)\in G$ such that
    $j(k)\neq i$ for every $i\in s_k$.

    We now define a node sequence
    $\angles{(i_0,t),\theta_1,..,\theta_{f(d-1)},(i_{f(d-1)+1},d)}$
    and show that it is a centipede for $S$ in $(r,0\ldots d)$.  For every
    $l=1..f(d-1)$, let $k=\lceil\frac{l}{g-1}\rceil$, and define
    $\theta_l=(h_{k,j(k)},k)$. Observe that $f(k-1)<l\leq f(k)$, and
    hence by choice of $j(k)$ that $\dist(h_{k,j(k)},i_l)=1$. Since
    $k\leq d-1$ we obtain that $\theta_l\tspc(i_l,d)$. Moreover, if
    $l<f(d-1)$ then $\theta_l\spc\theta_{l+1}$. For if
    $k>\frac{l}{g-1}$ then $\theta_l=\theta_{l+1}$ and the result
    stems from the reflexivity of $\spc$, while if $k=\frac{l}{g-1}$
    then, noting that $\theta_l\in H_k$ and $\theta_{l+1}\in H_{k+1}$,
    we get the result from clause (2) above. Finally, we note that
    $(i_0,t)\spc\theta_1=(h_{1,j(1)},1)$ since $h_{1,j(1)}\in H_1$ and
    using clause (1), and similarly that
    $\theta_{f(d-1)}=(h_{d-1,j(d-1)},d-1)\spc(i_{f(d-1)+1},d)$ since
    $h_{d-1,j(d-1)}\in H_{d-1}$ and from clause (3) above.
	Figure~\ref{fig:bounds2} shows a fragment of the described centipede.

    We have shown that there exists a centipede in $(r,0\ldots d)$ for every
    sequence 
     $\angles{i_0,i_1,..,i_{f(d-1)+1}}$ such that
    $\{i_1,..,i_{f(d-1)+1}\}\subseteq G$. By Theorem
    \ref{thm:centipede_suff} we get that $(\Rfip_{d,g},r,t')\sat
    K_{i_{f(d-1)+1}}K_{i_{f(d-1)}}\cdots K_{i_1}\occurred{e}$ for
    every such sequence. We thus obtain, considering that
    $f(d-1)+1=(g-1)(d-1)+1=\Mdg-1$, that $(\Rfip_{d,g},r,t')\sat
    E_G^{\Mdg-1}\occurred{e}$, by definition of $E$ operator.
\renewcommand{\toto}{lem:CK_bound_tight}
\end{proofL}

Theorem~\ref{thm:CK_fip} and Lemma~\ref{lem:CK_bound_tight} tightly bound the
levels of $E_G^k$ that can hold without common knowledge necessarily arising.
They draw an essential connection between this bound, the size of the set of
processes~$G$ in question, and the time that elapses since the ND event of
interest occurs. It is natural to ask whether this property is restricted to
$\fip$, or perhaps may be true in general. We now show that it is not true for
all protocols. In fact, there is a protocol that can attain arbitrary levels of
nested knowledge quickly, without giving rise to common knowledge.

\begin{example}
\label{ex:Conway}
    Let ${\gammax}'$ be a context with $\Proc=\{s,0,1\}$, where the network is V-shaped
    with $s$ at the base, and the communication bounds are $b_{s,0}=b_{s,1}=1$.
    The initial state of process~$s$ contains an {\em initial value} consisting
    of a natural number~$k\ge 0$. We assume that the protocol $P$ that~$s$ is
    following prescribes the following actions upon receiving an external input
    (an event that we denote by~$e$): If $k$ is odd, then~$s$ sends the message
    $\langle\occurred{e},k\rangle$ to process~$1$, and the message
    $\langle\occurred{e},k-1\rangle$ to process~0. If $k$ is even, then~$s$
    sends the message $\langle\occurred{e},k\rangle$ to process~$0$ and, {\em in
    case $k>0$} it also sends the message $\langle\occurred{e},k-1\rangle$ to
    process~1. Moreover, $s$ never sends a message of the form
    $\langle\occurred{e},d\rangle$ if~$e$ does not occur.

    Thus, if $k=0$ then only one process will receive a message, and in all
    other cases both of them will. Whenever an process receives the message
    $\langle\occurred{e},h\rangle$, it knows that $e$ occurred but does not
    know whether $k=h$ or~$h+1$. In particular, upon receiving
    $\langle\occurred{e},0\rangle$, process~0 considers it possible that~1
    received nothing and does not know that~$e$ occurred.
\end{example}

We now show that arbitrarily deeply nested knowledge can be obtained in this
setting within a single time step, without common knowledge arising:

\begin{lemma}
\label{lem:con}
    In the context of Example~\ref{ex:Conway}, let $r\in R=\Rrep(P,{\gammax}')$, let
    $G=\{0,1\}$ and assume that the event~$e$, consisting of  the receipt of an
    external input by~$s$ at time~$t$, in~$r$. If the initial value of~$s$
    in~$r$ is~$k$ then \[(R,r,t+1)\sat E_G^k\occurred{e}~\wedge~\neg
    C_G\occurred{e}.\]
\end{lemma}

\begin{proofL}
	We split the proof into two parts, handled by Lemmas~\ref{lem:ub}
	and~~\ref{lem:lb}. Assume that $e$ occurs in~$r$ at time~$t$ as stated, and
	that the initial value is~$k$. By Lemma~\ref{lem:ub} we have that $(R,r,t+1)\sat
	E_G^k\occurred{e}$ and by Lemma~\ref{lem:ub} that $(R,r,t+1)\sat \neg
	E_G^{k+1}\occurred{e}$. Since $\sat \neg  E_G^{k+1}\varphi\imp \neg
	C_G\varphi$ is a validity, the latter implies that $(R,r,t+1)\sat \neg
	C_G\occurred{e}$, and the claim holds.
	\renewcommand{\toto}{lem:con}
\end{proofL}

\begin{lemma}\label{lem:ub}
	The conditions of Lemma~\ref{lem:con} imply that $(R,r,t+1)\sat
	E_G^k\occurred{e}$.
\end{lemma}

\begin{proofL}
	Observe that by the structure of the protocol,
	$\langle\occurred{e},d\rangle$ messages are sent only if~$e$ indeed takes
	place. Thus, for both processes $i\in\{0,1\}$ it is the case that if~$i$
	receives a message of the form $\langle\occurred{e},d\rangle$ at time~$t+1$
	in~$r$ with any value $d\ge 0$, then $(R,r,t+1)\sat K_i\occurred{e}$.

	By convention, we define $E^0_G\varphi=\varphi$. We prove by induction on
	$k\ge 0$ that if the initial value of~$s$ in~$r$ is~$h\ge k$ and~$e$ occurs
	at~$(r,t)$, then $(R,r,t+1)\sat E_G^k\occurred{e}$. In particular, this
	implies that $(R,r,t+1)\sat E_G^k\occurred{e}$ in the case $h=k$,
	establishing the claim. We consider two case. 
	\begin{description}
		\item[$k=0:$] By assumption, $e$ occurs at time~$t$ in~$r$, and thus
		$(R,r,t+1)\sat \occurred{e}$, and by definition of $E_G^0$ also
		$(R,r,t+1)\sat E_G^0\occurred{e}$.
		\item[$k>0:$] In this case, process $i={\tt parity}(h)$ receives the
		message $\langle\occurred{e},h\rangle$, and the other process $j=1-i$
		receives $\langle\occurred{e},h-1\rangle$. According to the protocol, a
		message $\langle\occurred{e},d\rangle$ is received if the initial value
		is either~$d$ or~$d+1$, and hence at least as large as~$d$. Both
		processes thus know that the initial value is at least as large as
		$h-1$. Since $h\ge k$ by assumption, and by the inductive hypothesis we
		have that  $(R,r,t+1)\sat E_G^{k-1}\occurred{e}$ whenever $h\ge k-1$, it
		follows that both $(R,r,t+1)\sat K_i E_G^{k-1}\occurred{e}$ and
		$(R,r,t+1)\sat K_j E_G^{k-1}\occurred{e}$. Hence, $(R,r,t+1)\sat
		E_G^{k}\occurred{e}$ and we are done.
	\end{description}
	\renewcommand{\toto}{lem:ub}
\end{proofL}

\begin{lemma}\label{lem:lb}
	The conditions of Lemma~\ref{lem:con} imply $(R,r,t+1)\sat \neg
	E_G^{k+1}\occurred{e}$.
\end{lemma}

\begin{proofL}
	First notice that, in every run $r'\in R$, a process that does not receive
	a message of the form $\langle\occurred{e},d\rangle$ does not know that~$e$
	occurred, since there is another run~$r''\in R$ in which its local history
	is identical to then one in~$r'$, and where the event does not occur. We can
	now prove the claim by induction of~$k$. If $k=0$ then process~1 does not
	receive a $\langle\occurred{e},d\rangle$ by time~$t+1$. Thus, $(R,r,t+1)\sat
	\neg K_1\occurred{e}$ and so $(R,r,t+1)\sat \neg E_G^1\occurred{e}$, as
	claimed.

	Let $k>0$ and assume inductively that the claim holds for $k-1$. By
	definition of the protocol, process $i={\tt parity}(k)$ receives the message
	$\langle\occurred{e},k\rangle$, and the other process $j=1-i$ receives
	$\langle\occurred{e},k-1\rangle$. There is a run~$r'\in R$ in which
	$r'_j(t+1)=r_j(t+1)$ and the initial value is~$k-1$. It follows that
	$(R,r,t+1)\sat \neg K_jE_G^k\occurred{e}$, and thus $(R,r,t+1)\sat \neg
	E_G^{k+1}\occurred{e}$, and we are done.
	\renewcommand{\toto}{lem:lb}
\end{proofL}

We note that the epistemic structure obtained here is similar to that which
arises in the electronic mail game of Rubinstein~\cite{Rubinstein1989}, and in
the coordinated attack problem \cite{HM1}.  One distinguishing feature is that
in our example here the high degree of nested knowledge is obtained in one
step, with two messages, whereas a long interactive exchange of~$k$ messages is
required to achieve $k$ levels of nesting in the other cases. A similar
epistemic structure also arises in the analysis of the initial states of the
{\em muddy children} puzzle~\cite{FHMV}, or of the {\em Conway
paradox}~\cite{Con}.

\section{Conclusions}\label{sec:CKG_conclusions}
Taking a step beyond nested knowledge, this chapter develops the theory
needed in order to characterize common knowledge gain, an epistemic state
that is only possible in synchronous systems \cite{HM1}. We define the \emph{centibroom},
a simpler, tighter, communication structure than the centipede, and prove the Common Knowledge
Gain Theorem that validates the centibroom's causal nature. We then show that
the centibroom is also necessary in solution to the \emph{Simultaneous Response} problem.

Based on  the $\fip$, first introduced in Chapter~\ref{ch:NKG}, it is shown that centibrooms
are also sufficient for common knowledge gain. We then utilize this result to determine sharp 
thresholds regarding when nested knowledge becomes common knowledge under $\fip$. 
Finally, Example~\ref{ex:Conway} shows that this phenomenon is not universal to all protocols. 
A protocol exists in which no depth of nested knowledge must imply common knowledge.

\ifx \isEmbedded \undefined
\bibliography{../../z}


\maketitle

\else
\fi
\label{ch:generalized_ordering}

\section{Introduction}\label{sec:intro_GO}
Previous chapters have studied coordination under various restrictions:
linear ordering of responses, simultaneous responses, or a linear
ordering of sets of simultaneous responses. In this chapter
we remove all structural restrictions on ordering and consider
systems where the required ordering of the responses is given by  any
non-particularized partial order.

As we will see, this ultimate generalization does not spawn yet
more intricate causal structures and epistemic states. Rather,
solutions require multiple instances of the (already defined) 
generalized centipedes to exist in triggered runs.

We start with a concrete, if simple, example. Consider the 
following case, describing the production process for the
\emph{Munchy Crunchy} chocolate bar.

\begin{example}\label{ex:go_ex1}
	\emph{Charlie's Chocolate Factory} produces all kinds of chocolate, based on
	distributed processes that control various machines and manufacturing
	stages. The \emph{Munchy Crunchy} is one of the chocolate bars manufactured
	in the plant.  Those processes involved in its production are visualized in
	Figure~\ref{fig:munchyl}. There are 10 different distributed processes involved
	in the manufacturing, arranged into 2 initiating singleton clusters 
	and 3 multi-process clusters. Assume that the underlying network
	graph relating all of the processes is full, and that it contains many
	other processes besides those shown in the figure.

	\begin{figure}[!h]
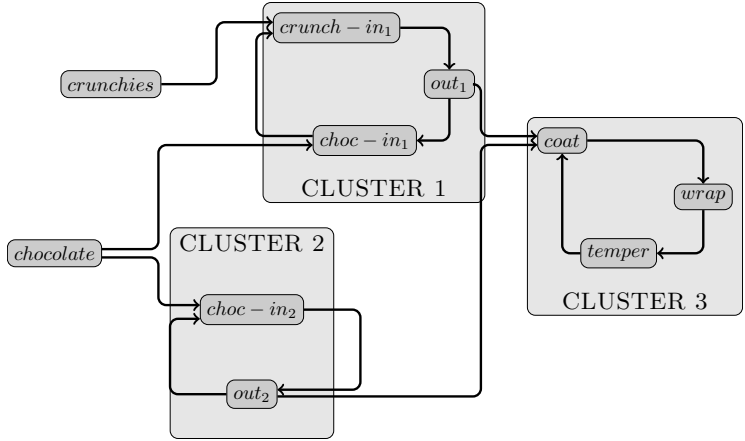

		\centering
			\newcommand{\tikzfname}{tikz_munchy}
			\tikzsetnextfilename{\cfcurrentfolder Figures/\tikzfname}
			\scalebox{0.75}{\cfinput{Tikz/\tikzfname}}	
		\caption{Production process for  \emph{Munchy Crunchy} chocolate bar}
		\label{fig:munchyl}
	\end{figure}
	
	The figure describes the required manufacturing process, with arrows
	signifying activation order. Two processes, each a singleton cluster,
	initiate manufacturing by sending chocolate and crunchies into
	the system. Each of these processes is controlled by a human operator.
	\emph{Cluster} 1 contains processes that control the input and output valves
	of a mixing bowel that mixes together chocolate and crunchies. These valves
	must all be opened simultaneously (this is visualized by a cycle in the
	ordering graph), but only if both chocolate and crunchies are being streamed
	into the system. \emph{Cluster} 2 controls another mixing bowl, with only
	one input and one output valve. Here pure chocolate for the bar's coating is
	blended with unhealthy chemicals. Again, both valves must operate together,
	but only if chocolate is being streamed in. Finally, the processes in
	\emph{Cluster} 3 control the coating, tempering and wrapping machines. These
	too must start to work simultaneously, but only if the mixing bowls are 
	sending out their blends.
\end{example}

The required manufacturing process described above goes beyond the problem formulations
we have thus far seen. We are seeing not only requirements for linear ordering and for 
simultaneity, but also multiple triggers, and events that are causally dependent
upon multiple independent causes.  

We now define a class of problems for which the requirements  
graph shown in Figure~\ref{fig:munchyl} would be an instance.
Note that the graph  can also be expressed as 
a partial order $\angles{V^{crunchy},\preceq^{crunchy}}$ defined on a set of events, where

\begin{itemize}
	\item[]$V^{crunchy}=\{chocolate, crunchies, choc-in_{1}, crunch-in_{1}, 
	out_{1}, choc-in_{2}, out_{2}, coat, temper, wrap\}$, and
	\item[] the partial order $\preceq^{crunchy}$ is defined for every pair edge-connected
	pair of processes.
\end{itemize}•

In order to fully express the graph, we need to add to the partial
order a distinction between those triggering events that are spontaneous
external inputs and those that are responses to such triggers. 
In Example~\ref{ex:go_ex1}, the set of triggers is ] $T^{crunchy}=\{chocolate, crunchies\}$.
We formalize requirements such as the one given above, in the following way .

\begin{definition}[Generalized Response Problem]
	An instance of the generalized response problem is defined by a tuple $\GR=\angles{V,T, \preceq}$
	where
	\begin{enumerate}
		\item $V$ is a set of events,
		\item $T\subseteq V$ is the set of ND external inputs in $V$, and
		\item $\preceq$ is a partial order on $V$, such that  every 
		$\tau\in T$ is $\preceq$ minimal (i.e. for every $\tau\in T$ and $e\in V$, 
		if $e\preceq \tau$ then $e=\tau$).
 	\end{enumerate}•
	
	A protocol~$P$ solves the instance $\GR=\angles{V,T,\preceq}$ of the {\em
	Generalized Response} problem if it guarantees that in every run $r$,
	\begin{enumerate} 
		\item if $e\preceq e'$ occur at $t$ and $t'$
		respectively, then $t\leq t'$. Moreover,
		\item for every $e'\in V$, $e'$ occurs in $r$ iff for every $e\preceq e'$, $e$
		occurs in $r$ .
	\end{enumerate} 
\end{definition}•

Consider a protocol $P^{crunchy}$, that solves the instance of $\GR$ defined
by the tuple $\angles{V^{crunchy},T^{crunchy},\preceq^{crunchy}}$. 
Our concern in this chapter is to characterize the necessary causal structures 
that must obtain in executions of protocols solving instances of $\GR$, such as
the protocol $P^{crunchy}$.

\section{Condensed Representation of $\GR$}\label{sec:GR_condensed}
Each of the clusters in Example~\ref{ex:go_ex1} contains a cycle, 
while the manufacturing requirements specify that the events in each cluster 
must occur simultaneously. 
The next lemma shows the existence of a cycle does indeed guarantee simultaneity
in protocols that solve $\GR$.

\begin{lemma}\label{lem:sim_strong_comp}	
	Let $P$ be a protocol solving $\GR=\angles{V,T,\preceq}$. Fix $e,e'\in V$
	such that $e\preceq e'$ and $e'\preceq e$. \\
	If $e$ occurs at $(r,t)$ then $e'$ occurs at $(r,t)$ too.
\end{lemma}

\begin{proofL}
	From $e'\preceq e$ and definition of $\GR$, there must be some $t'\leq t$
	such that $e'$ occurs at $(r,t')$. Again from definition and from $e\preceq
	e'$, we also get that $t\leq t'$. Hence $t'=t$.
\end{proofL}

The $\GR$ formalization can be used to designate any required order of events,
but often, as in the case of Example~\ref{ex:go_ex1}, it is more sensible to
consider the strongly connected components in the graph as single units. Thus
for every instance $\angles{V,T,\preceq}$ of $\GR$, we consider the condensed
form $\angles{V',T',\preceq'}$, which is derived by collapsing every strongly
connected component in $\angles{V,\preceq}$ into a single vertex, or component,
$C\in V'$. 

The partial order $\preceq'$ holds between $C$ and $C'$ iff there exist $e\in C$
and $e'\in C'$ such that $e\preceq e'$. Since for every $\tau\in T$ if $e\preceq
\tau$ for some $e\in V$ then $e=\tau$, the subset of triggering components is
$T'=\bigcup_{\tau\in T}\{t\}\subseteq V'$. We will blur the distinction between
$\{\tau\}\in T'$ and $\tau\in T$ freely and freely interchange between the two
forms.

Thus the condensed form $\angles{V',T',\preceq'}$ may be considered as an
instance of $\GR$ in its own right. In fact, we will even speak of a component
$C\in G'$ as ``occurring'' at time $t$, if all of its member events are
simultaneously occurring at that time. Condensed forms of directed graphs enjoy
the desirable property of containing no cycles. This property makes it easier
for us to work with them than with the original instance of the problem, and
following lemma shows that there is no harm done, as the two forms are
equivalent with respect to protocol solutions.

%

\begin{lemma}\label{lem:gr_iff_con}
	Let $\GR=\angles{V,T,\preceq}$ and let $\angles{V',T',\preceq'}$ be the condensed form of $\GR$.
	Protocol $P$ solves $\GR$ iff it solves the condensed form.
\end{lemma}

\begin{proofL}
	\begin{description}
		\item[$\Rightarrow$] Assume that $P$ solves $\GR$. Fix run $r$.
		\begin{enumerate}
			\item Fix $C\in V'$ and $e,e'\in C$ such that $e$ occurs at $(r,t)$.
			By definition of $C$ we have that $e'\preceq e \preceq e'$. 
			Hence it must be that $e'$ occurs at $(r,t)$ too.
			\item Fix $C,C'\in V'$ such that $C\preceq' C'$ and $C$ and $C'$ occur at $t$ and $t'$
			respectively. Then there exist $e\in C$ and $e'\in C'$ that occur at $(r,t)$ and $(r,t')$
			respectively. By definition of $\preceq'$ we get that $t\leq t'$.
			\item Fix $C'\in V'$ and $\{C_{h}\}_{h=1..k}$, the set of components
			such that $C_{h}\preceq C'$ for all $h$. Let $e'\in C'$ and
			$e_{h}\in C_{h}$ for all $h$. By definition of $\preceq'$
			$e_{h}\preceq e'$ for all $h$. Since $P$ solves $\GR$ we get that
			$e'$ occurs iff $e_{h}$ occurs for all $h$, and hence $C'$ occurs
			iff $C_{h}$ occurs for all $h$.
		\end{enumerate}•
		
		\item[$\Leftarrow$] The arguments pretty much repeat those in the
		other direction.
	\end{description}•
\end{proofL}

\section{Generalized Ordering Requires Multiple Generalized Centipedes}\label{sec:gr2gcenti}
The ordering requirement formalized by a generalized ordering problem
$\angles{V,T,\preceq}$ expresses the idea that an event $e\in V$ should occur
iff a set of prerequisite actions had already been performed. These in turn will
have their own set of prerequisites, etc. The next lemma reformulates the
prerequisites for the occurrence of such an event in terms of chains of linear
orderings. Focussing on condensed forms, we are assured that there are no
cycles in the graph, and hence no infinite chains to reckon with.
Of particular interest for us when considering solutions to $\GR$
are \emph{component chains}, defined below.

\begin{definition}[Component chains]
	Let $\angles{V',T',\preceq'}$ be the condensed form of $\GR$. A
	\emph{component chain} for  $C\in V'$ is a sequence
	$\angles{C_{0},C_{1},..,C_{k}}$ of alternating members of $V'$ such that
	$C_{0}\in T'$, $C_{h}\preceq C_{h+1}$ for all $h<k$, and $C_{k}=C$.
\end{definition}

Given $\GR=\angles{V,T,\preceq}$, we say that
$\angles{C_{0},C_{1},..,C_{k}}$ is a component chain for $e\in V$ if $e\in C$
for some $C\in V'$ where $\angles{V',T',\preceq'}$ is the condensed form of $\GR$,
and $\angles{C_{0},C_{1},..,C_{k}}$ is a component chain for $C$.

\begin{lemma}\label{lem:comp_chains}
	Suppose that protocol $P$ solves  $\angles{V',T',\preceq'}$, the condensed form of $\GR$.
	Fix $r\in \Rmax(P,\gammax)$ and let $C\in V'$ be a component that occurs at $(r,t)$.
	Then for every component chain $\angles{C_{0},C_{1},..,C_{k}}$ for $C$
	there exist $t_{0}\leq t_{1} \leq \cdots \leq t_{k}=t$ such that 
	for every  $h\leq k$, $C_{h}$ occurs at $t_{h}$ .
\end{lemma}

\begin{proofL}
	Suppose that there exists a component chain
	$\angles{C_{0},C_{1},..,C_{k}}$ for $C$ in which the condition does not
	hold. Then there exists some $h\leq k$ where one of the following hold
	
	\begin{description}
		\item[$C_{h}$ does not occur in $r$:] In this case, as $P$ solves 
		$\angles{V',T',\preceq'}$, component $C$ also does not occur in $r$,
		contrary to the assumption.

		\item[there exists some $h'>h$ such that $t_{h}>t_{h'}$:] In this case,
		as $C_{h}\preceq' C_{h'}$ it must be that $C_{h'}$ does not occur at all
		in $r$, and hence also that $C_{h}$ does not occur, and the case is
		reduced to the previous one, and thus to a contradiction.
	\end{description}•
\end{proofL}

Lemma~\ref{lem:comp_chains} reduces the necessary requirements for the
occurrence of a component $C$ into a set of linearly ordered requirements - one
for each component chain  leading back from $C$. Recalling the Ordered Group
Response problem from chapter~\ref{ch:ordered_group} and the notion of weakly
solving, we note that each of the linearly ordered requirements just mentioned
is in fact a requirement of the protocol that it weakly solve an instance of the
$\OGR$ problem that is specified by the component chain.

The following theorem formalizes this insight, giving us the necessary condition,
in causal terms, for correct solutions to the $\GR$ problem.

\begin{theorem}\label{thm:GR2gcenti}	
	Let $\GR=\angles{V,T,\preceq}$ and let $P$ be a protocol that solves $\GR$.
	Fix $r\in \Rmax(P,\gammax)$, and suppose that event $e\in V$ occurs at $(r,t)$.\\
	Then for each component chain $\angles{\{\tau\},C_{1},..,C_{k}}$ of $e$,
	there exists a generalized centipede $\angles{\theta_{0},I^{1},\ldots,I^{k}}$ in $(r,t_{0}..t_{k})$
	where $I^{h}$ and $t_{h}$ are the set of processes where events of $C_{h}$ occur and their time
	of occurrence respectively, and $\theta_{0}$ is the node	where $\tau$ occurs.
\end{theorem}•

\begin{proofT}
	By Lemma~\ref{lem:gr_iff_con} $P$ solves $\GR$ iff it solves the condensed form  
	$\angles{V',T',\preceq'}$. By Lemma~\ref{lem:comp_chains} in every run where $e$ occurs,
	for every initial component chain $\angles{C_{0}=\{\tau\},C_{1},..,C_{k}}$ for $C$,
	there exist $t_{0}\leq t_{1} \leq \cdots \leq t_{k}=t$ such that $C_{h}$
	occurs at $t_{h}$ for all $h\leq k$. As the occurrences of $C_{h}$ for all $h\leq k$
	are necessary whenever component $C$ occurs, we have by definition that
	$P$ weakly solves the $\OGR$ instance defined by $\angles{\tau, C_{1},..,C_{k}}$.
	By Theorem~\ref{thm:OGRgcpede}, there must exist a generalized centipede for
	$\angles{I^{1},\ldots,I^{k}}$ in $(r,t_{0}..t)$.
\renewcommand{\toto}{thm:GR2gcenti}
\end{proofT}

\begin{figure}[!h]
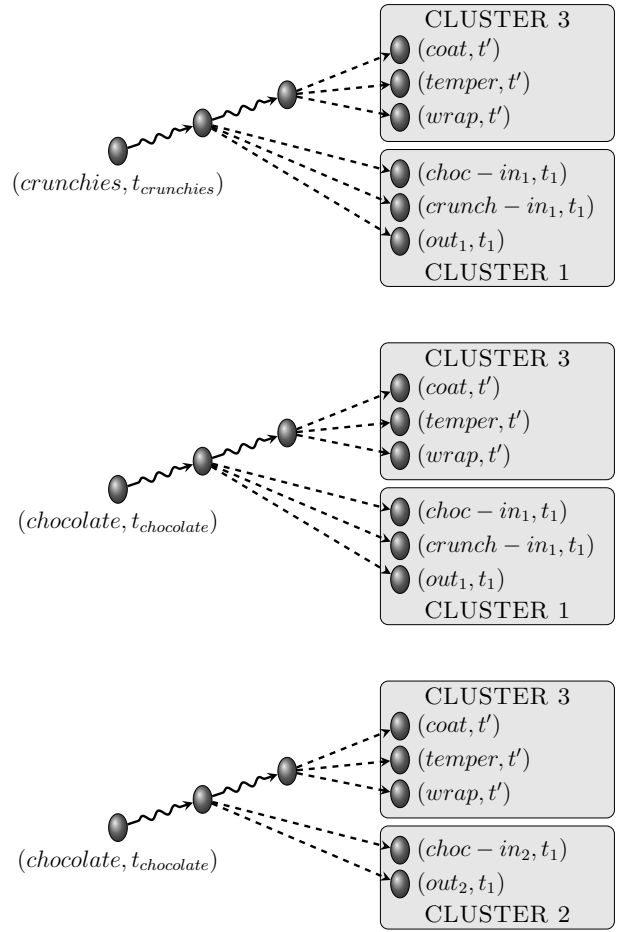

	\centering
		\newcommand{\tikzfname}{tikz_munchy2}
		\tikzsetnextfilename{\cfcurrentfolder Figures/\tikzfname}
		\scalebox{0.75}{\cfinput{Tikz/\tikzfname}}	
	\caption{Generalized centipedes in the \emph{Crunchy Munchy} production line}
	\label{fig:munchy2}
\end{figure}

In a protocol that solves an instance $\angles{V,T,\preceq}$ of $\GR$ then, whenever an event 
$e\in V$ occurs, there exists a set of generalized centipedes - one for each component
chain of $e$. Returning to Example~\ref{ex:go_ex1} this means that in a protocol that 
properly controls the production process for \emph{Crunchy Munchies}, whenever
both chocolate and crunchies are being pushed into the production system, the 
minimal communication between the distributed processes must contain all
three communication structures see in Figure~\ref{fig:munchy2}.

\section{Conclusions}\label{sec:GO_conclusions}
This chapter brings our investigation into the causal structures that 
underly coordination for purpose of event ordering to an end. Using the 
generalized centipede and the notion of weakly solving, we have 
shown that for any prescribed partial ordering on events, the communication
requirements in a protocol that solves the generalized ordering can be characterized
as a set of generalized centipedes. 

We use the \emph{condensed} representation of the ordering graph,
in order to avoid the loops that are created  whenever the ordering
prescribes simultaneous occurrences. Under this representation, vertices represent
the strongly connected components of the original graph. 

As we proved, each maximal path in the condensed graph represents
a requirement for the existence of a generalized centipede in every run where
the path's ultimate component, or cluster of simultaneous events, occurs.
The generalization of the ordering requirements to allow for multiple triggering
events is translated into communication requirements that include sets of 
such centipedes.

\ifx \isEmbedded \undefined


\maketitle

\else
\fi
\label{ch:ignorance}

\section{Introduction}\label{sec:intro}
This chapter takes a different, complementing, look at the way transmission
bounds affect knowledge and causality in distributed systems. In place of
studying the effects of upper bounds in such systems, we will now focus upon
lower bounds on transmission times, and how \emph{these} affect knowledge gain.
Intuitively, the existence of lower bounds makes it possible for one process to
gain knowledge of another process's ignorance respecting  a recent event.

Such considerations seem to make more sense in an environment that is motivated
by competition, rather than cooperation.  In his book {\em The Rothschilds}
\cite{Morton1998},  Frederic Morton gives an account of the events in the
London Stock exchange at the time of the Battle of Waterloo, in which knowledge
about lower bounds on transmission times supposedly played a major role.
Morton's (disputed) account can be summarized as follows: on the night of June
15, 1815. Nathan Rothschild, one of London's most prominent financiers at the
time, was informed by his special private couriers that the Battle of Waterloo
was won by the British. Official word by Wellington's men could only arrive on
the next day. On the next morning, Rothschild went to the London Stock exchange,
and signaled his agents to furiously sell consuls (government bonds). ``{\it He
knows...[who won]}$\,$" was the word among traders. The market crashed, and just
before Wellington's men arrived with the news of victory, Rothschild signaled
his agents to buy all available consuls, at a fraction of their original price.
He is said to have made a fortune on that day.

For the course of events described by Morton to be plausible, not only was it
necessary for Rothschild to know about the outcome before everyone else. He also
had to know that the others were {\em ignorant} of the outcome. Otherwise, he
would fear that one of his rivals could out-smart him, gradually buy his shares
and make out with a huge gain at Rothschild's expense. The epistemic circumstances 
in this example are based on Rothschild's courier system being known to have
lower minimal transmission times than that of Wellington's
communication lines.

The Battle of Waterloo example above illustrates the importance of knowledge
about other's ignorance in particular circumstances. For another example,
consider a sealed-bid first-price auction for mining rights. Suppose that near
the auction closing a potential bidder learns of a relevant event~$e$, say that
gold was found in an adjacent site. The bidder's valuation of the auctioned
rights may have changed. But the decision regarding if, and by what amount, to
alter her bid would depend on her knowledge about whether  her competitor knows
about~$e$. In particular, if she knows that he is ignorant of~$e$, then she
should not increase her bid by a significant amount. The analysis presented will serve to
show how our favored bidder can use her information about transmission times to
figure out whether her competitor is ignorant of~$e$.

Our analysis will start by presenting a novel view of how bounds on message
transmission times in a communication network induce  {\em causal cones} of
information flow among events in the system, in analogy with the light cones in
Einstein-Minkowski spacetime considered in physics \cite{Feynman,Naber2003}.%
\footnote{Our setting can be thought of as consisting of a single inertial
system, in which there is a single, non-relativistic,  notion of time for all
sites.}

Based on this probing into causal cones, we will develop the formal
theory of knowledge of ignorance. As mentioned above, such considerations
are more naturally understood in the context of  a competitive environment. For this
reason this chapter will not introduce any kind of cooperative ordering task
to motivate the analysis.

\section{Bounded Communication and Cones of Influence}\label{sec:cones}
Consider a fixed inertial system in which all sites are at rest with each other.
In such a setting, light rays carry information at a constant speed~$c$ in
Euclidean space. In terms of Einstein-Minkowski spacetime, the light rays outgoing from
an event (or a 4-dimensional point~$p$) form a surface in spacetime called the
event's future light cone. The light rays converging on an event form a surface
called the event's past light cone. The spacetime points within $p$'s future
light cone make up its {\em absolute future} and those within its past light
cone make up its {\em absolute past}: the former are spacetime points  that
events at~$p$ can influence and the latter are the points  can influence $p$.
Events at points outside both light cones of $p$ can neither influence nor be
influenced by events at~$p$. Such events are considered {\em independent of}, or
sometimes called {\em concurrent with} events at~$p$. Observe that the absolute
future and absolute past cones of a point~$p$ are fixed and depend only on the
coordinates of~$p$.

In analogy, consider a computer network based on a specific context $\gammab$
where for every channel $i\mapsto j$, there is a fixed transmission time:
$min_{ij}=max_{ij}$. Moreover, assume that the processes follow the {\em
full-information} protocol $\fip$ in which, at every instant, they send a
message describing their whole history to all neighbors. 
With fixed transmission rates and constant message sending, we would
get that in every run $\theta\spc\theta'$ iff $\theta\tspc\theta'$.
Just as in the case of light traveling in Einstein-Minkowski spacetime, in this
setting every node~$\theta$ would define a future cone
$\fut(\theta)=\{\theta'|\theta\tspc\theta'\}$ and a past cone
$\pas(\theta)=\{\theta'|\theta'\tspc\theta\}$, as well as nodes that are
causally concurrent with respect to~$\theta$. In this section we focus primarily
upon future causality, where an intricate dynamics transforms potentiality into
necessity, as we shall soon see.

\begin{figure}[h!]
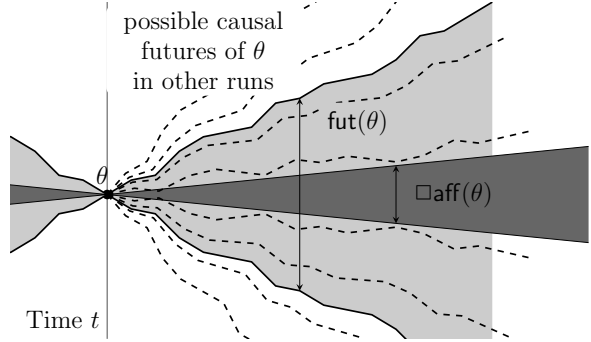

	\begin{center}
		\newcommand{\tikzfname}{causal_cones1}
		\tikzsetnextfilename{\cfcurrentfolder Figures/\tikzfname}
		\scalebox{0.8}{\cfinput{Tikz/\tikzfname}}
		\caption{The future causal cones of $\theta$ in asynchronous systems and $\gammax$}
		\label{fig:cones-1}
	\end{center}
\end{figure}

What happens when transmission times are {\em not} fixed? In purely asynchronous
settings, where $max_{ij}=\infty$ and thus messages can take arbitrarily long to
be delivered, a node $\theta'$ can be influenced by $\theta=(i,t)$ only if
$\theta\Lam\theta'$. Thus, Lamport's $\Lam$ relation defines a future cone (and
a past cone) for every given node. As opposed to the fixed-transmission system
described above, however, here the  cone may differ significantly  between
different runs due to the varying transmission times. Figure~\ref{fig:cones-1}
shows  the future cone of node $\theta$ in a specific run, for an  observer with
complete information about the future. The alternative futures that remain
unrealized in the current run are shown in outline. Observe that a ``core'' cone
can be made out in the center of $\fut(\theta)$, of nodes that are guaranteed 
{\em a priori} to be within $\fut(\theta)$, and will thus necessarily be
affected by~$\theta$. We denote this cone by $\necaff(\theta)$. In an
asynchronous context, this core consists of the set of nodes $(i,t')$ such that
$t'\geq t$.

The picture becomes more interesting in the presence of upper bounds $max_{ij}$ on message
transmission times. Recall that we denote by $\Dis{ih}$ the shortest distance
between vertices~$i$ and~$h$ in the $max$-weighted network graph. Under the
$\fip$ described above, we are guaranteed to have $(i,t)\Lam(h,t')$ whenever
$t'\ge t+\Dis{ih}$. Thus, maximal transmission times extend the inner cone into
$\necaff(\theta)=\{\theta'|\theta\tspc\theta'\}$. As in the asynchronous case, for
every run $r\in\Rrep(\fip,\gammax)$ and node~$\theta$, necessarily
$\fut(\theta)\supseteq\necaff(\theta)$, as messages that are delivered earlier
than at the upper bounds on a channel introduce into $\fut(\theta)$ nodes that
were not guaranteed  {\em a priori} to be in  $\necaff(\theta)$.
  
We may also consider the set $\necunaff(\theta)$, counterbalancing 
$\necaff(\theta)$, and consisting of nodes that are necessarily unaffected causally by
$\theta=(i,t)$. As long as no lower bounds are defined, this set consists of all nodes
in $\theta$'s \emph{temporal} past, as well the nodes $(j,t)$ where $j\neq
i$.\footnote{We assume that messages between processes are not instantaneous and
spend at least one time unit in transmission.} 

\begin{figure}[h!]
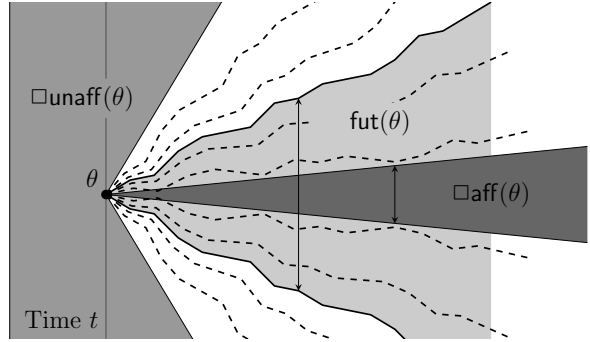

	\begin{center}
		\newcommand{\tikzfname}{causal_cones2}
		\tikzsetnextfilename{\cfcurrentfolder Figures/\tikzfname}
		\scalebox{0.8}{\cfinput{Tikz/\tikzfname}}
		\caption{The future causal cones of $\theta$ in  $\gammab$}
		\label{fig:cones-2}
	\end{center}
\end{figure}

When we move to the contexts $\gammin$ or based on $\gammab$, in which 
there are lower bounds on transmission times, $\necunaff(\theta)$ gets a richer structure. 
Lower bounds on transmission play a related, albeit somewhat different role than
that of upper bounds. Suppose that a spontaneous event~$e$ takes place at
$\theta=(i,t)$ and that, based on the lower bounds, the fastest that
communication from~$i$ can reach~$j$ is $\dis{ij}$.%
\footnote{In analogy to the definition of the $\Dis{ij}$ values, 
$\dis{ij}$ is defined as the shortest distance between $i$ and~$j$ in the $min$-weighted network graph.}  
If $\theta'=(j,t')$ where
$t'<t+\dis{ij}$, then events at $\theta'$ cannot be causally influenced by~$e$. It
follows that that  the $\necunaff(\theta)$ region is now defined as the set
$\{(j,t'):t'< t+\dis{ij}\}$. Figure~\ref{fig:cones-2} shows the
causal cones of $\theta$ in $\gammab$.

We have considered the sets $\fut(\theta), \necaff(\theta)$ and
$\necunaff(\theta)$, which are all easily determined 
given complete information regarding the run's infinite execution.
To be of practical use however, we should consider whatever
it is that can be made known about causal influence, given a specific
``present'' point in time $t'$  and assuming that future events in the run 
are as yet undetermined. We define $\fut(\theta,t')$ as the set
$\{(j,t'')|\theta\Lam (j,t'')\text{ and }t''\leq t'\}$, the set of nodes that have, by time $t'$,
already been realized as a part of  $\theta$'s future. 

The portion of the run realized  by time~$t'$ determines the sets of necessarily
affected and unaffected nodes  relative to the current time, in a way that
extends them beyond $\necaff(\theta)$ and $\necunaff(\theta)$, respectively. The
set $\necaff(\theta, t')$ of all nodes that are guaranteed to be causally
affected by $\theta$ given $\fut(\theta,t')$, is the union of the
$\necaff(\theta')$ cones of all $\theta'\in\necaff(\theta,t')$. As we already
have $\theta\Lam(k,\bar t)$, this suffices to ensure that $(j,t'')\in
fut(\theta)$. We denote with  $\posunaff(\theta,t')$ the set of nodes that are
potentially unaffected by $\theta$ relative to current time $t'$. This set is
the complement of the set  $\necaff(\theta,t')$.

A more challenging definition is that of the set $\posaff(\theta,t')$ of nodes
that, at time $t'$,  are potentially affected by  $\theta$. A node $\theta'$ is
potentially affected if it is possible, given $\fut(\theta,t')$, that the
current run will evolve so as to include $\theta'$ in $\fut(\theta)$. This set
is inductively defined: If $\theta=(i,t)$ then $\posaff(\theta,t)=\{(j,t'')|
t''\ge t+\dis{ij}\}$, and    $\posaff(\theta,t')=\bigcup_{(k,{\bar t})\in
\posaff(\theta,t'-1)}\{(j,t'')| t''\ge \bar t+\dis{kj}\}$ for $t'>t$. The set of
necessarily unaffected nodes $\necunaff(\theta,t')$ is  the complement of
$\posaff(\theta,t')$.

\begin{figure}[h!]
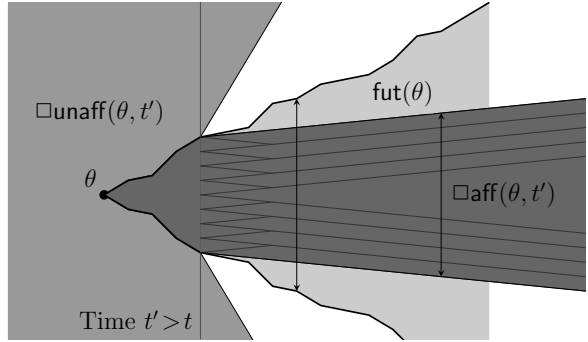

	\begin{center}
		\newcommand{\tikzfname}{causal_cones3}
		\tikzsetnextfilename{\cfcurrentfolder Figures/\tikzfname}
		\scalebox{0.8}{\cfinput{Tikz/\tikzfname}}
		\caption{The necessarily affected and unaffected regions by $\theta$ in  $\gammab$, w.r.t.\  time $t'>t$}
		\label{fig:cones-3}
	\end{center}
\end{figure}

Observe that with time, as larger portions of the run get realized, the set of nodes that are 
neither necessarily affected by $\theta$ nor necessarily unaffected by it, 
given by $\posaff(\theta,t')\cap \posunaff(\theta,t')$, monotonically shrinks.
This can visualized by comparing the state of the cones in Figure~\ref{fig:cones-2} with 
that of Figure~\ref{fig:cones-3}, that displays the same run at a later point in time.
It is the case that at time $t'$
every node in the time interval $[t,t']$ is either in  $\necaff(\theta,t')$ or in $\necunaff(\theta,t')$.
Moreover, the $\necaff(\theta,t')$ cone and $\necunaff(\theta,t')$ region grow monotonically grow with $t'$.

In summary, while light cones define fixed regions of influence and concurrency, 
communication dynamically determines the cones of influence and their complements.

\section{Transmission Guarantees and Knowledge of Ignorance}
\label{sec:igno}

Cones of influence and information flow as discussed in the previous section are
clearly closely related to knowledge about knowledge and to knowledge about
ignorance. In this section we build on the cones interpretation to analyze the
dynamics of what would probably be best termed as ``knowledge gain about
ignorance''.

For the following analysis we introduce some variations in the formal language
that is used in the proofs. In place of logical operators whose validity is
dependent upon system, run and time (the $(\Rrep,r,t)$ at the left hand side of
the \emph{satisfies} operator $\sat$), this chapter utilizes \emph{timestamped}
operators that are dependent only upon a system and a specific run.

The set~$\Phi$ of primitive propositions consists of the propositions
$\tocc{t}{e}$ for all events~$e$ and times $t$, and the propositions
$\theta\forLam\theta'$ for all pairs of process-time nodes. The logical
language~$\Lang$ is obtained by closing $\Phi$ under propositional connectives
and knowledge formulas. We write $\theta\not\forLam\theta'$ instead of
$\neg(\theta\forLam\theta')$. Our knowledge operators are indexed by a node
$\theta=(i,t)$, and so are time stamped. Thus, $\Phi\subset\Lang$, and if
$\varphi\in\Lang$, $i\in\Proc$ and $t$ is a time, then
$K_{(i,t)}\varphi\in\Lang$. The formula $K_{(i,t)}$ is read {\em process $i$ at
time $t$ knows~$\varphi$}.

We write ~$(R,r)\sat\varphi$ ~to state that $\varphi$ holds in the run~$r$, with
respect to system~$R$. We write $r\sim_{(i,t)}r'$ whenever process~$i$'s local
state at time~$t$ in~$r$ is identical to it's local state at time $t$ in run
$r'$, and inductively define

\begin{description} 
\item[$(R,r)\sat{\theta\forLam\theta'}$] \qquad  ~iff ~$\theta\Lam\theta'$ in the run~$r$; 
\item[$(R,r)\sat\tocc{t}{e}$]  iff the event~$e$ 
occurs in~$r$ by time $t$;
and
\item[$(R,r)\sat K_{(i,t)}\varphi$]\qquad  \:iff~$(R,r')\sat \varphi$
  for every run~$r'$ satisfying $r\sim_{(i,t)}r'$;
\end{description}

Propositional connectives are handled in the standard way, and their
clauses are omitted above. Despite the slight variance in nomenclature,
$K_{(i,t)}\varphi$ still follows~\cite{FHMV} in being  satisfied if
$\varphi$ holds at all points at which~$i$ has the same local state as
it does at time $t$. Thus, given~$R$, the local state determines what
processes know. Note that $(R,r)\sat\tocc{t}{e}$ holds iff 
$(R,r,t')\sat\atn{t}{\occurred{e}}$ for any time $t'$. Similarly, 
$(R,r)\sat K_{(i,t)}\varphi$ iff $(R,r,t')\sat\atn{t}{K_{i}\varphi}$ for any time $t'$.
So for the most part, this chapter's formal semantics is but an adaptation of those
introduced in Chapter~\ref{ch:model}.

The motivation here is twofold. First, the timestamped language allows us 
to simplify the presentation. Second, The use of timestamped epistemic operators
and the introduction of the Lamport relation into the formal language implies 
greater expressivity that we hope will ferment new insights into the study of 
causation in distributed systems.

Our analysis here will be performed within the context $\gammin$, in which lower bounds on 
message transmission times are available. 
Recall that in $\gammin$ there are 
no upper bounds on message transmission times;
messages can take an arbitrarily long 
amount of time to be delivered. In these settings,
Lemma~\ref{lem:async_cpede} tells us that nested knowledge implies a message
chain linking the processes. The converse, shown below in Lemma~\ref{lem:gamin_suff_KG},
states that under $\fip$, such a message chain implies nested knowledge.

\begin{lemma}\label{lem:gamin_suff_KG}
	Let~$\Rrep=\Rrep(\fip,\gammin)$.
        Assume
	$e$ is an ND event occurring at $(i_0,t)$ in~$r$.
	If there is a chain
        $(i_0,t)\Lam(i_1,t_1)\Lam\cdots\Lam(i_k,t_k)$ in $\rmmp$,
	then 
	$(\Rrep,r,t')\sat K_{i_k}K_{i_{k-1}}\cdots K_{i_1}\ndocc{e}$.
\end{lemma}

\begin{proofL}
	 The proof is arrived at by first noting that Lemma~\ref{lem:fip1} implies
	 the following for $\gammin$: if $(\Rrep,r,t)\sat K_i\varphi$
	 and~$(i,t)\Lam(j,t')$ in~$r$, then $(\Rrep,r,t')\sat
	 K_j(\atn{t}{K_{i}\varphi})$. Then, repeated applications of this result
	 give us the required outcome.
\end{proofL}

In the rest of this section, we will focus on how different cones of influence
combine to determine when an process knows that another process is ignorant
about an event of interest. We will give a complete characterization of this
question for the $\fip$ and draw implications from this to the general case of
arbitrary protocols.

Recall the sealed-bid first-price auction described in the Introduction. Our
bidder is named $i_2$, her competitor is $i_1$, and the bids need to be in by
time~$t_1$. Moreover, $i_2$ must decide on her bid at time~$t_2$. Finally, the
event~$e$ in which information about a newly found gold mine was disclosed
occurred at $\theta_0=(i_0,t_0)$. The goal, then, is to determine whether
$K_{\theta_2}\neg K_{\theta_1}\tocc{t_0}{e}$.

Given that processes following $\fip$ have the perfect recall property, the
following lemma shows that the knowledge state of a process determines its
causal past.
 
\begin{lemma}\label{lem:causal_tr}
	Fix $r\in \Rrep(\fip,\gamma^{min})$ and nodes $\theta_{0},\theta_{1}, \theta_{2}$ such that $i_{0}\neq i_{1}$.\\
	$(\Rrep,r)\sat K_{\theta_{2}}\forLam(\theta_{0},\theta_{1})$ \qquad iff  \qquad
	 $\theta_{0}\Lam\theta_{1}\Lam\theta_{2}$ in $r$.
\end{lemma}•

\begin{proofL}
\begin{description}
	\item[$\Rightarrow$] If $\theta_{0}\not\Lam\theta_{1}$ then $(\Rrep,r)\nsat
	K_{\theta_{2}}\forLam(\theta_{0}, \theta_{1})$ by the Knowledge Axiom,
	immediately contradicting the lemma's assumptions.

	Suppose now that $\theta_{0}\Lam\theta_{1}$ but that
	$\theta_{1}\not\Lam\theta_{2}$. Since  $i_{0}\neq i_{1}$, agent $i_{1}$ must
	receive a message at some point $t'_{1}\in(t_{0},t_{1}]$. Since in
	$\gamma^{min}$ every message receive is a nondeterministic event, using
	Lemma~\ref{lem:async_cpede} we get that $(\Rrep,r)\nsat
	K_{\theta_{2}}\forLam(\theta_{0},\theta_{1})$. By definition of $\sat$ there
	exists a run $r'\sim_{\theta_{2}}r$ such that $(\Rrep,r')\sat \neg
	\forLam(\theta_{0},\theta_{1})$, and hence
	$\theta_{0}\notin\pas(r',\Lam,\theta_{1})$, again leading to contradiction.
	
	\item[$\Leftarrow$] $(\theta_{0},\theta_{1})$ is an edge in
	$\pas(r,\Lam,\theta_{2})$. From perfect recall we get that
	$(\theta_{0},\theta_{1})$ is an edge in $\pas(r',\Lam,\theta_{2})$ for all
	$r\sim_{\theta_{2}}r'$. By definition of $\pas$, $\theta_{0}\Lam\theta_{1}$
	in every $r'$, and therefore $(\Rrep,r)\sat K_{\theta_{2}}\forLam
	(\theta_{0},\theta_{1})$ .
\end{description}•
\end{proofL}

As discussed in Chapter~\ref{ch:model}, the meaning of the lower bounds
$min_{ij}$  in the $\Net$ labeled graph component of the context is that certain
message chains, in which messages travel faster than the lower bounds specify,
are impossible. We say that a sequence of nodes
$\theta_0,\theta_1,\ldots,\theta_m$ is a {\em legal message chain} with respect
to $\Net$ if for every $h<m$ we have (i) $t_h<t_{h+1}$ and (ii) if $i_h\ne
i_{h+1}$ then $i_h\mapsto i_{h+1}$ is a channel in~$\Net$, and $(t_{h+1}-t_h)\ge
\min_{i_hi_{h+1}}$. Clearly, for every legal message chain, there is a run of
$\gammin$ with network $\Net$ in which this message chain is realized, and
$\theta_0\Lam\theta_1\Lam\cdots\Lam\theta_m$. Conversely, if $\theta\Lam\theta'$
in a run~$r$, then there is a legal message chain starting at~$\theta$ and
ending at~$\theta'$, that is a causal chain in~$r$.

By Lemma~\ref{lem:async_cpede} we have that $K_{\theta_2}\neg
K_{\theta_1}\tocc{t_0}{e}$ will hold if $\theta_2$ knows that
$\theta_0\not\forLam\theta_1$ in the {\em current run}. We now formalize the
required conditions, based on causal cones and legal message chains.

\begin{definition}[Set of legal paths]
	We denote by $\Legal{\theta_0}{\theta_1}$ the set of legal message chains
	starting at~$\theta_0$ and ending at $\theta_1$.
\end{definition}

\begin{figure}[h!]
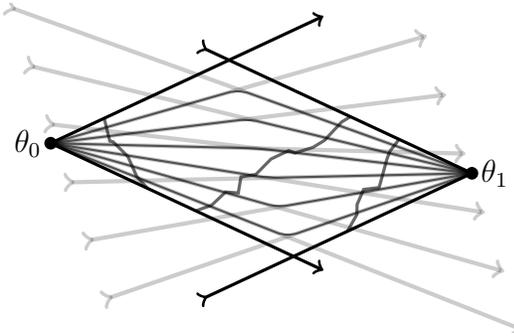

	\begin{center}
		\newcommand{\tikzfname}{pot_graph}
		\tikzsetnextfilename{\cfcurrentfolder Figures/\tikzfname}
		\scalebox{1}{\cfinput{Tikz/\tikzfname}}
		\caption{The set $\Legal{\theta_0}{\theta_1}$ of possible chains from $\theta_0$ to $\theta_1$}
		\label{fig:pg}
	\end{center}
\end{figure}

$\Legal{\theta_0}{\theta_1}$ consists of all message chains that are both
within $\posaff(\theta_0)$ of nodes that can possibly be affected
by~$\theta_0$, and in the analogous region of nodes that can possibly
affect~$\theta_1$.  See Figure~\ref{fig:pg}.

\begin{definition}[Cut]\label{def:cut}
	A $\theta_{0}\theta_{1}$-cut is a set of nodes $C$ that appear in the paths of 
	$\Legal{\theta_{0}}{\theta_{1}}$ that intersects every path 
	in $\Legal{\theta_{0}}{\theta_{1}}$.
		
	\noindent The cut $C$ is called $\theta_0${\em -clean} in run $r$
	if $\theta_0\not\Lam c$ in~$r$, for every $c\in  C$.
	\end{definition}	

Figure~\ref{fig:pg} depicts three different $\theta_{0}\theta_{1}$ cuts. 

Lemma~\ref{lem:causal_tr} tells us that in order for an process to be
informed of any communication link between two distinct sites, the site
on the receiving side must be in the process's past causal cone. The above
discussion suggests that the existence of a clean cut, in this cone, on
the paths between these sites is of importance. Moreover, we should be
looking for cuts that are somehow more ``recent''. The following
definition picks up on this intuition.
\begin{definition}[Causal front]
	Fix  nodes $\theta_{0},\theta_{1}, \theta_{2}$. 
	The causal front of $\theta_{0}\theta_{1}$ with respect to $\theta_{2}$ in run~$r$, denoted by 
	$\cf{\theta_{2}}{\theta_{0}}{\theta_{1}}$, 	is the set  of nodes 
	$$
	\{\,\phi\,|~ \mbox{$\phi$ is on some chain in $\Legal{\theta_0}{\theta_1}$ and }
	\exists \Psi\in \Legal{\phi}{\theta_{1}}\text{ s.t. }\Psi\cap \pas(\theta_{2})=\{\phi\}\}$$
\end{definition}

Let $\Psi$ be a legal message chain connecting between $\theta_{0}$ and
$\theta_{1}$ that is also, at least in part, within the scope of
$\pas(\theta_{2})$. By definition of $\cf{\theta_{2}}{\theta_{0}}{\theta_{1}}$,
it will contain a ``latest contact point'' of $\theta_{2}$ with the nodes of
$\Psi$.  So, as far as $i_{2}$ knows at time $t_{2}$, it is possible that
$\psi\Lam\theta_{1}$. Now if it is also the case that $\theta_{0}\Lam\psi$, then
a communication path between $\theta_{0}$ and $\theta_{1}$ has been established.
There is a certain subtlety involved in the definition. The fact that  $(i,t)$
is in $\cf{\theta_{2}}{\theta_{0}}{\theta_{1}}$ does not mean that $(i,t')$ is
not in the front for $t'>t$. We can still have $(i,t')\in
\cf{\theta_{2}}{\theta_{0}}{\theta_{1}}$ for some $t'> t$, if each of the nodes
$(i,t)$ and $(i,t')$ constitutes a latest contact point for some potential path
to $\theta_{1}$.

\begin{figure}[h!]
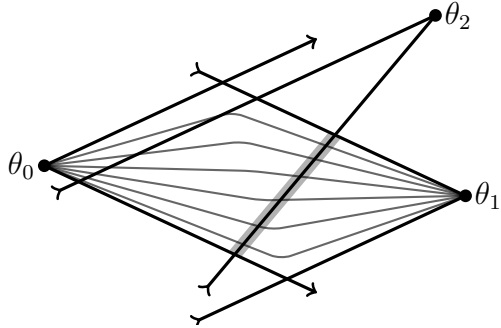

	\begin{center}
		\newcommand{\tikzfname}{causal_front}
		\tikzsetnextfilename{\cfcurrentfolder Figures/\tikzfname}
		\scalebox{1}{\cfinput{Tikz/\tikzfname}}
		\label{fig:causal_front}
		\caption{Thick marking gives a schematic view of  $\cf{\theta_{2}}{\theta_{0}}{\theta_{1}}$}
	\end{center}
\end{figure}

We are now ready to characterize knowledge of ignorance in $\gammin$, by showing 
that it reduces to existence of a ``$\theta_0$-clean'' cut in the causal front:

\begin{theorem}\label{thm:IG_suff}
	Let $r\in \Rrep(\fip,\gammin)$ and 
	denote $F=\cf{\theta_{2}}{\theta_{0}}{\theta_{1}}$. 
	Then $(\Rrep,r)\sat K_{\theta_{2}} (\theta_0\not\forLam\theta_1)$ 
	iff both (a) $F$ is $\theta_0$-clean, and (b) $F$ is a $\theta_{0}\theta_{1}$-cut.
\end{theorem}

\begin{proofT}
	\begin{description}
		\item[$\Rightarrow:$] Suppose, to the contrary, that $F$ is either not 
		$\theta_0$-clean, or is not a  $\theta_{0}\theta_{1}$-cut.
		Choose a run $r'\in \Rrep$ such that 	
		\begin{itemize}
			\item $\pas(r',\theta_{2})=\pas(r,\theta_{2})$, and where
			\item all messages sent and delivered outside $\pas(r,\theta_{2})$
			have minimal transmission times. 
		\end{itemize}
	
		That such a run exists is given by $\Rrep$ being a representing
		system and by the non-dependence of
		nondeterministic events on the past of the run in which they occur.
		By Lemma~\ref{lem:past-local} we get that $r\sim_{\theta_{2}}r'$,
		and hence that $(\Rrep,r')\sat \theta_0\not\forLam\theta_1$.
		Moreover, as $\pas(r',\theta_{2})=\pas(r,\theta_{2})$ we also
		have that $\cf{r'}{\theta_{2}}{\theta_{0}}{\theta_{1}}=\cf{\theta_{2}}{\theta_{0}}{\theta_{1}}=F$.
		We now have two choices:
		\begin{description}
			\item[$F$ is not a  $\theta_{0}\theta_{1}$-cut:] Then there exists
			$$\Psi=\angles{\theta_{0}=\psi_{0}\ntLam\psi_{1}\cdots\ntLam\psi_{n}=
			\theta_{1}}\in \Legal{\theta_0}{\theta_1}$$ such that
			$\Psi\cap \pas(r',\theta_{2})=\emptyset$. By definition of
			$r'$ we get that $\psi_{0}\Lam\psi_{1}\cdots\Lam\psi_{n}$ in $r'$,
			and hence that $(\Rrep,r')\sat \theta_{0}\forLam\theta_{1}$,
			contradiction.
			\item[$F$ is not $\theta_0$-clean:] In this case there exists
			$$\Psi=\angles{\theta_{0}=\psi_{0}\ntLam\psi_{1}\cdots\ntLam\psi_{n}
			=\theta_{1}}\in \Legal{\theta_0}{\theta_1}$$ and some $k< n$
			such that $\psi_{0}\Lam \psi_{k}$ and $\angles{\psi_{k+1}\cdots
			\psi_{n}}\cap \pas(r',\theta_{2})=\emptyset$. Again by
			definition of $r'$ we get that $\psi_{k+1}\Lam\cdots\Lam\psi_{n}$
			in $r'$. We obtain that $\psi_{0}\Lam \psi_{n}$ and hence that
			$(\Rrep,r')\sat \theta_{0}\forLam\theta_{1}$, again contradicting
			the assumption.
		\end{description}•

		\item[$\Leftarrow:$] Suppose that $(\Rrep,r)\nsat K_{\theta_{2}}
		\theta_0\not\forLam\theta_1$. Then there exists a run $r'$ such that 
		$r\sim_{\theta_{2}}r'$, where $(\Rrep,r')\sat \theta_0\forLam\theta_1$.
		Let $\Psi=\angles{\psi_{0},\psi_{1},..,\psi_{n}}$ be a sequence such
		that $\theta_{0}=\psi_{0}\ntLam\psi_{1}\cdots\ntLam\psi_{n}=\theta_{1}$
		in $r'$.

		It follows that $\Psi\in\ \Legal{\theta_0}{\theta_1}$. Since $F$ is a
		$\theta_{0}\theta_{1}$-cut, there must exist some $\phi\in F\cap \Psi$.
		Since $\phi\in \Psi$ we get that $\theta_{0}\in \pas(r',\phi)$. Since
		$F$ is a $\theta_{2}$ causal front we have that
		$\phi\in\pas(r,\theta_{2})$, and as $r'\sim_{\theta_{2}}r$ and
		$\gamma^{min}$ is causally traced  we also obtain that
		$\pas(r',\phi)=\pas(r,\phi)$. This gives us that $\theta_{0}\Lam \phi$
		in $r$ too, contradicting the assumption that $F$ is $\theta_0$-clean in
		$r$.
	\end{description}•
\renewcommand{\toto}{thm:IG_suff}
\end{proofT}

Theorem~\ref{thm:IG_suff} characterizes knowledge of non-causality 
under $\fip$ in a system with lower bounds on transmission times. 
Based on the knowledge gain theorem, we can translate this into conditions 
on when an process will know that another process is ignorant of the occurrence of an 
event of interest. Consider an event $e_0$ that can occur only at~$i_0$. 
We are interested in when $K_{\theta_{2}}\neg K_{\theta_{1}}\tocc{t_{1}}{e_{0}}$ holds.
Clearly, if $i_2$ knows that $e_0$ did not take place, then it would know that $i_1$ 
does not know that $e_0$ took place. 

Theorem~\ref{thm:IG_suff} provides a condition enabling knowledge at $\theta_2$
that $\theta_0\not\forLam\theta_1$. Suppose that $\theta_0=(i_0,t_0)$. Since
$\Lam$ is transitive, however, $\theta_0\not\Lam\theta_1$ implies that
$\theta'\not\Lam\theta_1$ for all $\theta'=(i_0,t')$ with $t'>t_0$. So, by
Lemma~\ref{lem:async_cpede}, $\theta_1$ could not have knowledge that $e_0$
happened at any time after $t_0$ too. Combining these observations, we are able
to obtain a tight characterization of knowledge about ignorance regarding the
occurrence of a nondeterministic event:

\begin{theorem}[Knowledge of Ignorance Theorem]
\label{thm:IG_nec}
	Let $r\in \Rrep(\fip,\gammin)$, fix a node $\theta_2$, and let $e_0$ 
	be a nondeterministic $i_0$-event.  
	Let $t'$ be the latest time for which   
	$(\Rrep,r)\sat K_{\theta_{2}} \neg \tocc{t'}{e_{0}}$ holds, and denote $\theta_{0}=(i_{0},t'+1)$. Then\\
	 $(\Rrep,r)\sat K_{\theta_{2}}\neg K_{\theta_{1}}\tocc{t_{1}}{e_{0}}$ ~~iff~~ 
	$\cf{\theta_{2}}{\theta_{0}}{\theta_{1}}$ is a $\theta_0$-clean, $\theta_{0}\theta_{1}$-cut.
	\end{theorem}

\begin{proofT}
	\begin{description}
		\item[$\Rightarrow:$] We will prove the counter-position. Suppose that
		$F$ is not a $\theta_{0}\theta_{1}$-cut or it is not is
		$\theta_{0}$-clean. Note that in particular, this means that
		$\theta_{0}\pot\theta_{1}$ and hence that $t_{0}\leq t_{1}$. As $t_{0}$
		is the latest time for which $(\Rrep,r)\sat K_{\theta_{2}} \neg
		\tocc{t_{0}-1}{e_{0}}$ holds, there must exist a run
		$r'\sim_{\theta_{2}}r$ where $e_{0}$ occurs at $(i_{0},t_{0})$. As
		$\gamma^{min}$ is causally traced we get that
		$\pas(r',\theta_{2})=\pas(r,\theta_{2})$ and hence that
		$\cf{r'}{\theta_{2}}{\theta_{0}}{\theta_{1}}=\cf{\theta_{2}}{\theta_{0}}
		{\theta_{1}}=F$. Theorem~\ref{thm:IG_suff} now shows that
		$(\Rrep,r')\nsat K_{\theta_{2}}\theta_0\not\forLam\theta_1$. So there
		must exist a run $r''\in \Rrep$ such that $r''\sim_{\theta_{2}}r'$,
		where $(\Rrep,r'')\sat \theta_0\forLam\theta_1$. As the processes are
		following $\fip$ we get, using Lemma~\ref{lem:gamin_suff_KG}, that
		$(\Rrep,r')\sat K_{\theta_{1}}\tocc{t_{0}}{e_{0}}$. Since $t_{0}\leq
		t_{1}$ we get $(\Rrep,r')\sat K_{\theta_{1}}\tocc{t_{1}}{e_{0}}$.
		Finally,  since $r''\sim_{\theta_{2}}r'\sim_{\theta_{2}}r$, we get that
		$(\Rrep,r)\nsat K_{\theta_{2}}\neg K_{\theta_{1}}\tocc{t_{1}}{e_{0}}$,
		contradicting our assumptions.
		\item[$\Rightarrow:$] We will prove the counter-position. Suppose that
		$F$ is not a $\theta_{0}\theta_{1}$-cut or it is not is
		$\theta_{0}$-clean. Note that in particular, this means that
		$\theta_{0}\pot\theta_{1}$ and hence that $t_{0}\leq t_{1}$. As $t_{0}$
		is the latest time for which $(\Rrep,r)\sat K_{\theta_{2}} \neg
		\tocc{t_{0}-1}{e_{0}}$ holds, there must exist a run
		$r'\sim_{\theta_{2}}r$ where $e_{0}$ occurs at $(i_{0},t_{0})$. As
		$\gamma^{min}$ is causally traced we get that
		$\pas(r',\theta_{2})=\pas(r,\theta_{2})$ and hence that
		$\cf{r'}{\theta_{2}}{\theta_{0}}{\theta_{1}}=\cf{\theta_{2}}{\theta_{0}}
		{\theta_{1}}=F$. Theorem~\ref{thm:IG_suff} now shows that
		$(\Rrep,r')\nsat K_{\theta_{2}}\theta_0\not\forLam\theta_1$. So there
		must exist a run $r''\in \Rrep$ such that $r''\sim_{\theta_{2}}r'$,
		where $(\Rrep,r'')\sat \theta_0\forLam\theta_1$. As the processes are
		following $\fip$ we get, using Lemma~\ref{lem:gamin_suff_KG}, that
		$(\Rrep,r')\sat K_{\theta_{1}}\tocc{t_{0}}{e_{0}}$. Since $t_{0}\leq
		t_{1}$ we get $(\Rrep,r')\sat K_{\theta_{1}}\tocc{t_{1}}{e_{0}}$.
		Finally,  since $r''\sim_{\theta_{2}}r'\sim_{\theta_{2}}r$, we get that
		$(\Rrep,r)\nsat K_{\theta_{2}}\neg K_{\theta_{1}}\tocc{t_{1}}{e_{0}}$,
		contradicting our assumptions.
		
		\item[$\Leftarrow:$] Choose an arbitrary $r'\in\Rrep$ such that
		$r\sim_{\theta_{2}}r'$. We consider three options for the occurrence of
		event $e_{0}$: \begin{itemize} \item•$e_{0}$ does not occur in run $r'$:
		in this case we get, in particular, that $(\Rrep,r')\sat \neg
		K_{\theta_{1}}\tocc{t_{1}}{e_{0}}$. \item $e_{0}$ occurs before time
		$t_{0}$: in this case we obtain a contradiction to the theorem's
		assumption that $(\Rrep,r)\sat K_{\theta_{2}} \neg
		\tocc{t_{0}-1}{e_{0}}$. \item $e_{0}$ occurs at some time $t'\geq
		t_{0}$: from $r\sim_{\theta_{2}}r'$ and Lemma~\ref{lem:causal_tr} we get
		that
		$\cf{r'}{\theta_{2}}{\theta_{0}}{\theta_{1}}=\cf{\theta_{2}}{\theta_{0}}
		{\theta_{1}}=F$. Theorem~\ref{thm:IG_suff} is now used to show that
		$(\Rrep,r')\sat K_{\theta_{2}} \theta_0\not\forLam\theta_1$, and thus
		that $(\Rrep,r')\sat  \theta_0\not\forLam\theta_1$. By definition of
		$\Lam$ we also get that $(i_{0},t')\not\Lam\theta_{1}$. Using
		Lemma~\ref{lem:async_cpede} we conclude that $(\Rrep,r')\sat \neg
		K_{\theta_{1}}\tocc{t_{1}}{e_{0}}$. \end{itemize}•

		We showed that $(\Rrep,r')\sat \neg K_{\theta_{1}}\tocc{t_{1}}{e_{0}}$
		for all $r\sim_{\theta_{2}}r'$. By definition  of $\sat$ we get that
		$(\Rrep,r)\sat K_{\theta_{2}}\neg K_{\theta_{1}}\tocc{t_{1}}{e_{0}}$, as
		required.
	\end{description}•
\renewcommand{\toto}{thm:IG_nec}
\end{proofT}

\section{Conclusions}\label{sec:conc}
While in timing-based algorithms such as clock-synchronization algorithms
\cite{Halpern1985, Attiya1996} lower bounds on transmission times are typically
of limited impact, our thesis in the current chapter is that lower bounds play a
crucial role in determining knowledge about ignorance. This, in turn, can be of
value in player's considerations in non-cooperative settings. In this chapter we
characterized when knowledge about ignorance is obtained in runs of the
full-information protocol, in the presence of lower bounds. A natural question
involves characterizing knowledge of ignorance for general protocols, or in
strategic settings in which a player has uncertainty concerning other players'
strategies. Our results have natural implications about more general settings:
if Alice knows that even under the full-information protocol Bob cannot know
about Charlie, then she may be able to conclude the same even under lesser
communication. But the analysis required for the general question is more
subtle, since Alice could hear from intermediate points without having full
knowledge of what information they have. There is considerable room for further
exploration of this point.

This chapter also draws an analogy between the causal cones that are formed by
information in synchronous systems with bounds, and the notion of causal
light-cones in physics. The invariance of the speed of light causes the causal
cone of a given point in 4-dimensional Einstein-Minkowski spacetime to be fixed
{\em a priori} and not change as time proceeds. In contrast, in the digital
space of communication networks, upper bounds induce a region of points that are
definitely affected by a spontaneous event occurring at a given point, while
lower bounds define a region of points guaranteed to {\em not} be affected. For
a given point $\theta=(i,t)$, these regions grow with time, converging at the
end of time to form the set of points actually affected by~$\theta$. We used
this view to motivate our analysis of knowledge of ignorance. We believe that
further study of the causal cones and their evolution over time will provide
insights into the fundamental properties of synchronous environments.

\ifx \isEmbedded \undefined
\bibliography{../../z}


\maketitle

\else
\fi
\label{ch:conclusions}
This thesis investigates causality and coordination in distributed systems. It
extends Lamport's original paper on causality \cite{Lamclocks} by looking into
causality in synchronous systems. As we saw in Chapter~\ref{ch:NKG}, our results
provide a generalization of Lamport's work, in the sense that asynchronous
networks are modeled as systems with infinite upper transmission bounds.

Our results show that while the dissemination pattern of asynchronous causality
is a fairly straightforward extension of the happened-before relation into
message chains, synchronous causality spreads in rather complex patterns that
combine message deliveries and timing guarantees.

Formally, our study is based on  knowledge-based analysis~\cite{FHMV}. As it
turns out, the notion of knowledge provides a very close  formal approximation
of causal influence, which also underlies temporal precedence. In fact, the
various forms of temporal event orderings that we examine are each reduced to
corresponding epistemic conditions. The two basic kinds being linear ordering,
extended to nested knowledge, and simultaneous ordering, which is reduced to a
common knowledge requirement. In Chapter~\ref{ch:generalized_ordering} these two
basic ordering types are combined, providing a characterization of the causal
pattern requirement for any given partial ordering on events.
Chapter~\ref{ch:ignorance} introduces a complementing approach, studying the
minimal requirements needed for a process to an ensure that an event at a remote
site has \emph{not} yet transpired.

The current study opens up many possible venues for further research. An
immediate concern would be to further extend the study of causality into weaker
systems, such as systems with clock drifts, or systems where the processes have
only partial knowledge of the bounds on communication. In such networks two
learning dynamics intertwine. First, the dynamics covered in this thesis, that
explain how information about events in the current run are spread in the
system. A second process of information flow concerns the gradual learning of
the processes about the communication characteristics of the underlying network.
Relevant information to be gained here for a process is not only what the actual
bounds are, but also what other processes may have learned about these bounds.

More generally, the model suggested here makes pretty strong synchrony assumptions.
Weakening the model by removing the global clock, reducing the available knowledge
about network characteristics, and allowing failures, is necessary in order to bring
our results closer to real world applications.

Another salient extension to system characteristics would be to consider mobile
networks. Here, given the graveness of energy considerations and the multi-hop
nature of communication, our characterizations of minimal communication
requirements may be highly relevant. In such systems though, we expect not only
the bounds to vary throughout a run, but also the number and identity of
participating processes. Some causal analyses have been
suggested~\cite{Prakash2004, Baquero98}, but they follow the classical
asynchronous paradigm.

Another direction for further investigation would be to consider the 
finer-grained patterns that are introduced if we consider protocol-specific
knowledge. The thesis follows the approach of Chandy and Misra~\cite{ChM},
in charting out the communication that is \emph{universally} necessary for knowledge gain.
It is clear that once a specific protocol is considered, the set of communication patterns 
that lead to knowledge gain is reduced. First, it may be that process $i$
does not gain any information regarding an occurrence at $j$ despite the existence
of a causal connection, simply because the messages relating the two processes 
do not convey this fact (a typical case is a NULL message, that carries very limited 
information content). Moreover, processes may delay message relaying, or may be
committed to a specific communication channel despite the existence of several
alternatives. A protocol-specific characterization of knowledge gain, and hence
of temporal ordering, would take all such protocol-specified limitations  on
communication into account. We would expect  tighter necessity conditions 
for knowledge gain here. 

Lamport's definition for the \emph{concurrency} of two events $e$ and $e'$ 
in~\cite{Lamclocks} is that $e\not\Lam e'$ and $e'\not\Lam e$. Further inquiry
about  concurrency in synchronous systems is also in place. Of course, given
the global clock, the obvious candidate for concurrency is that both $e$ and $e'$
occur at exactly the same time. However, a more subtle approach may be 
appropriate here. Despite the possibility of actually confirming the exact time of
occurrence of events, in many cases the concurrency  of two events is actually
accidental. What we are really after is a notion of ``temporal independence'':
that events $e$ and $e'$ may occur at any temporal ordering in relation to each other.
Such an inquiry may serve to extend the measure of parallelism in a protocol
by identifying and uncoupling simultaneous occurrences that should
really be temporally independent.

Several possible extensions of the thesis may be of less relevance to the
distributed systems community, referring instead to the study of 
multi-agent systems under other disciplines. As we mentioned in Chapter~\ref{ch:ignorance},
one such case is the study of ``knowledge of ignorance''. For one process to be able to
tell that another process is unaware of a certain occurrence seems to be highly
relevant for systems which are of a competitive nature, as studied in game theory.
The existing results presented here are by and large incomplete, as we do not know 
what more complex epistemic states involving ignorance, such as
$K_{i_{3}}\neg K_{i_{2}}\neg K_{i_{1}}\ndocc{e_{0}}$, would require in terms
of (dis)communication. Another possible research direction is the relaxation of
knowledge into belief. Formal systems involving belief are ubiquitous in game theory
and also in the general study of multi-agent systems~\cite{Baltag99, Stalnaker96}.
Do the causal patterns that characterize knowledge gain also characterize the spread 
of belief? We do not currently know.

Finally, our results as presented give a ``snapshot'' of the communication 
pattern that must exist, if knowledge has been gained. They insinuate that the
dynamics of information flow is such that information moves ``outward'' from the
site of occurrence of an ND event. Yet we do not prove that such is the case.
We have began to extend our results in this direction too. Here some subtle
considerations must be made in order to accommodate information ``flow''.
Also, as it turns out, one must consider the possibility that certain information,
say fact $\vphi$, is actually spread as separate packets, each containing a fraction of the
required information. 

All in all, we believe (and hope) that synchronous causality, along with the new
concepts and methodology presented in this thesis, will turn out to be a fruitful
advancement in the field of distributed computing, as well as for multi-agent systems in general.

\ifx \isEmbedded \undefined